\documentclass[12pt,a4paper]{article}
\usepackage{setspace}
\usepackage[english]{babel}
\usepackage{amsfonts,amsmath,amssymb,graphicx,xcolor,here,hyperref,accents,color,float,subfigure,booktabs,multirow,rotating,ragged2e,tabularx,threeparttable}
\usepackage{natbib} \bibpunct{(}{)}{;}{a}{,}{,} 
\usepackage{tikz}
\usepackage{enumerate}

\newtheorem{theorem}{Theorem}

\newtheorem{definition}[theorem]{Definition}
\newtheorem{example}[theorem]{Example}

\newtheorem{lemma}[theorem]{Lemma}

\newtheorem{proposition}[theorem]{Proposition}
\newtheorem{remark}[theorem]{Remark}

\newenvironment{proof}[1][Proof]
{\textbf{#1.} }{\ \rule{0.5em}{0.5em}}

\def\qed{\hfill \vrule height 6pt width 6pt depth 0pt}

\allowdisplaybreaks[3]
\newcommand{\dif}{\mathrm{d}}
\def\EE{\mathbb{E}}
\def\PP{\mathbb{P}}

\def\Lbrb#1{\lbrace #1 \rbrace}

\DeclareMathAccent{\wtilde}{\mathord}{largesymbols}{"65}
\hypersetup{
    colorlinks,
    linkcolor={red!50!black},
    citecolor={blue!50!black},
    urlcolor={blue!80!black},
    pdfborder={0 0 0}
}

\numberwithin{theorem}{section}

\usepackage{footnote}
\usepackage[affil-it]{authblk}
\usepackage[english]{babel}
\usepackage{blindtext}

\begin{document}

\title{\vskip -1cm \textbf{On Joint Marginal Expected Shortfall and Associated Contribution Risk Measures}}
\author[a]{Tong Pu}%
\author[a]{Yifei Zhang}
\author[a]{Yiying Zhang\thanks{Email: \texttt{zhangyy3@sustech.edu.cn}. The corresponding author.}$^,$}
\affil[a]{Department of Mathematics, Southern University of Science and Technology, Shenzhen 518055, P.R. China.}
\date{Accepted in Quantitative Finance}
\maketitle

\begin{abstract}
Systemic risk is the risk that a company- or industry-level risk could trigger a huge collapse of another or even the whole institution. Various systemic risk measures have been proposed in the literature to quantify the domino and (relative) spillover effects induced by systemic risks such as the well-known CoVaR, CoES, MES and CoD risk measures, and associated contribution measures. This paper proposes another new type of systemic risk measure, called the joint marginal expected shortfall (JMES), to measure whether the MES of one entity's risk-taking adds to another one or the overall risk conditioned on the event that the entity is already in some specified distress level. We further introduce two useful systemic risk contribution measures based on the difference function or relative ratio function of the JMES and the conventional ES, respectively. Some basic properties of these proposed measures are studied such as monotonicity, comonotonic additivity, non-identifiability and non-elicitability. For both risk measures and two different vectors of bivariate risks, we establish sufficient conditions imposed on copula structure, stress levels, and stochastic orders to compare these new measures. We further provide some numerical examples to illustrate our main findings. A real application in analyzing the risk contagion among several stock market indices is implemented to show the performances of our proposed measures compared with other commonly used measures including CoVaR, CoES, MES, and their associated contribution measures.

    \noindent
    \\[1mm]
\end{abstract}
\baselineskip17.5pt
\thispagestyle{empty}


\section{Introduction}\label{introduction}
The Financial Stability Board (FSB) defines systemic risk as ``\textit{the risk of disruption to the flow of financial services that is (i) caused by an impairment of all or parts of the financial system; and (ii) has the potential to have serious negative consequences for the real economy}''. In other words, systemic risks refer to the potential threats that can disrupt or destabilize an entire financial system, often originating from interconnectedness and interdependencies among various institutions or markets.

Examples of systemic risks include financial market crashes, widespread bank failures, contagion effects that spread across multiple sectors, and disruptions in critical infrastructure or technology systems. For example, the global financial crisis of 2008, triggered by the collapse of Lehman Brothers, is an explicit event of systemic risk. The interconnectedness of financial institutions and the proliferation of complex financial products led to a widespread crisis that affected economies worldwide. Another example of systemic risk is the European sovereign debt crisis that began in 2010. The high levels of debt in several European countries, combined with the interconnectedness of European banks, created a contagion effect that threatened the stability of the entire Eurozone and had far-reaching implications for global financial markets.

Since the financial crisis, international policymakers have diligently collaborated to establish a regulatory framework aimed at addressing systemic risk. Initially focusing on the banking industry, their attention later expanded to include the insurance sector. The International Association of Insurance Supervisors (IAIS), entrusted by the FSB, undertook the responsibility of devising a methodology for identifying global systemically important insurers (G-SIIs) and formulating policy measures to mitigate their systemic significance.

Systemic risk is more prevalent in stock markets, encompassing the transmission of risks between individual stocks and the interconnectedness of risks across different markets. Analyzing stock prices grouped into regional indices can provide valuable insights into the influence of one stock market on the other markets, considering regional stock indices serve as indicators of economic activities within countries. Readers with interest may refer to \cite{rodriguez2013systemic} and \cite{abuzayed2021systemic} for relevant studies. In this paper, we shall employ the proposed novel systemic risk measures as well as some well-known ones to quantify and compare the influence and the risk spillover effect of the US stock market on other major stock markets.

To proceed, let us present some useful notations and definitions used in the sequel. We denote by $X$ a random variable (r.v.) with cumulative distribution function (c.d.f.) $F$, and survival function (s.f.) $\overline{F}$. Let $F^{-1}(p)=\inf \{x \in \mathbb{R} \mid F(x) \geq p\}$ be the quantile function (or the generalized inverse) of the c.d.f. $F$, for any value $p \in(0,1)$, and $F^{-1}(0)$ and $F^{-1}(1)$ are defined as the left and right extremes of the support, respectively, that is $F^{-1}(0)=\mbox{ess}\inf (F):=\inf\{x:F(x)>0\}$ and $F^{-1}(1)=\mbox{ess}\sup (F):=\sup\{x:F(x)<1\}$. The quantile function is also known as the \textit{Value-at-Risk} (VaR), and is denoted by $\mbox{\rm VaR}_p[X]\equiv F^{-1}(p)$, for $p \in(0,1)$.

For a fixed $p$, the $\mbox{VaR}_p[X]$ does not provide information about the thickness of the upper tail of the distribution, which is of crucial interest, and for this purpose some other measures have been considered. For example, the \textit{Expected Shortfall} (ES) of $X$  at confidence level $p\in(0,1)$ is defined as
\begin{equation}\label{def:TVaR}
{\rm ES}_{p}[X]= \frac{1}{1-p}\int_{p}^{1}{\rm VaR}_{t}[X]\dif t,
\end{equation}
provided that the integral exists. In some context, if $X$ is a continuous r.v., the ${\rm ES}_{p}[X]$ is also termed as the \textit{Conditional Tail Expectation} (CTE) or the \textit{Tail Value-at-Risk} (TVaR), denoted by ${\rm CTE}_{p}[X]$ and ${\rm TVaR}_{p}[X]$, respectively. VaR and ES are commonly used in determining companies' solvency status and capital requirements under international solvency standards such as Basel III and Solvency II. Recently, the Basel IV\footnote{See detailed in BCBS (2019). \textit{Minimum Capital Requirements for Market Risk. February 2019}. Basel Committee on Banking Supervision. Basel: Bank for International Settlements. https://www.bis.org/bcbs/publ/d457.pdf} revises the internal/advanced model approach by replacing the VaR measure with the ES measure, which further addresses the importance of ES in solvency regulation.

However, conventional risk measures such as VaR and TVaR perform badly in monitoring the spillover effect of systemic risks as they evaluate risks in an isolated way and cannot take into account the interdependence of different entities in the system. In this regard, many systemic risk measures have been proposed in the literature for quantifying and measuring the (relative) spillover effect induced by systemic risks. For example, \cite{Adrian2016} defined the \textit{Conditional Value-at-Risk} (${\rm CoVaR}$) as follows:
\begin{equation*}
    {\rm CoVaR}_{\alpha,\beta}[Y|X]={\rm VaR}_{\beta}[Y|X>{\rm VaR}_{\alpha}[X]],
\end{equation*}
for $\alpha,\beta\in(0,1)$, where $X$ and $Y$ denote the risks of two financial institutions (or one is an institution and the other is the whole system). Later on, the \textit{Conditional Expected Shortfall} (${\rm CoES}$) was introduced by \cite{mainik2014} as follows:
\begin{equation*}
    {\rm CoES}_{\alpha,\beta}[Y|X]=\frac{1}{1-\beta}\int_{\beta}^1{\rm CoVaR}_{\alpha,t}[Y|X]\dif t.
\end{equation*}
\cite{Acharya2017} defined the \textit{Marginal Expected Shortfall} (${\rm MES}$) as follows:
\begin{equation}\label{MES:def}
    {\rm MES}_{\alpha}[Y|X]=\mathbb{E}[Y|X>{\rm VaR}_{\alpha}[X]].
\end{equation}
With regard to extreme risks, some asymptotic studies of ${\rm MES}$ were carried out in \cite{Asimit2011}, \cite{HuaandJoe2014}, and \cite{Asimit2018b}. Besides, numerical methods for calculating ${\rm MES}$ can be found in \cite{Kalkbrener2004}, \cite{glasserman2005}, \cite{targino2015} and the references therein. Recently, \cite{dhaene2022systemic} defined the general class of \textit{Conditional Distortion} (${\rm CoD}$) risk measures, which contains the above three explicit classes of systemic risk measures. \cite{Waltz2022VulnerabilityCoVaRIT} proposed an interesting extension to CoVaR termed the \textit{Vulnerability-CoVaR} (VCoVaR), which is defined as the VaR of a financial system/institution, given that at least one other institution is equal to or below its VaR. They further investigated its properties on the cryptocurrency market and demonstrated that the VCoVaR captures domino effects better than other CoVaR extensions. Interested readers can refer to \cite{Reboredo2015IsTD}, \cite{Hespeler2016MonitoringSR}, \cite{Reboredo2016DownsideAU}, and \cite{Xu2017SystemicRI} for more investigations on applications of various systemic risk measures in insurance and finance.

To measure the effect of $X$'s magnitude on $Y$, some risk contribution measures including ${\rm \Delta CoVaR}$, ${\rm \Delta CoES}$, ${\rm \Delta MES}$ and ${\rm \Delta CoD}$ have been proposed correspondingly; see \cite{Girardi2013},  \cite{Adrian2016} and  \cite{dhaene2022systemic}. As for comparisons of ${\rm CoD}$ and ${\rm \Delta CoD}$ measures, \cite{dhaene2022systemic} provide sufficient conditions to stochastically order two random vectors, which can also be correspondingly applied to ${\rm MES}$, ${\rm CoES}$, ${\rm CoVaR}$, ${\rm \Delta CoES}$ and ${\rm \Delta CoVaR}$. Very recently, \cite{ZhangIME2023} defined three classes of ${\rm CoD}$-based contribution ratio measures and established sufficient conditions for comparing these new measures of two different bivariate risk vectors. It is worth noting that the above-mentioned research works focus on the setting where the stress event is generated by only one risk. Along this line, \cite{ortega2021stochastic} employed stochastic orders to investigate several multivariate measures of risk contagion, including the generalizations of CoVaR and CoES based on multivariate stress events. Interestingly but not surprisingly, some multivariate versions of stochastic orders are adopted for carrying out the comparisons.

As noted in \cite{ji2021tail}, one disadvantage of MES is that when the dependence between $X$ and $Y$ becomes weaker but not as weak as the independent case, the information provided by $X$ does not fully reflect the behavior of $Y$. It was addressed that measures such as ES and MES may underestimate the underlying risk since they ignore the joint effect of dependence and heavy-tailedness. Interested readers can refer to \cite{ji2021tail} for more detailed discussions. To compensate for this, \cite{ji2021tail} proposed a new risk measure called the \textit{Joint Expected Shortfall} (JES) as an alternative to quantifying extreme risks. The JES can be viewed as a consolidation of both ES and MES and has the desirable property of measuring risk by jointly capturing both tail dependence and heavy-tailedness. Specifically, for two risks $X$ and $Y$ and a fixed confidence level $\alpha\in[0,1)$, \cite{ji2021tail} defined JES as
\begin{equation}\label{Def:JES}
    \mathrm{JES}_{\alpha}[Y|X]=\mathbb{E}\left[Y \mid X>\operatorname{VaR}_{\alpha}(X), Y>\operatorname{VaR}_{\alpha}(Y)\right],
\end{equation}
which combines both ES and MES in the sense that the expected loss of $X$ is calculated given the joint information of both $X$ and $Y$ exceeding their corresponding VaR limits. It should be addressed that the JES can only be used in quantifying the effects of systemic risks when the regulator sets the same regulation levels for both risks; however, in many practical scenarios, different regulations might be employed by the regulator to carry out stress tests and backtesting for different risks, especially in the finance sector. For instance, when assessing the level of contribution of the G-SIIs to systemic risk and designing a regulatory framework capable of ensuring financial stability, the international financial regulatory institutions would like to set different confidence levels (e.g., common choices are $\alpha=1\%$, $\alpha=5\%$, or $\alpha=10\%$) for the whole system and a specified individual financial institution. Similar phenomenon can be also noted in the insurance sector. Interested readers can refer to \cite{jackson2002regulatory}, \cite{Girardi2013}, \cite{zimper2014minimal}, and \cite{karimalis2018measuring} for more details.


As a generalization of (\ref{Def:JES}), we propose a similar but new systemic risk measure called the \textit{Joint Marginal Expected Shortfall} (JMES), with its explicit expression given by
\begin{equation*}\label{equa:JMES0}
    {\rm JMES}_{\alpha,\beta}[Y|X]=\mathbb{E}[Y|X>{\rm VaR}_{\alpha}[X],Y>{\rm VaR}_{\beta}[Y]],
\end{equation*}
where $\alpha\in[0,1)$ and $\beta \in [0,1)$. Such a measure can be employed to quantify the joint effect of dependence and heavy-tailedness for two risks with different stress levels. Based on JMES, we propose two new associated contribution risk measures termed as ${\rm \Delta JMES}_{\alpha,\beta}[Y|X]$ and ${\rm \Delta^R JMES}_{\alpha,\beta}[Y|X]$ (see Definitions \ref{def:contributionJCoVaRESJES} and \ref{defcontribution}), where the former is location invariant and the latter is scale invariant. As will be seen later, ${\rm JMES}_{\alpha,\beta}[Y|X]$ contains (\ref{def:TVaR}) (when $\alpha = 0$) and (\ref{MES:def}) (when $\beta = 0$) as two special cases. Some basic properties of these three new types of risk measures are analyzed and quantile function-based integrals are presented. We then establish sufficient conditions for comparing the spillover effects of $X$ on $Y$ and the reversed spillover effects of $Y$ on $X$ on account of these measures. We further investigate sufficient conditions under which these measures are ordered for two sets of bivariate risks with (i)  common marginal distributions but different copulas, (ii) common copula but different marginal distributions, or (iii) different copulas and different marginal distributions.

Compared with \cite{ji2021tail}, \cite{ortega2021stochastic}, and \cite{ZhangIME2023}, the novel contribution of the present paper is as follows:
\begin{enumerate}[(i)]
    \item The proposed JMES measure is a natural generalization of the JES (\ref{Def:JES}) introduced in \cite{ji2021tail} in the sense that the stress events have different confidence levels. In contrast with the asymptotic analysis of JES to study the interplay between tail dependence and heavy-tailedness, we focus on investigating the comparisons of JMES for paired risks as well as different bivariate risk vectors based on stochastic orders and copula functions.
    \item Several types of JMES-based contribution measures, which were not touched in  \cite{ji2021tail}, are proposed and their relationships with stochastic orders among marginal distributions are investigated.
    \item Different from the multivariate systemic risk measures studied in \cite{ortega2021stochastic} with the conditional events generated by more than two different risks, our proposed JMES in Definition \ref{def:JMES} is to measure the conditional expectation of $Y$ provided that $Y$ exceeds some stress level and the other risk $X$ also exceeds some interested stress level. The stochastic orders used here still fall within the class of univariate stochastic orders.
    \item The CoD-based contribution ratio measures introduced in \cite{ZhangIME2023} follow the classical line of defining systemic risk measures like CoVaR, CoES and their contribution measures since the conditional systemic event is generated by only one risk, while our proposed JMES-based contribution measures have different meanings and logistics; see the detailed expressions in Definitions \ref{def:contributionJCoVaRESJES} and \ref{defcontribution}. Besides, the interplay between the tail dependence and marginal distributions' tail behaviours is illustrated by some numerical examples and a real application in analyzing the risk contagion and spillover effect among several stock market indices.
\end{enumerate}

The remainder of the paper is organized as follows. Section \ref{preliminaries} recalls some useful notions, definitions, and lemmas used in the sequel. In Section \ref{JMESriskmeasures}, we introduce JMES, $\Delta$JMES, and $\Delta^{\rm R}$JMES measures and provide the integral expressions. Some interesting properties are also explored. Section \ref{pairedrisks} studies sufficient conditions for the comparisons of the above proposed three new types of risk measures for paired risks. Section \ref{twobivariateriskvec} establishes sufficient conditions for comparing these JMES-based measures for two sets of bivariate vectors with either different marginals or different copulas. In Section \ref{Sec:numerical}, we provide some numerical examples under certain marginal distributions and copulas to illustrate our main findings. Section \ref{realapplication} presents a real application in quantifying the spillover effect among several stock market indices for showing the usage of these three new types of systemic risk measures compared with some existing ones. Section \ref{conclusion} concludes this paper. Some useful definitions and notations used in the paper and all proofs of the main results are relegated to the Appendix.

\section{Preliminaries}\label{preliminaries}
Throughout this paper, the terms ``increasing'' and ``decreasing'' are used in a mild sense. Expectations and integrals are assumed to exist whenever they appear. Let ``$\overset{\rm sign}{=}$'' denote both sides of the equality have the same
sign. We adopt either ``$\stackrel{\rm d}{=}$'' or ``$\stackrel{\rm st}{=}$'' to denote that both sides of the equality have the same c.d.f.. For a $n$-dimensional real-valued function $f$, we shall use ``$\partial_i f$'' to denote the partial derivative w.r.t. its $i$th variable, for $i=1,\ldots,n$.

\subsection{Stochastic orders}
Stochastic orders are partial orders defined on sets of c.d.f.'s and serve as a powerful tool for comparing different interested random variables. 
Recall that the \textit{excess wealth}\footnote{With a little bit of abusive notation, we express this function as the excess wealth instead of calling it as the expected shortfall since we have denoted it by (\ref{def:TVaR}) in the previous statements and it will be seen later that our proposed systemic risk measure is also based on the expected shortfall defined as in (\ref{def:TVaR}).} of $X$ at confidence level $p\in(0,1)$ is defined as
\begin{equation}\label{def:ewfunc}
  W_p(X)=\mathbb{E}[(X-\mbox{\rm VaR}_p[X])_+].
\end{equation}
The excess wealth function measures the thickness of the upper tail from a fixed VaR, and it is also known as the \emph{right spread} function. Interested readers can refer to \cite{shaked2007sto} and \cite{belzunce2015introduction} for more details.

\begin{definition} Let $X$ and $Y$ be two random variables (r.v.'s) with respective c.d.f.'s $F$ and $G$, survival functions (s.f.'s) $\overline{F}$ and $\overline{G}$, and probability density functions (p.d.f.'s) $f$ and $g$, respectively. $X$ is said to be smaller than $Y$ in the
\begin{enumerate}[(i)]
    \item likelihood ratio order (denoted by $X \leq_{\operatorname{lr}} Y$) if $g(x) / f(x)$ is increasing in $x \in \mathbb{R}$;
    \item usual stochastic order (denoted by $X \leq_{\rm st} Y$) if $\overline{F}(x) \leq \overline{G}(x)$ for all $x \in \mathbb{R}$, or equivalently, $\mbox{\rm VaR}_p[X]\leq \mbox{\rm VaR}_p[Y]$ for all $p\in(0,1)$;
    \item increasing convex order\footnote{In the context of actuarial science, it is also termed as the stop-loss order since it can be equivalently written as $\mathbb{E}[(X-d)_+]\leq \mathbb{E}[(Y-d)_+]$, for all $d\in\mathbb{R}$ if $X \leq_{\mathrm{icx}} Y$.} (denoted by $X \leq_{\rm icx} Y$) if $\mathbb{E}[\phi(X)] \leq \mathbb{E}[\phi(Y)]$ for any increasing and convex function $\phi: \mathbb{R} \rightarrow \mathbb{R}$, or equivalently, $\mbox{\rm ES}_p[X]\leq \mbox{\rm ES}_p[Y]$ for all $p\in(0,1)$;
    \item dispersive order (denoted by $X \leq_{\rm disp} Y$) if $\mbox{\rm VaR}_v[X]-\mbox{\rm VaR}_u[X] \leq \mbox{\rm VaR}_v[Y]-\mbox{\rm VaR}_u[Y] $, for all $0<u \leq v<1$;
    \item excess proportional wealth\footnote{This order is introduced in \cite{belzunce2012comparison} and called as the \emph{expected proportional shortfall} (in short \emph{EPS}) order. For the sake of defining the joint marginal expected shortfall order introduced below, we instead denote this interesting order as the \emph{excess proportional wealth} (in short \emph{EPW}) order in this paper.} (denoted by $X \leq_{\rm epw} Y$) if
    \begin{equation}\label{def:EPW}
      \mbox{\rm EPW}_p[X]=\mathbb{E}\left[\left(\frac{X-\mbox{\rm VaR}_p[X]}{\mbox{\rm VaR}_p[X]}\right)_+\right]\leq \mathbb{E}\left[\left(\frac{Y-\mbox{\rm VaR}_p[Y]}{\mbox{\rm VaR}_p[Y]}\right)_+\right]=\mbox{\rm EPW}_p[Y],
    \end{equation}
    for all $p\in\{q\in(0,1):\mbox{\rm VaR}_q[X]\neq0~\mbox{\rm and}~\mbox{\rm VaR}_q[Y]\neq0\}$.
\end{enumerate}
\end{definition}

It is well known that $X \leq_{\rm lr} Y$ implies $X \leq_{\rm st} Y$, which further implies $X \leq_{\rm icx} Y$. All of these three orders imply that $\mathbb{E}[X]\leq \mathbb{E}[Y]$. If the left extreme of the support of $X$ equals to the left extreme of the support of $Y$, i.e., $F^{-1}(0)=G^{-1}(0)>-\infty$, it then holds that $X \leq_{\rm disp} Y$ implies $X \leq_{\rm st} Y$; see Theorem 3.B.13 in \cite{shaked2007sto}. As shown in \cite{belzunce2012comparison}, the excess proportional wealth order is a partial order with the property of scale-invariance, and, with the partial property of  location-invariance in the sense that if $X \leq_{\rm epw} Y$, then $X+c \leq_{\rm epw} Y$ for all $c>0$. They also showed that several situations of parametric families of distributions are ordered in the excess proportional wealth order according to some parameters such as the Uniform, Exponential and Pareto distributions. Moreover, if $\mathbb{E}[X]<\infty$ and $\mathbb{E}[Y]< \infty$, then $X \leq_{\rm \star} Y$ implies $X \leq_{\rm epw} Y$, where `$\leq_{\rm \star}$' is the star order, which is widely used in economics, actuarial science and applied probability.

\cite{sordo2007} provided useful characterizations of the usual stochastic order and the increasing convex order as follows, which will be employed to prove part of the main results.
\begin{lemma}\label{lem:icxicv} Let $X$ and $Y$ be two r.v.'s with continuous and strictly increasing c.d.f.'s $F$ and $G$, respectively. Then, $X \leq_{\rm st[icx]} Y$ if and only if
$$
\int_0^1 F^{-1}(t) \mathrm{d} \phi(t) \leq \int_0^1 G^{-1}(t) \mathrm{d} \phi(t),
$$
for all increasing [increasing convex] $\phi:[0,1] \rightarrow[0,1]$.
\end{lemma}

The following definition provides some well-known positive and negative dependence notions widely used in actuarial science to model the dependence structure of interested risks.
\begin{definition}
    \begin{enumerate}[(i)]
        \item The bivariate random vector $(X, Y)$ is said to be totally positive of order 2 [reverse regular of order 2] (written as $T P_2\left[R R_2\right]$) if $\left[X \mid Y=y_1\right] \leq_{\mathrm{lr}}$ $\left[\geq_{\mathrm{lr}}\right]\left[X \mid Y=y_2\right]$, for all $y_1 \leq y_2$, and $\left[Y \mid X=x_1\right] \leq_{\operatorname{lr}}\left[\geq_{\operatorname{lr}}\right]\left[Y \mid X=x_2\right]$, for all $x_1 \leq x_2$.
        \item $X$ is said to be stochastically increasing [decreasing] in $Y$ (written as $X \uparrow_{\mathrm{SI}}\left[\uparrow_{\mathrm{SD}}\right] Y$) if $\left[X \mid Y=y_1\right] \leq_{\mathrm{st}}\left[\geq_{\mathrm{st}}\right]\left[X \mid Y=y_2\right]$, for all $y_1 \leq y_2$.
        \item The bivariate random vector $(X, Y)$ is said to be positively [negatively] dependent through stochastic ordering (written as PDS [NDS]) if $X \uparrow_{\mathrm{SI}}\left[\uparrow_{\mathrm{SD}}\right] Y$ and $Y \uparrow_{\mathrm{SI}}\left[\uparrow_{\mathrm{SD}}\right] X$ at the same time.
        \item  $X$ is said to be right tail increasing [decreasing] in $Y$ (written as $X \uparrow_{\mathrm{RTI}}\left[\uparrow_{\mathrm{RTD}}\right] Y$ ) if $\mathbb{P}(X>x \mid Y>y)$ is increasing [decreasing] in $y \in \mathbb{R}$, for all $x \in \mathbb{R}$.
    \end{enumerate}
\end{definition}
\par It is well known that the $\mathrm{TP}_2$ property of $(X, Y)$ implies $X \uparrow_{\mathrm{SI}} Y$ and $Y \uparrow_{\mathrm{SI}} X$, which further  imply $X \uparrow_{\mathrm{RTI}} Y$ and $Y \uparrow_{\mathrm{RTI}} X$, and thus $(X, Y)$ is PQD. In a similar manner, the condition $(X, Y)$ is $\mathrm{RR}_2$ implies $X \uparrow_{\mathrm{SD}} Y\left[Y \uparrow_{\mathrm{SD}} X\right]$, which further implies $X \uparrow_{\mathrm{RTD}} Y\left[Y \uparrow_{\mathrm{RTD}} X\right]$, and thus $(X, Y)$ is NQD. Hence, if $(X, Y)$ is $\mathrm{TP}_2\left[\mathrm{RR}_2\right]$, then it must be PDS [NDS]. Moreover, $(X, Y)$ is $\mathrm{TP}_2\left[\mathrm{RR}_2\right]$ if and only if its copula is $\mathrm{TP}_2\left[\mathrm{RR}_2\right]$. Interested readers may refer to \cite{Muller2002}, \cite{Denuit2005}, and \cite{cai2012optimal} for more discussions.

Recall that a distortion function $h:[0,1] \mapsto[0,1]$ is an increasing function such that $h(0)=0$ and $h(1)=1$. The set of all distortion functions is henceforth denoted by $\mathcal{H}$. The class of distortion risk measures is then defined as follows.
\begin{definition} For a distortion function $h \in \mathcal{H}$ and a r.v. $X$ with c.d.f. $F$, the distortion risk measure $\mathrm{D}_h$ is defined as
$$
\mathrm{D}_h[X]=\int_0^{+\infty} h(\overline{F}(t)) \dif t-\int_{-\infty}^0[1-h(\overline{F}(t))] \dif t.
$$
\end{definition}
If $h$ is continuous, the transformation of the tail function $\overline{F}$ of $X$ given by $h\left(\overline{F}(x)\right)=h \circ \overline{F}(x)$ defines a new tail function associated to a certain random variable $X_h$, which is the induced by $h$ distorted r.v. of $X$.

The following lemma, borrowed from Lemma 14 of \cite{sordo2018IME}, will be adopted to prove some of our main results.
\begin{lemma} \label{lem:ineq_delta_invF}
    Let $X$ and $Y$ be two continuous r.v.'s with c.d.f.'s $F_X$ and $F_Y$, respectively. Let $h$ be a concave distortion function and let $g$ be another distortion function such that $h(t) \leq g(t)$, for all $t \in[0,1]$. Denote by $X_h$ and $Y_g$ the distorted random variables induced from $X$ and $Y$ by the distortion functions $h$ and $g$, respectively. If $X \leq_{\rm disp} Y$, then
\begin{enumerate}[(i)]
    \item $F_{X_h}^{-1}(p)-F_X^{-1}(p) \leq F_{Y_g}^{-1}(p)-F_Y^{-1}(p)$, for all $p \in(0,1)$;
    \item $F_{X_g}^{-1}(p)-F_{X_h}^{-1}(p) \leq F_{Y_g}^{-1}(p)-F_{Y_h}^{-1}(p)$, for all $p \in(0,1)$.
\end{enumerate}
\end{lemma}
\subsection{Copula}
Let $(X, Y)$ be an absolutely continuous random vector with joint c.d.f. $H$ and marginal c.d.f.'s $F$ and $G$. The joint c.d.f. $H$ can be expressed as
\begin{equation} \nonumber
    H(x,y)=C(F(x),G(y)),
\end{equation}
where $C$ is the copula of $(X, Y)$, that is, the joint c.d.f. of $(U,V)$, where $U=F(X)$ and $V=G(Y)$ \citep[cf.][]{Nelsen2007}. The copula contains information about the dependence of the random vector $(X, Y)$ apart from the behaviour of the marginal distributions. We denote by $ \overline{C}$ the \textit{joint tail function} for two uniform r.v.'s whose joint c.d.f. is the copula $C$, that is
\begin{equation} \nonumber
    \begin{aligned}
        \overline{C}(u, v) & =\PP\Lbrb{X>F^{-1}(u), Y>G^{-1}(v)} \\
        & =1-u-v+C(u, v), \quad \mbox{$(u,v)\in[0,1]^2$}.
        \end{aligned}
\end{equation}
The joint tail function $\overline{C}$ is different with the \emph{survival copula} of $U$ and $V$, which is defined as
\begin{equation*}
\hat{C}(u,v)=u+v-1+C(1-u,1-v),\quad\mbox{$(u,v)\in[0,1]^2$}.
\end{equation*}
The survival copula $\hat{C}$ couples the joint survival function to its univariate margins (i.e. survival functions)
in a manner completely analogous to how a copula links the joint c.d.f. to its margins.
Clearly, $$\overline{C}(u,v)=\hat{C}(1-u,1-v).$$
\par The well-known Farlie-Gumbel-Morgenstern (FGM) copula, as a tractable example of copula, is defined as
\begin{equation} \nonumber
    C_{\theta}(u,v) = uv(1+\theta(1-u)(1-v)), \quad-1 \leq \theta \leq 1.
\end{equation}
The FGM copula degenerates to the independence copula when $\theta = 0$. Furthermore, it is $\mathrm{TP}_2$ [$\mathrm{RR}_2$] when $\theta \geq [\leq] 0$. Another frequently-used copula is the Gumbel copula, belonging to the well-known class of Archimedean copulas, which is defined as
\begin{equation}\label{copula:Gumbel}
    C_{\theta}(u,v)=\exp\left\{-((-\log(u))^{\theta} + (-\log(v))^{\theta})^{1/{\theta}} \right\}, \quad\theta \in [1,\infty).
\end{equation}
The Gumbel copula becomes the independence copula when $\theta = 1$, and approaches the Fr\'echet upper bound copula (comonotonicity case) when $\theta \to \infty$. More studies on the properties of these two types of copulas can be found in \cite{Nelsen2007}.

\section{JMES risk measures}\label{JMESriskmeasures}
\subsection{Definitions of JMES and related contribution measures}
As a direct generalization of the JES (\ref{Def:JES}) proposed by \cite{ji2021tail}, we define the so-called \textit{Joint Marginal Expected Shortfall}\footnote{One of the main reasons why we incorporate the term ``Marginal'' into ``JES'' is that the definition is a combination of ES and MES marginally under joint tail events induced by two risks. Another reason is that the JES proposed by \cite{ji2021tail} is defined under the setting $\alpha=\beta$, and we adopt JMES to make a significant difference with JES.} (in short JMES) risk measures as follows when $X$ and $Y$ exceed their threshold levels with different confidence levels.

\begin{definition}\label{def:JMES}
For any $\alpha,\beta \in [0,1)$, the joint marginal expected shortfall of $Y$ given that $X$ and $Y$ exceed their threshold levels with respective confidence levels $\alpha$ and $\beta$, denoted by ${\rm JMES}_{\alpha,\beta}[Y|X]$, is defined as follows:
\begin{equation}\label{equa:JMES}
    {\rm JMES}_{\alpha,\beta}[Y|X]=\mathbb{E}[Y|X>{\rm VaR}_{\alpha}[X],Y>{\rm VaR}_{\beta}[Y]].
\end{equation}
\end{definition}

\begin{remark} It is worth mentioning that our proposed \eqref{equa:JMES} also covers the JES-based generalization definition (3.1) in \cite{ji2021tail}. More specifically, the generalization of JES introduced in \cite{ji2021tail} is given as follows:
\begin{equation}\label{equa:JESJigene}
    {\rm JES}_{\alpha,\xi}^{G}[Y|X]=\mathbb{E}[Y|X>{\rm VaR}_{\alpha}[X],Y>\xi{\rm VaR}_{\alpha}[Y]],
\end{equation}
where $\xi\in(0,1]$ is a predetermined constant. Clearly, ${\rm JES}_{\alpha,\xi}^{G}[Y|X]$ degenerates into the JES when $\xi=1$. 

Assume that $\xi\in(0,1)$ and $\alpha\in(0,1)$. To explore the connection and difference between \eqref{equa:JMES} and \eqref{equa:JESJigene}, without loss of generality, we assume that both $X$ and $Y$ have continuous c.d.f.'s. Since $0<\xi{\rm VaR}_{\alpha}[Y]<{\rm VaR}_{\alpha}[Y]$ and ${\rm VaR}_{p}[Y]$ is continuous in $p\in(0,1)$, it immediately follows that there must exist some $\beta\in(0,\alpha)$ such that ${\rm VaR}_{\beta}[Y]=\xi{\rm VaR}_{\alpha}[Y]$. Therefore, under the setting of $\xi\in(0,1]$, the definition in \eqref{equa:JESJigene} can be also treated as a special case of \eqref{equa:JMES}, and it only considers the case when $\beta\leq\alpha$.

In spite of the connection between these two definitions, our main task for the subsequent theoretical findings is to introduce several types of JMES-based contribution measures, analyze their basic properties and then apply stochastic orders and copulas to compare different sets of bivariate risks. In contrast to our results, the main focus of Subsection 3.3 of \cite{ji2021tail} is to establish asymptotic expansion of ${\rm JES}_{\alpha,\xi}^{G}[Y|X]$ for the Fr\'echet distribution case in the max-domain of attraction from the perspective of the extreme value theory. Hence, one natural question arises whether the asymptotic expansion of ${\rm JES}_{\alpha,\xi}^{G}[Y|X]$ could be established or not whenever $1<\xi<\infty$, corresponding to $\beta>\alpha$ under the framework of ${\rm JMES}_{\alpha,\beta}[Y|X]$. We leave this problem to future research.
\end{remark}
\par Note that, if both $X$ and $Y$ are non-negative and such that $\mathbb{P}(X=0)=\mathbb{P}(Y=0)=0$, then  ${\rm JMES}_{\alpha,\beta}[Y|X]$ degenerates to conventionally expected shortfall ${\rm ES}_{\beta}[Y]$ when $\alpha = 0$, while degenerates to the marginal expected shortfall ${\rm MES}_{\alpha}[Y\vert X]$ when $\beta = 0$, which is also a special case of the CoD-risk measures developed in \cite{dhaene2022systemic}. However, JMES has no relationship with CoD in general.

Following Definition \ref{def:JMES}, we can define their associated risk contribution measures in terms of the difference function and the ratio function as follows.
\begin{definition}\label{def:contributionJCoVaRESJES}
    For $\alpha_1, \alpha_2, \beta \in [0,1)$ such that $\alpha_1\leq\alpha_2$, we define the ${\rm JMES}$-based difference contribution measure ${\rm \Delta JMES}_{\alpha_1,\alpha_2,\beta}[Y|X]$ as follows:
\begin{equation} \nonumber
    {\rm \Delta JMES}_{\alpha_1,\alpha_2,\beta}[Y|X]={\rm JMES}_{\alpha_2,\beta}[Y|X]-{\rm JMES}_{\alpha_1,\beta}[Y|X].
\end{equation}
In particular, if $\alpha_1 = 0$, then ${\rm JMES}_{\alpha_1,\beta}[Y|X]$ degenerates into ${\rm ES}_{\beta}[Y]$, and thus we denote ${\rm \Delta JMES}_{0,\alpha_2,\beta}[Y|X]$ by ${\rm \Delta JMES}_{\alpha,\beta}[Y|X]$ for brevity. In other words,
\begin{equation}\nonumber
    {\rm \Delta JMES}_{\alpha,\beta}[Y|X]={\rm JMES}_{\alpha,\beta}[Y|X]-{\rm ES}_{\beta}[Y],
\end{equation}
for $\alpha, \beta \in [0,1)$.
\end{definition}

To quantify the relative spillover effect induced by systemic risks, we defined the ${\rm JMES}$-based ratio contribution measure as follows.
\begin{definition}\label{defcontribution}
    For $\alpha_1, \alpha_2, \beta \in [0,1)$ such that $\alpha_1\leq\alpha_2$, we define the ${\rm JMES}$-based contribution ratio  measure ${\rm \Delta^R JMES}_{\alpha_1, \alpha_2, \beta}[Y|X]$ as follows:
    \begin{equation}\label{delta-JMES-diff}
        {\rm \Delta^R JMES}_{\alpha_1,\alpha_2,\beta}[Y|X]=\frac{{\rm JMES}_{\alpha_2,\beta}[Y|X]-{\rm JMES}_{\alpha_1,\beta}[Y|X]}{{\rm JMES}_{\alpha_1,\beta}[Y|X]},
    \end{equation}
    provided that ${\rm JMES}_{\alpha_1,\beta}[Y|X]\neq0$. Furthermore, when $\alpha_1 = 0$, we shall denote (\ref{delta-JMES-diff}) by ${\rm \Delta^R JMES}_{\alpha,\beta}[Y|X]$ for brevity, that is
\begin{equation} \nonumber
    {\rm \Delta^R JMES}_{\alpha,\beta}[Y|X]=\frac{{\rm JMES}_{\alpha,\beta}[Y|X]-{\rm ES}_{\beta}[Y]}{{\rm ES}_{\beta}[Y]},
\end{equation}
 for $\alpha, \beta \in [0,1)$, provided that ${\rm ES}_{\beta}[Y]\neq0$.
\end{definition}

The ${\rm \Delta JMES}$ risk measure satisfies the location-invariant property, while the ${\rm \Delta^R JMES}$ risk measure has the scale-invariant property. As will be seen later, the dispersive order between marginal risks can be translated into the ordering between their corresponding  ${\rm \Delta JMES}$ risk measures, and the excess proportional wealth order can be translated into the ordering between their corresponding  ${\rm \Delta^R JMES}$ risk measures.

It is worth noting that \cite{dhaene2022systemic} introduced a wide class of CoD-risk measures and some kinds of CoD-risk contribution measures. The measures defined in Definitions \ref{def:JMES} and \ref{def:contributionJCoVaRESJES} cannot be incorporated by the CoD-risk measures since there are two stress events in JMES and $\Delta$JMES. Besides, the ratio contribution measure  ${\rm \Delta^R JMES}$ is relatively new compared with other known ratio-type systemic measures such as the CoVaR-based ratio contribution measure studied in \cite{Adrian2016}.

The next theorem presents the integral-based expressions for the aforementioned two types of systemic risk measures.
\begin{lemma}\label{lem:int}
Let $(U,V)\sim C$ where $C$ is a copula of the joint c.d.f. $H$. Suppose $F$ and $G$ are continuous and strictly increasing. Then, for $\alpha,\beta \in [0,1)$, we have
$${\rm JMES}_{\alpha,\beta}[Y|X]=\int_{0}^1G^{-1}(\overline{h}^{-1}_{\alpha,\beta}(t)) \dif t = \int_{0}^{1} G^{-1}(t)\dif \overline{h}_{\alpha,\beta}(t)= \int_{\beta}^{1} G^{-1}(t)\dif \overline{h}_{\alpha,\beta}(t),$$
where
\begin{eqnarray}\label{eq:JMES-calclt}
\overline{h}_{\alpha,\beta}(t) &:= & F_{V|U>\alpha,V>\beta}(t)\nonumber\\
&=&\frac{\overline{C}(\alpha,\beta)-\overline{C}(\alpha,\max\{\beta,t\})}{\overline{C}(\alpha,\beta)}=
\begin{cases}
1-\frac{\overline{C}(\alpha,t)}{\overline{C}(\alpha,\beta)}, & ~ \mbox{for}~~ \beta < t \leq 1;\\
0, & ~ \mbox{for}~~0\leq t\leq \beta.
\end{cases}
\end{eqnarray}
Further, we have
\begin{equation*}
  \Delta {\rm JMES}_{\alpha_1,\alpha_2,\beta}[Y|X]  =\int_{\beta}^{1} G^{-1}(t)\dif \overline{h}_{\alpha_2,\beta}(t)-\int_{\beta}^{1} G^{-1}(t)\dif \overline{h}_{\alpha_1,\beta}(t),
\end{equation*}
\begin{equation}\label{eq:DJMES-representation}
  \Delta {\rm JMES}_{\alpha,\beta}[Y|X]  =\int_{\beta}^{1} G^{-1}(t)\dif \overline{h}_{\alpha,\beta}(t)-\int_{\beta}^{1} G^{-1}(t)\dif \overline{h}_{\beta}(t),
\end{equation}
\begin{equation*} \nonumber
    {\rm \Delta^R JMES}_{\alpha_1,\alpha_2,\beta}[Y|X] = \frac{\int_{\beta}^{1} G^{-1}(t)\dif \overline{h}_{\alpha_2,\beta}(t)-\int_{\beta}^{1} G^{-1}(t)\dif \overline{h}_{\alpha_1,\beta}(t)}{\int_{\beta}^{1} G^{-1}(t)\dif \overline{h}_{\alpha_1,\beta}(t)},
\end{equation*}
and
\begin{equation*} \nonumber
    {\rm \Delta^R JMES}_{\alpha,\beta}[Y|X] = \frac{\int_{\beta}^{1} G^{-1}(t)\dif \overline{h}_{\alpha,\beta}(t)-\int_{\beta}^{1} G^{-1}(t)\dif \overline{h}_{\beta}(t)}{\int_{\beta}^{1} G^{-1}(t)\dif \overline{h}_{\beta}(t)},
\end{equation*}
where $\overline{h}_{\beta}(t):=\max\left\{\frac{t-\beta}{1-\beta},0\right\}$.
\end{lemma}

One can observe from the proof of Lemma \ref{lem:int} that the survival function of the variable $[Y|X>{\rm VaR}_{\alpha}[X],Y>{\rm VaR}_{\beta}[Y]]$ can be written as
\begin{equation} \nonumber
    \overline{F}_{Y|X>{\rm VaR}_{\alpha}[X],Y>{\rm VaR}_{\beta}[Y]}(y) = 1-\overline{h}_{\alpha,\beta}(1-\overline{G}(y)).
\end{equation}
Hence, $[Y|X>{\rm VaR}_{\alpha}[X],Y>{\rm VaR}_{\beta}[Y]]$ can be seen as a distorted r.v.  induced from $Y$ by the distortion function
\begin{equation}\label{eq:h-def}
    h_{\alpha,\beta}(t)=1-\overline{h}_{\alpha,\beta}(1-t) = \min\left\{ \frac{\overline{C}(\alpha,1-t)}{\overline{C}(\alpha,\beta)}, 1\right\}.
\end{equation}
Similarly, $[Y|Y>{\rm VaR}_{\beta}[Y]]$ can be treated as a distorted r.v. induced from $Y$ by the distortion function
\begin{equation}\label{eq:h0-def}
    h_{\beta}(t)=1-\overline{h}_{\beta}(1-t)=\min\left\{\frac{t}{1-\beta}, 1\right\}.
\end{equation}

\subsection{Basic properties of JMES and related contribution measures}
The following lemma establishes sufficient conditions to justify the sign of the JMES-based difference contribution measures.
\begin{lemma}\label{deltapositive}
    If $X\uparrow_{\rm RTI}[\uparrow_{\rm RTD}]Y$, then ${\rm \Delta JMES}_{\alpha,\beta}[Y|X]\geq[\leq] 0$, for all $\alpha,\beta \in [0,1)$.
    \end{lemma}

The following result provides sufficient conditions for the monotonicity of ${\rm JMES}_{\alpha,\beta}[Y|X]$ w.r.t. the two confidence levels $\alpha,\beta\in[0,1)$.
\begin{theorem}\label{thm:monot-property} Suppose $(X,Y)$ admits a copula $C$.
    \begin{enumerate}[(i)]
        \item For any fixed $\alpha\in[0,1)$, ${\rm JMES}_{\alpha,\beta}[Y|X]$ increases w.r.t. $\beta \in [0,1)$.
        \item If $\overline{C}$ is ${\rm TP}_2$ $[{\rm RR}_2]$, then for any fixed $\beta\in[0,1)$, ${\rm JMES}_{\alpha,\beta}[Y|X]$ increases [decreases] w.r.t. $\alpha \in [0,1)$.
    \end{enumerate}
\end{theorem}

One can easily observe that the monotonicity of ${\rm JMES}_{\alpha,\beta}[Y|X]$ w.r.t. $\alpha$ implies the same monotonicity of $\Delta {\rm JMES}_{\alpha,\beta}[Y|X]$ and ${\rm \Delta ^R JMES}_{\alpha,\beta}[Y|X]$. Hence, the result of Theorem \ref{thm:monot-property}(ii) provides the respect monotonicity of $\Delta {\rm JMES}_{\alpha,\beta}[Y|X]$ and ${\rm\Delta ^R JMES}_{\alpha,\beta}[Y|X]$ w.r.t. $\alpha$ as well. In addition, it is worth mentioning that the monotonicity of $\Delta {\rm JMES}_{\alpha,\beta}[Y|X]$ and ${\rm \Delta ^R JMES}_{\alpha,\beta}[Y|X]$ w.r.t. $\beta$ cannot be established in general, let alone for the general measures $\Delta {\rm JMES}_{\alpha_1,\alpha_2,\beta}[Y|X]$ and ${\rm \Delta ^R JMES}_{\alpha_1,\alpha_2,\beta}[Y|X]$. Related examples are provided in Section \ref{subsec:numeexam} to illustrate this point.
\par The JMES enjoys the properties of translation invariance and positive homogeneity. Furthermore, it has comonotonic additivity when the c.d.f.'s of marginals are continuous and strictly increasing, which is summarized in the following proposition.
\begin{proposition}\label{prop:comono} Consider two bivariate risks $(X,Y_1)$ and $(X,Y_2)$ with continuous and strictly increasing marginal c.d.f.'s. If $Y_1$ and $Y_2$ are comonotonic, then
\begin{equation*}
    {\rm JMES}_{\alpha,\beta}[Y_1+Y_2|X]={\rm JMES}_{\alpha,\beta}[Y_1|X]+{\rm JMES}_{\alpha,\beta}[Y_2|X].
\end{equation*}
\end{proposition}

\par Given the significance of evaluating systemic risk measures, it is imperative to establish rigorous statistical assessments of the predictive efficacy of various models, particularly in scenarios where multiple forecasting models for systemic risk measures are available. Such evaluations are commonly referred as ``backtests'' or ``backtestings'' within the realm of finance. The practice of backtesting systemic risk measures has garnered considerable attention from researchers during the past several years. For instance, \cite{girardi2013systemic} conducted backtests on CoVaR estimates derived from their constructed GARCH model and the proposed three-step estimation procedure. \cite{banulescu2021backtesting} presented an innovative approach for backtesting systemic risk measures, including MES and $\Delta$CoVaR. 

In the sequel, we explore the feasibility of conducting backtests on JMES, adopting the perspective delineated by \cite{fissler2023backtesting} on backtesting methodologies. There exist two distinct approaches to backtesting risk measures, ``traditional'' and ``comparative'' backtests, each serving different objectives. ``Traditional backtests'' check how well a sequence of risk measure forecasts aligns with corresponding observations of losses. Traditional backtests rely on the \textit{identifiability} (Definition \ref{def:identifiable}) of the underlying risk measure, which ensures the existence of a (possibly multivariate) function that uniquely identifies the true report. ``Comparative backtests'' aim to evaluate and compare the predictive accuracy of multiple forecasting models in relation to each other. These comparative backtests leverage the \textit{elicitability} (Definition \ref{def:elicitable}) of the underlying risk measure, which implies the existence of a real-valued loss (or scoring) function, that minimizes the expected loss (or score) of the optimal forecast. 
\par Take VaR as an example and denote by $y_t$ the observed losses at time $t\geq0$. On the one hand, it can be verified that the ``pinball loss'' $$S(r, y)=(\mathbb{I}\{y>r\}-1+\beta)(r-y),$$ or specifically $$S({\rm VaR}_{\beta}(y_t), y_t)=(\mathbb{I}\{y_t>{\rm VaR}_{\beta}(y_t)\}-1+\beta)({\rm VaR}_{\beta}(y_t)-y_t),$$ is strictly $\mathcal{F}$-consistent. Hence, VaR is elicitable. On the other hand, $$V(r, y)=\mathbb{I}\{y>r\}-(1-\beta)$$ is a strict $\mathcal{F}$-identification function for VaR. Therefore, VaR is identifiable. Considering these two properties, ``traditional'' and ``comparative'' backtests can be done on the basis of VaR. However, both properties do not hold for the ES. Interested readers can refer to \cite{fissler2016expected} and \cite{fissler2016higher} for more details.
\par The following proposition shows that JMES, like CoVaR, CoES and MES, generally fails to be identifiable or elicitable on sufficiently rich classes of bivariate distributions. Denote by $\mathcal{F}^0(\mathbb{R}^2)$ the set of all bivariate distributions on $\mathbb{R}^2$.
\begin{proposition} \label{prop:nine}
    For $\alpha, \beta\in(0,1)$, ${\rm JMES}_{\alpha,\beta}$ is neither identifiable nor elicitable on any class $\mathcal{F}\subseteq \mathcal{F}^0(\mathbb{R}^2)$ containing all bivariate normal distributions along with their finite mixtures. 
\end{proposition}
\par The practical implication of Proposition \ref{prop:nine} is that neither traditional nor comparative backtests can be carried out on the basis of JMES alone \citep[cf. Page 1 in][]{fissler2023backtesting}. However, JMES, combined with other risk measures, might be jointly identifiable or jointly elicitable, which is left for future research.

\section{Comparison results for paired risks}\label{pairedrisks}
For a pair of risks $(X,Y)$ with marginal c.d.f.'s $F$, $G$ and copula $C$, one might be interested in comparing ${\rm JMES}_{\alpha,\beta}[Y|X]$ with
    \begin{equation*}
    {\rm JMES}_{\alpha,\beta}[X|Y]=\mathbb{E}[X|Y>{\rm VaR}_{\alpha}[Y],X>{\rm VaR}_{\beta}[X]],
\end{equation*}
which can be used to infer which risk has more spillover effect than the other one. The following result establishes sufficient conditions for comparing ${\rm JMES}_{\alpha,\beta}[Y|X]$ and ${\rm JMES}_{\alpha,\beta}[X|Y]$.
\begin{theorem}\label{PairRisks}
If $X\leq_{{\rm icx}}Y$, $X\uparrow_{{\rm SI}}Y$ and $C$ is symmetric\footnote{A bivariate copula is said to be (exchange or permutation) \textit{symmetric} if $C(u,v)=C(v,u)$ for all $(u,v)\in[0,1]^2$. Many commonly used copulas are symmetric such as the family of Archimedean copulas and the FGM copula with only one dependence parameter.}, then we have ${\rm JMES}_{\alpha,\beta}[X|Y]\leq{\rm JMES}_{\alpha,\beta}[Y|X]$ for any $\alpha,\beta\in [0,1)$.
\end{theorem}

The following result establishes sufficient conditions for comparing ${\rm \Delta JMES}_{\alpha,\beta}[Y|X]$ and ${\rm \Delta JMES}_{\alpha,\beta}[X|Y]$ to quantify the relative spillover effect from one risk on the other one.
\begin{theorem}\label{contributionpair}
If $X\leq_{\rm disp}Y$, and $C$ is PDS and symmetric, then we have ${\rm \Delta JMES}_{\alpha,\beta}[X|Y] \leq {\rm \Delta JMES}_{\alpha,\beta}[Y|X]$ for any $\alpha,\beta \in [0,1)$.
\end{theorem}

Note that the results in Theorems \ref{PairRisks} and \ref{contributionpair} have the same spirit in describing the marginal effects of one risk on the other. Suppose that the copula $C$ is PDS and symmetric, more effect induced by $X$ is generated on $Y$ in terms of the JMES risk measures if $X\leq_{{\rm icx}}Y$, and the absolute risk contribution induced by $X$ on $Y$ in terms of the $\Delta$JMES measures is more if $X\leq_{\rm disp}Y$. Moreover, under the same setting of Theorem \ref{PairRisks}, it can be concluded that ${\rm MES}_{\alpha}[X|Y]\leq {\rm MES}_{\alpha}[Y|X]$ for any $\alpha \in [0,1)$ by considering $\beta=0$.


As a generalization of Theorem \ref{contributionpair}, the following result establishes sufficient conditions for comparing ${\rm \Delta JMES}_{\alpha_1,\alpha_2,\beta}[X|Y]$ and ${\rm \Delta JMES}_{\alpha_1,\alpha_2,\beta}[Y|X]$.
\begin{theorem}\label{JMESconta1a2beta}
If $X\leq_{\rm disp}Y$, and $\overline{C}$ is ${\rm TP}_2$\footnote{It should be noted that the condition ``$\overline{C}$ is ${\rm TP}_2$'' is very different with the requirement that ``$C$ is ${\rm TP}_2$''. But for the case of FGM copula, it can be shown from (\ref{equa:FGMtailcopula}) that they are equivalent by saying that the dependence parameter $\theta\geq0$.} and symmetric, then we have ${\rm \Delta JMES}_{\alpha_1,\alpha_2,\beta}[X|Y] \leq {\rm \Delta JMES}_{\alpha_1,\alpha_2,\beta}[Y|X]$ for any $0\leq\alpha_1\leq\alpha_2<1$ and $\beta \in [0,1)$.
\end{theorem}

For a r.v. $X$ with c.d.f. $F$, we define
\begin{equation}\label{IABfunc}
  I_{A, B}(X)=\frac{\int_0^1 F^{-1}(t) \dif A(t)}{\int_0^1 F^{-1}(t) \dif B(t)}-1,
\end{equation}
where both $A(t)$ and $B(t)$ are distortion functions, that is, $A(t)\in\mathcal{H}$ and $B(t)\in\mathcal{H}$. Let $$\mathcal{C}_1=\{I_{A, B}:A(t)\in\mathcal{H},~B(t)\in\mathcal{H},~\mbox{$A\circ B^{-1}(t)$ is convex}\}$$
and
$$\mathcal{C}_2=\{I_{A, B}:A(t)\in\mathcal{H},~B(t)\in\mathcal{H},~\mbox{both $A\circ B^{-1}(t)$ and $B(t)$ are convex}\}.$$
Clearly, $\mathcal{C}_2$ is a subset of $\mathcal{C}_1$. Next, we compare the JMES-based ratio contribution measures for paired risks $(X,Y)$. The following lemma is due to Theorem 3.25 of \cite{belzunce2012comparison}, which establishes an equivalent characterization for the excess proportional wealth order in terms of the class of well-defined ratio integrals (\ref{IABfunc}) within $\mathcal{C}_2$.

\begin{lemma}{\rm\citep[Theorem 3.25 of][]{belzunce2012comparison}}\label{lemma:epw} Let $X$ and $Y$ be two random variables with c.d.f.'s $F$ and $G$, respectively. Then, $X \leq_{\rm epw} Y$ if and only if $I_{A,B}(X)\leq I_{A,B}(Y)$ for all $I_{A,B}\in \mathcal{C}_2$.
\end{lemma}

Now, we investigate the sufficient conditions under which the JMES-based contribution ratio measures are ordered for paired risks $(X, Y)$.
\begin{theorem}\label{thm:ratiopair} Suppose that $X\leq_{\rm epw}Y$, ${C}$ is symmetric and PDS. Then, we have $${\rm \Delta^R JMES}_{\alpha,\beta}[X|Y]\leq {\rm \Delta^R JMES}_{\alpha,\beta}[Y|X],\quad\mbox{for any $\alpha,\beta\in [0,1)$.}$$
\end{theorem}

The following result presents sufficient conditions for  comparing ${\rm \Delta^R JMES}_{\alpha_1,\alpha_2,\beta}[X|Y]$ and ${\rm \Delta^R JMES}_{\alpha_1,\alpha_2,\beta}[Y|X]$. The proof can be easily reached by applying Lemma \ref{lemma:epw} with a similar method with that of Theorem \ref{thm:ratiopair}.
\begin{theorem}\label{thm:ratiopairgeneral} Suppose that $X\leq_{\rm epw}Y$ and $C$ is symmetric and PDS. If $\overline{h}_{\alpha_2,\beta}(\overline{h}^{-1}_{\alpha_1,\beta}(t))$ is convex in $t\in[0,1]$, for $0\leq\alpha_1\leq\alpha_2<1$ and $\beta\in [0,1)$,  then we have $${\rm \Delta^R JMES}_{\alpha_1,\alpha_2,\beta}[X|Y]\leq {\rm \Delta^R JMES}_{\alpha_1,\alpha_2,\beta}[Y|X].$$
\end{theorem}

The PDS property of $C$ required in Theorem \ref{thm:ratiopairgeneral} ensures that both functions $\overline{h}_{\alpha_1,\beta}(t)$ and $\overline{h}_{\alpha_2,\beta}(t)$ are increasing and convex in $t\in[0,1]$, which naturally leads to the fact that $\overline{h}^{-1}_{\alpha_1,\beta}(t)$ is increasing and concave in $t\in[0,1]$. However, the convexity of $\overline{h}_{\alpha_2,\beta}(\overline{h}^{-1}_{\alpha_1,\beta}(t))$ cannot be ensured under this setting.

\section{Comparison results for two bivariate risk vectors}\label{twobivariateriskvec}
In this section, we study sufficient conditions under which the JMES and associated contribution measures can be ordered for two bivariate risk vectors. First, we consider the random vectors $(X_1,Y_1)$ and $(X_2,Y_2)$ have different marginals c.d.f.'s $(F_1,G_1)$ and  $(F_2,G_2)$, and they possess a common copula $C$. Along with the previous studies, it is assumed that all of $F_1$, $F_2$, $G_1$, and $G_2$ are continuous and strictly increasing to avoid cumbersome subtle technical discussions.
\begin{theorem}\label{the:twoJMES}
    Let $(X_1,Y_1)$ and $(X_2,Y_2)$ be two bivariate random vectors with c.d.f.'s $(F_1,G_1)$ and $(F_2,G_2)$, respectively. Suppose $(X_1,Y_1)$ and $(X_2,Y_2)$ admit a common copula $C$ which is such that $X_1\uparrow_{\rm SI}Y_1$. Then
    \begin{enumerate}[(i)]
        \item $Y_1\leq_{\rm icx}Y_2$ implies that ${\rm JMES}_{\alpha,\beta}[Y_1|X_1]\leq{\rm JMES}_{\alpha,\beta}[Y_2|X_2]$, for any $\alpha,\beta\in [0,1)$;
        \item $Y_1\leq_{\rm disp}Y_2$ implies that ${\rm \Delta JMES}_{\alpha,\beta}[Y_1|X_1]\leq{\rm \Delta JMES}_{\alpha,\beta}[Y_2|X_2]$, for any $\alpha,\beta\in [0,1)$;
        \item $Y_1\leq_{\rm epw}Y_2$ implies ${\rm \Delta^R JMES}_{\alpha,\beta}[Y_1|X_1]\leq{\rm \Delta^R JMES}_{\alpha,\beta}[Y_2|X_2]$, for any $\alpha,\beta\in [0,1)$.
    \end{enumerate}
\end{theorem}

The following theorem provides sufficient conditions for comparing ${\rm \Delta JMES}_{\alpha_1,\alpha_2,\beta}[Y_1|X_1]$ and ${\rm \Delta JMES}_{\alpha_1,\alpha_2,\beta}[Y_2|X_2]$, and ${\rm \Delta^R JMES}_{\alpha_1,\alpha_2,\beta}[Y_1|X_1]$ and ${\rm \Delta^R JMES}_{\alpha_1,\alpha_2,\beta}[Y_2|X_2]$ when $(X_1,Y_1)$ and $(X_2,Y_2)$ have different marginals c.d.f.'s $(F_1,G_1)$ and  $(F_2,G_2)$, but possess a common copula $C$. The proofs are omitted since they are very similar to those of Theorems \ref{JMESconta1a2beta} and \ref{thm:ratiopairgeneral}.
\begin{theorem}\label{the:twoJMESgeneral}
    Let $(X_1,Y_1)$ and $(X_2,Y_2)$ be two bivariate random vectors with c.d.f.'s $(F_1,G_1)$ and $(F_2,G_2)$, respectively. Suppose $(X_1,Y_1)$ and $(X_2,Y_2)$ admit a common copula $C$.
    \begin{enumerate}[(i)]
        \item If $Y_1\leq_{\rm disp}Y_2$ and $\overline{C}$ is ${\rm TP}_2$, then ${\rm \Delta JMES}_{\alpha_1,\alpha_2,\beta}[Y_1|X_1]\leq{\rm \Delta JMES}_{\alpha_1,\alpha_2,\beta}[Y_2|X_2]$, for any $0\leq\alpha_1\leq\alpha_2<1$ and $\beta\in [0,1)$.
        \item If $Y_1\leq_{\rm epw}Y_2$, $C$ is PDS, and $\overline{h}_{\alpha_2,\beta}(\overline{h}^{-1}_{\alpha_1,\beta}(t))$ is convex in $t\in[0,1]$, for $0\leq\alpha_1\leq\alpha_2<1$ and $\beta\in [0,1)$, then ${\rm \Delta^R JMES}_{\alpha_1,\alpha_2,\beta}[Y_1|X_1]\leq{\rm \Delta^R JMES}_{\alpha_1,\alpha_2,\beta}[Y_2|X_2]$.
    \end{enumerate}
\end{theorem}

We next provide sufficient conditions for comparing the JMES and associated contribution measures for two sets of bivariate random vectors $(X_1,Y_1)$ and $(X_2,Y_2)$ with common marginal c.d.f.'s $(F,G)$ but have different copulas $C_1$ and $C_2$. In the following discussions,  we denote
\begin{equation} \label{eq:lalphax}
    l_{\alpha}(t):=\frac{\overline{C}_2(\alpha,t)}{\overline{C}_1(\alpha,t)},\quad \mbox{for $t\in[\beta,1]$}.
\end{equation}
The following lemma, provided in \cite{barlow1975statistical}, is required to develop the main results.
\begin{lemma}\label{lemma:BP}{\rm \citep[Lemma 7.1 in][]{barlow1975statistical}} Let $W$ be a measure on the interval $[a, b]$, not necessarily nonnegative, where $-\infty < a<b <\infty$. Let $g$ be a nonnegative function defined on $[a, b]$. If $\int_t^b \dif W(u) \geq 0$, for all $t \in[a, b]$, and $g$ is increasing, then
$$
\int_t^b g(u) \dif W(u) \geq 0, \quad \mbox{for all $t \in[a, b]$}.
$$
\end{lemma}

\begin{theorem}\label{thm:diff_copula_same_margin}
    Suppose $X_1\stackrel{\rm d}{=}X_2$ and $Y_1\stackrel{\rm d}{=}Y_2$ with marginal c.d.f.'s $F$ and $G$, respectively. Assume that $(X_1,Y_1)$ and $(X_2,Y_2)$ admit different copulas $C_1$ and $C_2$. For the fixed $\alpha\in[0,1)$, if $l_{\alpha}(t) \geq l_{\alpha}(\beta)$ for all $t \in [\beta,1]$, then we have ${\rm JMES}_{\alpha,\beta}[Y_1|X_1]\leq{\rm JMES}_{\alpha,\beta}[Y_2|X_2]$, which immediately implies that $\Delta {\rm JMES}_{\alpha,\beta}[Y_1|X_1]\leq\Delta {\rm JMES}_{\alpha,\beta}[Y_2|X_2]$ and $ {\rm\Delta^R JMES}_{\alpha,\beta}[Y_1|X_1] \leq {\rm \Delta^R JMES}_{\alpha,\beta}[Y_2|X_2]$ under the setting.
\end{theorem}


It is worthy noting that the requirement ``$l_{\alpha}(t) \geq l_{\alpha}(\beta)$ for all $t \in [\beta,1]$'' in Theorem \ref{thm:diff_copula_same_margin} is quite general and will be automatically satisfied if $l_{\alpha}(t)$ increases w.r.t. $t \in [\beta,1]$ \footnote{The function ``$l_{\alpha}(t)$ increases w.r.t. $t \in [\beta,1]$'' is equivalent to saying  ``$\overline{C}_i(\alpha,t)$ is ${\rm TP_2}$ over $(i,t) \in \lbrace 1, 2\rbrace \times [\beta, 1]$''.}. Let us take the FGM copula as an example. Let $C_i$ be the bivariate FGM copula with parameters $\theta_i$, for $i=1,2$, which implies that, for any fixed $\alpha\in[0,1)$,
\begin{eqnarray}\label{equa:FGMtailcopula}
  \overline{C}_i(\alpha,t)  &=&  1-\alpha-t+C_i(\alpha,t) \nonumber\\
    &=&  1-\alpha-t+\alpha t[1+\theta_i(1-\alpha)(1-t)] \nonumber\\
    &=& (1-\alpha)(1-t)(1+\theta_i\alpha t),\quad\mbox{for $t\in[0,1]$.}
\end{eqnarray}
Hence, we have
$$l_{\alpha}(t):=\frac{\overline{C}_2(\alpha,t)}{\overline{C}_1(\alpha,t)}=\frac{1+\theta_2\alpha t}{1+\theta_1\alpha t}.$$
If $\theta_1\leq\theta_2$, then it is easy to compute that $l'_{\alpha}(t)\stackrel{\rm sign}{=}\alpha(\theta_2-\theta_1)\geq0$, which means that ``$l_{\alpha}(t)$ increases w.r.t. $t \in [\beta,1]$'' equals to ``$\theta_1\leq\theta_2$'' under the setting of FGM copula.

Combining the results in Theorem \ref{the:twoJMES} and Theorem \ref{thm:diff_copula_same_margin}, the following general results which consider bivariate random vectors with different marginals and different copulas can be reached.
\begin{theorem} \label{thm:dc-dm3combine}
    Let $(X_1,Y_1)$ and $(X_2, Y_2)$ be two bivariate random vectors with c.d.f.'s $(F_1,G_1)$ and $(F_2,G_2)$ and copulas $C_1$ and $C_2$, respectively. Suppose either $X_1\uparrow_{\rm SI}Y_1$ or $X_2\uparrow_{\rm SI}Y_2$ is satisfied and $l_{\alpha}(t) \geq l_{\alpha}(\beta)$ for all $t \in [\beta,1]$,  for the fixed $\alpha\in[0,1)$. Then
    \begin{enumerate}[(i)]
        \item $Y_1\leq_{\rm icx}Y_2$ implies ${\rm JMES}_{\alpha,\beta}[Y_1|X_1]\leq{\rm JMES}_{\alpha,\beta}[Y_2|X_2]$;
        \item $Y_1\leq_{\rm disp}Y_2$ implies $\Delta {\rm JMES}_{\alpha,\beta}[Y_1|X_1]\leq\Delta {\rm JMES}_{\alpha,\beta}[Y_2|X_2]$;
        \item $Y_1\leq_{\rm epw}Y_2$ implies ${\rm \Delta^R JMES}_{\alpha,\beta}[Y_1|X_1]\leq {\rm \Delta^R JMES}_{\alpha,\beta}[Y_2|X_2]$.
    \end{enumerate}
\end{theorem}

\section{Examples}\label{Sec:numerical}
This section is composed of two parts with Subsection \ref{subsec:JMESformula} providing the explicit expressions of JMES-based measures for some commonly used bivariate distributions, and Subsection \ref{subsec:numeexam} showing some illustrative numerical examples to demonstrate our theoretical findings.
\subsection{Expressions of JMES-based measures for some bivariate distributions}\label{subsec:JMESformula}
In this part, we present explicit expressions of the three newly introduced systemic risk measures based on the bivariate normal, bivariate log-normal and bivariate $t$-distributions.
\begin{example}\label{exampleBinorm}
{\rm (Bivariate normal distribution)} Assume that $(X,Y)\sim N(\boldsymbol{\mu},\Sigma)$, where $\boldsymbol{\mu}=(\mu_{X},\mu_{Y})$, $\mu_{X},\mu_{Y}\in\mathbb{R}$,
$
    \Sigma=\left(\begin{array}{cc}
       \sigma_{X}^2  & \rho\sigma_{X}\sigma_{Y} \\
        \rho\sigma_{X}\sigma_{Y} & \sigma_{Y}^2
    \end{array}\right)
$, $\sigma_{X},\sigma_{Y}>0$, and $\rho\in(-1,1)$. Then, the copula of $(X,Y)$ is given by
\begin{equation}\label{copulanormal}
    C_{\rho}(u,v)=\int_{-\infty}^{\Phi^{-1}(u)}\int_{-\infty}^{\Phi^{-1}(v)}\frac{1}{2\pi\sqrt{1-\rho^2}}\exp\left\{-\frac{s_{1}^2-2\rho s_{1}s_{2}+s_{2}^2}{2(1-\rho^2)}\right\}\dif s_{1}\dif s_{2},
\end{equation}
where $\Phi^{-1}(\cdot)$ is the inverse of the standard normal c.d.f. Firstly, it can be easily verified that $p=G(\mu_{Y}+\sigma_{Y}\Phi^{-1}(p))$, for any $p\in(0,1)$. Particularly,
\begin{equation}
    G^{-1}(\overline{h}^{-1}_{\alpha,\beta}(t))=\mu_{Y}+\sigma_{Y}\Phi^{-1}(\overline{h}^{-1}_{\alpha,\beta}(t)),
\end{equation}
where $\overline{h}_{\alpha,\beta}(t)$ is defined as (\ref{eq:JMES-calclt}), and $\overline{h}^{-1}_{\alpha,\beta}(t)=\inf\{x\in[0,1]|\overline{h}_{\alpha,\beta}(x)\geq t\}$, $t\in [0,1]$. Therefore, according to Lemma \ref{lem:int}, it follows that
\begin{eqnarray*}
    {\rm JMES}_{\alpha,\beta}[Y|X]&=&\int_{0}^1\left[\mu_{Y}
    +\sigma_{Y}\Phi^{-1}(\overline{h}^{-1}_{\alpha,\beta}(t))\right]\dif t\\
    &=&\mu_{Y}+\sigma_{Y}\int_{\beta}^1\Phi^{-1}(t)\dif \overline{h}_{\alpha,\beta}(t).
\end{eqnarray*}
Furthermore, it is known that
\begin{eqnarray*}
{\rm ES}_\beta(Y) &=&\frac{1}{1-\beta} \int_\beta^1 \operatorname{VaR}_t(Y) \dif t=\frac{1}{1-\beta} \int_\beta^1\left(\Phi^{-1}(t) \sigma_Y+\mu_Y\right) \dif t \\
& =&\frac{1}{1-\beta}\left(\int_\beta^1 \Phi^{-1}(t) \sigma_Y \dif t+\int_\beta^1 \mu_Y \dif t\right)=\frac{\sigma_Y}{1-\beta} \int_\beta^1 \Phi^{-1}(t)  \dif t+\mu_Y\\
&=& \sigma_Y\frac{\Phi'(\Phi^{-1}(\beta))}{1-\beta}+\mu_Y.
\end{eqnarray*}
Thus,
\begin{equation*}
  {\rm\Delta JMES}_{\alpha,\beta}[Y|X] = \sigma_{Y}\left[\int_{\beta}^1\Phi^{-1}(t)\overline{h}'_{\alpha,\beta}(t)\dif t-\frac{\Phi'(\Phi^{-1}(\beta))}{1-\beta}\right],
\end{equation*}
where $\Phi'$ is the p.d.f. of the standard normal r.v.
\end{example}


\begin{example}\label{Bilognorm}
    {\rm(Bivariate log-normal distribution)} Assume that $(\log X,\log Y)\sim N(\boldsymbol{\mu},\Sigma)$, where $\boldsymbol{\mu}$ and $\Sigma$ are defined as in Example \ref{exampleBinorm}. Then, the copula of $(X,Y)$ is again given in (\ref{copulanormal}) since $(\log X,\log Y)$ are monotone transformations of $(X,Y)$. Moreover, for any $p\in[0,1]$, it can be verified that $p=G(e^{\mu_{Y}+\sigma_{Y}\Phi^{-1}(p)})$, which leads to
    \begin{equation*}
   G^{-1}(\overline{h}^{-1}_{\alpha,\beta}(t))=e^{\mu_{Y}+\sigma_{Y}\Phi^{-1}(\overline{h}^{-1}_{\alpha,\beta}(t))}.
   \end{equation*}
Here, $\overline{h}_{\alpha,\beta}(t)$ is defined in  (\ref{eq:JMES-calclt}) and $\overline{h}^{-1}_{\alpha,\beta}(t)$ is defined the same as in Example \ref{exampleBinorm}. By Lemma \ref{lem:int}, it follows that
   \begin{equation*}
       {\rm JMES}_{\alpha,\beta}[Y|X]=\int_{0}^1e^{\mu_{Y}+\sigma_{Y}\Phi^{-1}(\overline{h}^{-1}_{\alpha,\beta}(t))}\dif t=e^{\mu_{Y}}\int_{\beta}^1e^{\sigma_{Y}\Phi^{-1}(t)}\dif \overline{h}_{\alpha,\beta}(t).
\end{equation*}
On the other hand, according to Exercise 2.7.17 in \cite{Denuit2005} we have
   \begin{eqnarray*}
       {\rm ES}_{\beta}[Y]&=&\frac{1}{1-\beta}\int_{\beta}^1{\rm VaR}_{t}(Y)\dif t=\frac{1}{1-\beta}\int_{\beta}^1e^{\Phi^{-1}(t)\sigma_{Y}+\mu_{Y}}\dif t\\
       &=&\frac{e^{\mu_{Y}}}{1-\beta}\int_{\beta}^1e^{\Phi^{-1}(t)\sigma_{Y}}\dif t=e^ {\mu_{Y}+\frac{\sigma_{Y}^2}{2}} \frac{\Phi\left(\sigma_{Y}-\Phi^{-1}(p)\right)}{1-p}.
   \end{eqnarray*}
As a result, one can obtain
\begin{equation*}
  {\rm\Delta JMES}_{\alpha,\beta}[Y|X]= e^{\mu_{Y}}\int_{\beta}^1e^{\sigma_{Y}\Phi^{-1}(t)}\dif \overline{h}_{\alpha,\beta}(t)-e^ {\mu_{Y}+\frac{\sigma_{Y}^2}{2}} \frac{\Phi\left(\sigma_{Y}-\Phi^{-1}(p)\right)}{1-p}.
\end{equation*}
\end{example}



\begin{example}
    {\rm(Bivariate Student $t$ distribution)} Assume that $(X,Y)\sim t_{\nu} (\boldsymbol{0},\Sigma)$, where $\Sigma=
\left(\begin{array}{cc}
   1  & \rho \\
    \rho &  1
\end{array}\right)$, $\rho\in(-1,1)$, and $\nu>0$ is the common
degrees of freedom of the marginals. The copula of $(X,Y)$ is then given by
\begin{equation} \label{eq:tcopula}
    C_{\rho,\nu}(u,v)=\int_{-\infty}^{t_{\nu}^{-1}(u)}\int_{-\infty}^{t_{\nu}^{-1}(v)}\frac{1}{2\pi\sqrt{1-\rho^2}}\left\{1+\frac{s_{1}^2-2\rho s_{1}s_{2}+s_{2}^2}{\nu(1-\rho^2)}\right\}^{-\frac{\nu+2}{2}}\dif s_{1}\dif s_{2},
\end{equation}
where $t_{\nu}^{-1}(\cdot)$ is the inverse of the Student $t$ c.d.f. with $\nu$ degrees of freedom. The expressions for the {\rm JMES}-based measures are hard to obtain for general $\nu>0$. While for $\nu=1$ and $\nu=2$, we can reach the following results:
\begin{itemize}
    \item[(i)] For $\nu=1$, we have
\begin{equation*}
    {\rm JMES}_{\alpha,\beta}[Y|X]=\int_{0}^1 \tan\left(\pi\left(\overline{h}_{\alpha,\beta}^{-1}(p)-\frac{1}{2}\right)\right)\dif p=\int_{\beta}^1 \tan\left(\pi\left(t-\frac{1}{2}\right)\right)\dif \overline{h}_{\alpha,\beta}(t).
\end{equation*}
\item[(ii)] For $\nu=2$, we have
\begin{equation*}
    {\rm JMES}_{\alpha,\beta}[Y|X]=\int_{0}^1 \frac{\sqrt{2}(\overline{h}_{\alpha,\beta}^{-1}(p)-\frac{1}{2})}{\sqrt{\overline{h}_{\alpha,\beta}^{-1}(p)(1-\overline{h}_{\alpha,\beta}^{-1}(p))}}\dif p=\int_{\beta}^1 \frac{\sqrt{2}(t-\frac{1}{2})}{\sqrt{t-t^2}}\dif \overline{h}_{\alpha,\beta}(t).
\end{equation*}
\end{itemize}
The {\rm JMES}-based risk contribution measures can be obtained similarly for the above two cases, which are omitted here for brevity.
\end{example}

\begin{figure}[htbp!]
    \subfigure[]{
      \begin{minipage}[t]{0.45\textwidth}
        \centering
        \includegraphics[width=\textwidth]{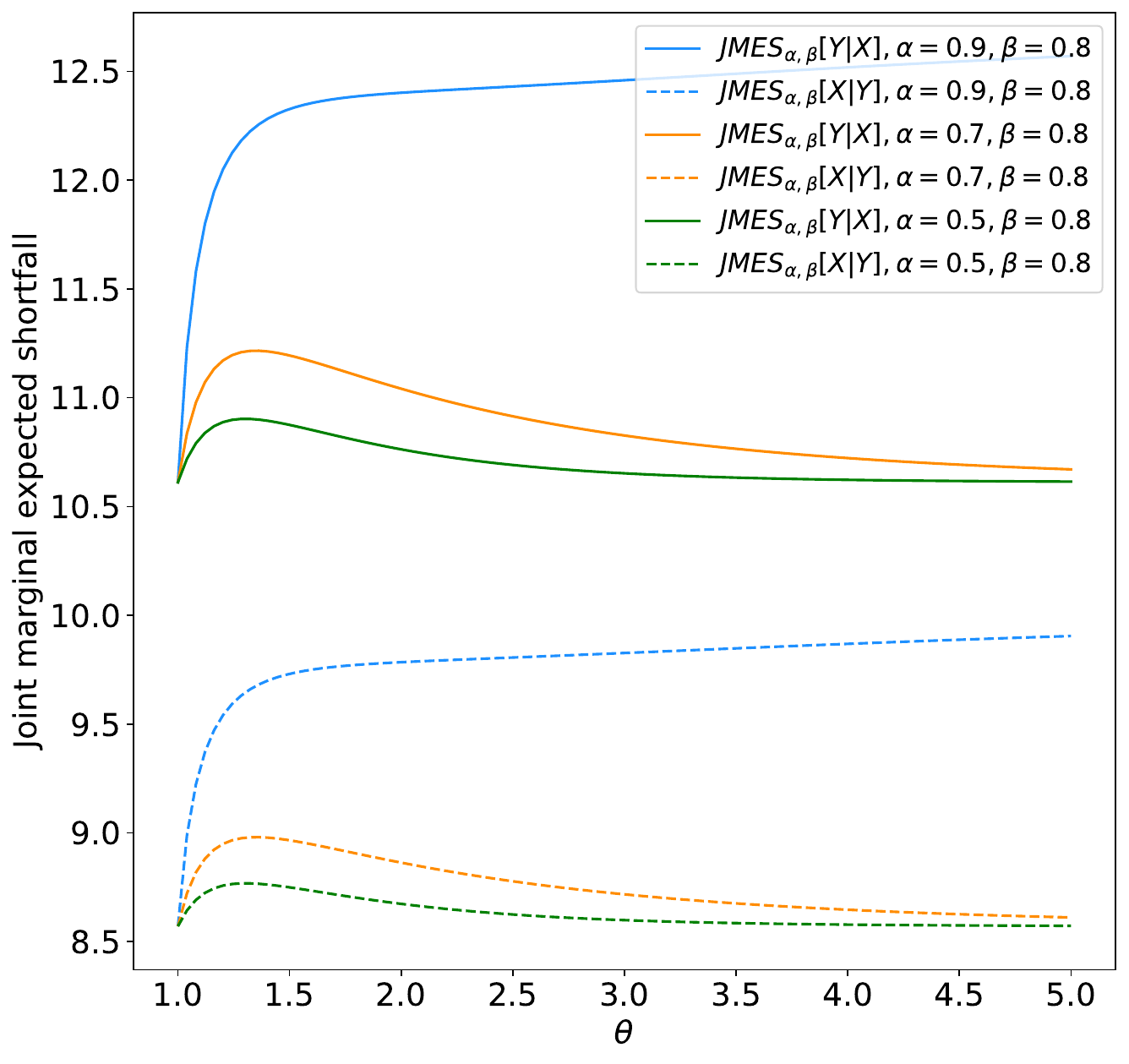}
        \label{fig:Thm4.1}
      \end{minipage}
    }
    \hfill
    \subfigure[]{
      \begin{minipage}[t]{0.45\textwidth}
        \centering
        \includegraphics[width=\textwidth]{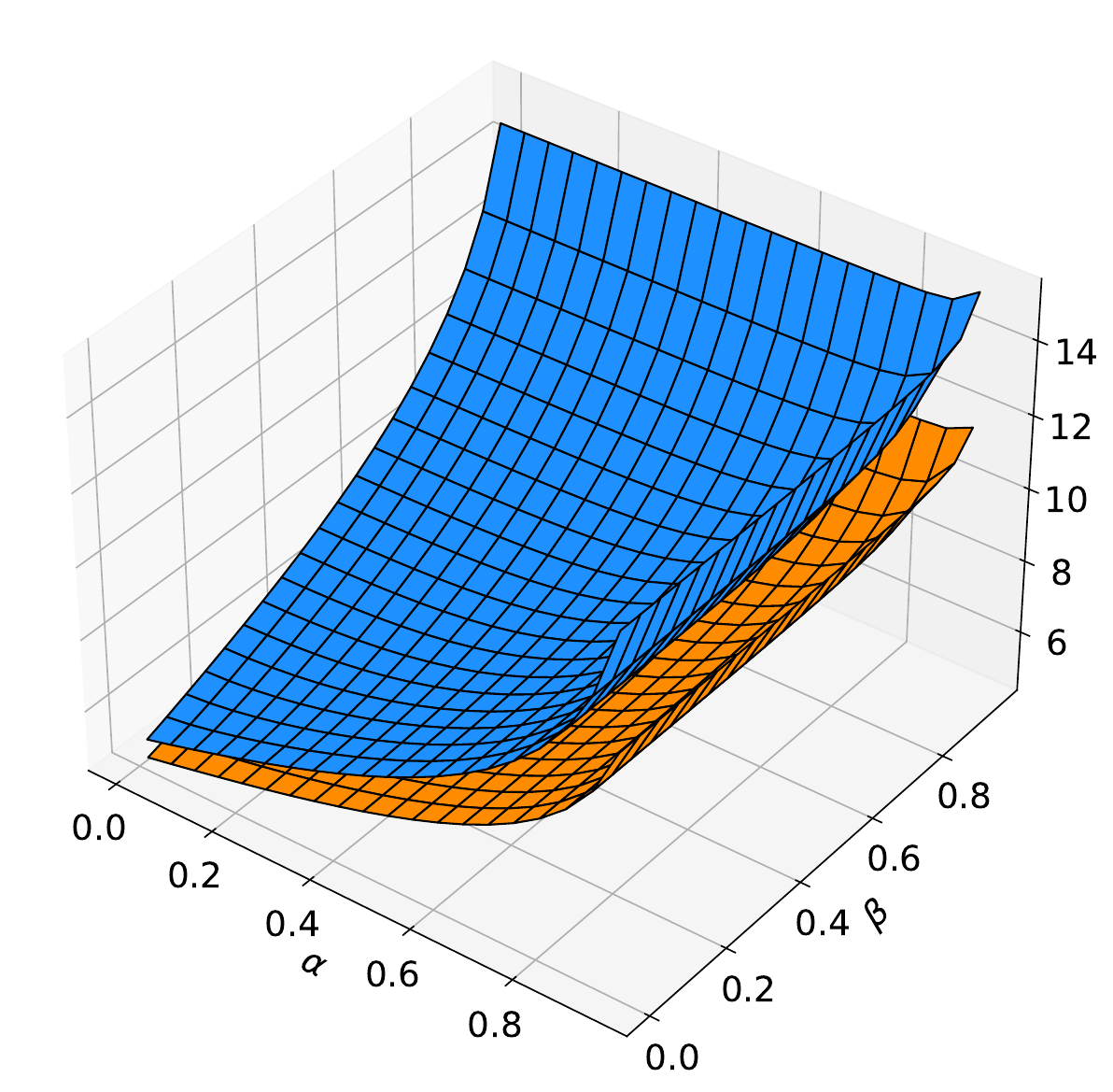}
        \label{fig:Thm4.1(2)}
      \end{minipage}
    }
    \caption{(a) Plot of ${\rm JMES}_{\alpha,\beta}[X | Y]$ and ${\rm JMES}_{\alpha,\beta}[Y | X]$ w.r.t. $\theta$ under different values of $\alpha$ and $\beta$. (b) The graphic plots ${\rm JMES}_{\alpha,\beta}[X | Y]$ and ${\rm JMES}_{\alpha,\beta}[Y | X]$ as functions of $(\alpha, \beta) \in [0,1) \times [0,1)$. The brown surface corresponds to ${\rm JMES}_{\alpha,\beta}[X | Y]$ and the blue one corresponds to ${\rm JMES}_{\alpha,\beta}[Y | X]$.}
    \label{fig:Thm4.1ab}
  \end{figure}

\subsection{Numerical examples}\label{subsec:numeexam}
In this subsection, we provide some numerical examples to illustrate our main findings. A r.v. $X$ is said to follow a gamma distribution with shape parameter $\alpha>0$ and scale parameter $\beta>0$, denoted by $X \sim {\rm Gam}(\alpha, \beta)$, if its p.d.f. is given by
\begin{equation}\nonumber
    f(x)=\frac{x^{\alpha-1} \exp \left\{-\frac{x}{\beta}\right\}}{\beta^\alpha \Gamma(\alpha)}, \quad \text { for all } x >0.
\end{equation}
Suppose $Y_1 \sim {\rm Gam}(\alpha_1, \beta_1)$ and $Y_2 \sim {\rm Gam}(\alpha_2, \beta_2)$. According to Tables 2.1-2.2 in \cite{belzunce2015introduction}, the following statements hold:
\begin{enumerate}[(a)]
  \item $Y_1 \leq_{\rm icx} Y_2$ when $\alpha_1 > \alpha_2$ and $\alpha_1 \beta_1 \leq \alpha_2 \beta_2$;
  \item $Y_1 \leq_{\rm disp} Y_2$ when $\alpha_1 \leq \alpha_2$ and $\beta_1 \leq \beta_2$;
  \item $Y_1 \leq_{\rm epw} Y_2$ when $\alpha_1 \geq \alpha_2$.
\end{enumerate}


  \begin{figure}[htbp!]
    \subfigure[]{
      \begin{minipage}[t]{0.45\textwidth}
        \centering
        \includegraphics[width=\textwidth]{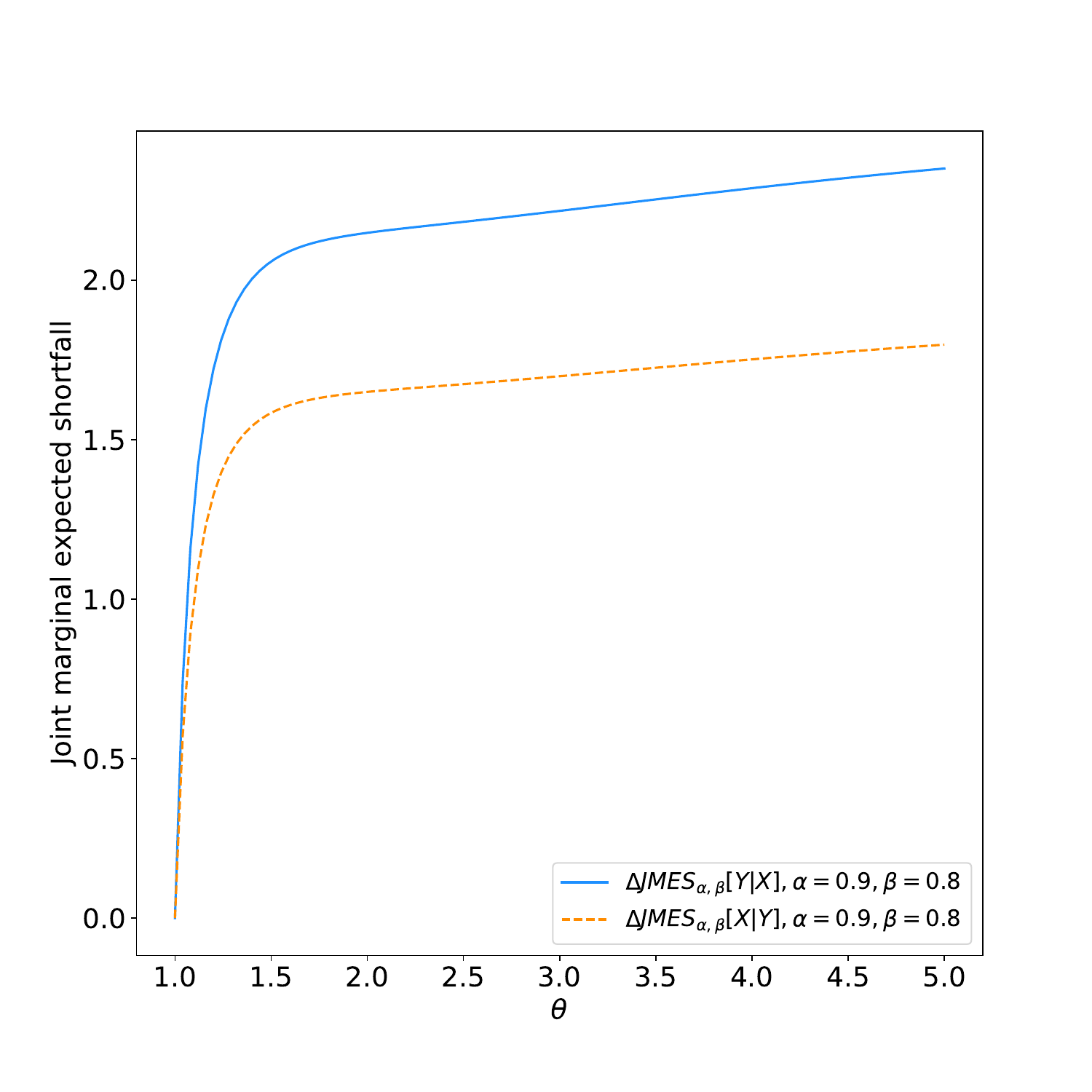}
        \label{fig:Thm4.2}
      \end{minipage}
    }
    \hfill
    \subfigure[]{
      \begin{minipage}[t]{0.45\textwidth}
        \centering
        \includegraphics[width=\textwidth]{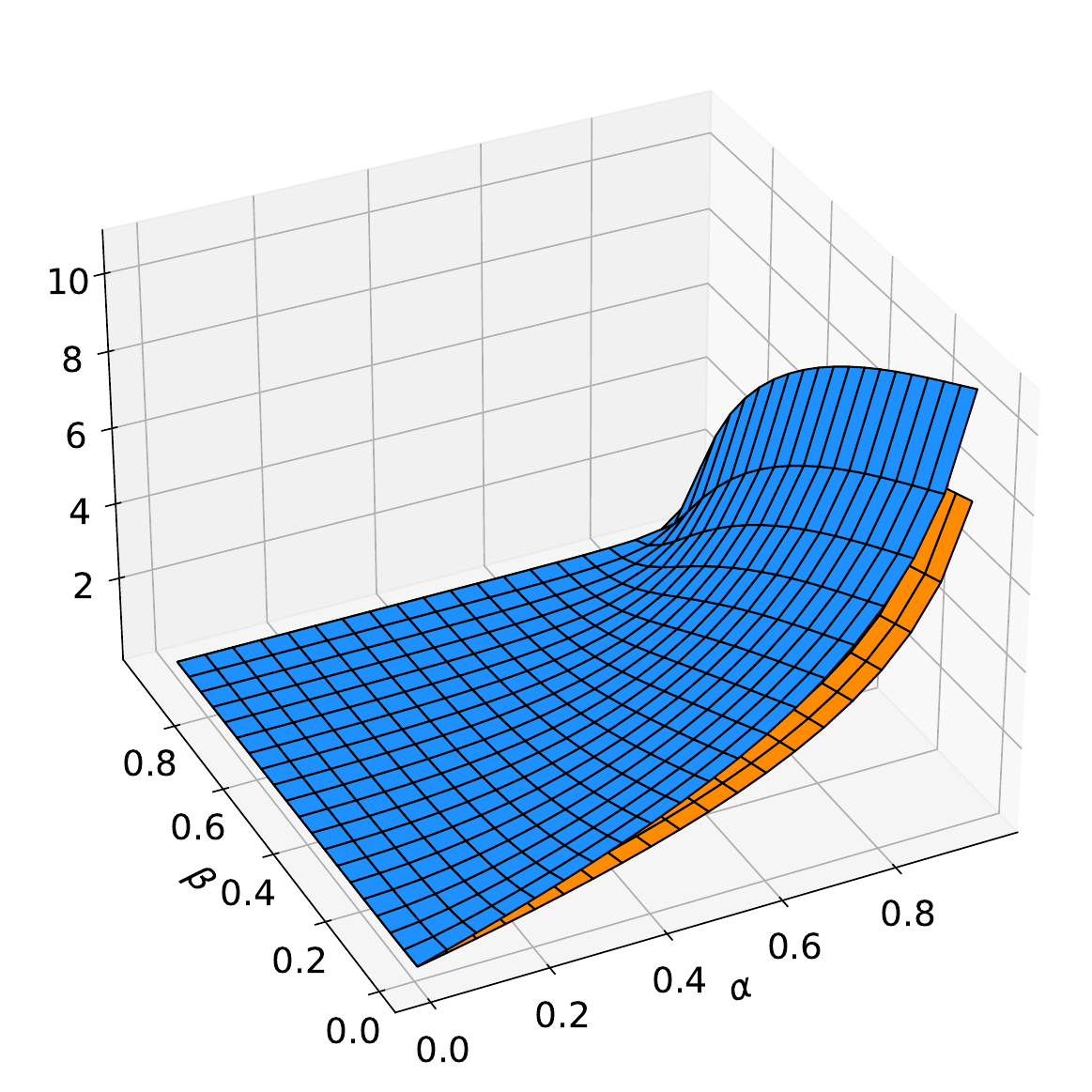}
        \label{fig:Thm4.2(2)}
      \end{minipage}
    }
    \caption{(a) Plot of $\Delta {\rm JMES}_{\alpha,\beta}[X | Y]$ and $\Delta {\rm JMES}_{\alpha,\beta}[Y | X]$ as functions of $\theta$ when $\alpha=0.9, \beta = 0.8$. (b) The graphic plots $\Delta {\rm JMES}_{\alpha,\beta}[X | Y]$ and $\Delta {\rm JMES}_{\alpha,\beta}[Y | X]$ as functions of $(\alpha, \beta) \in [0,1) \times [0,1)$. The orange surface corresponds to $\Delta  {\rm JMES}_{\alpha,\beta}[X | Y]$,  while the blue one corresponds to $\Delta {\rm JMES}_{\alpha,\beta}[Y | X]$.}
    \label{fig:pairriskDJMES}
  \end{figure}

The following two examples illustrate the main findings in Sections \ref{pairedrisks} and \ref{twobivariateriskvec}.
\begin{example}\label{exp:pairedrisks} In this example, we assume that the random vector $(X,Y)$ admits a Gumbel copula $C$ with parameter $\theta$ as given in (\ref{copula:Gumbel}).
    \begin{enumerate}[(i)]
        \item Set $X \sim {\rm Gam}(3,1.5)$ and $Y \sim {\rm Gam}(2,2.5)$. Thus, it holds that $X \leq_{\rm icx} Y$ but $X \nleq_{\rm st} Y$. It is readily seen that ${\rm JMES}$ increases w.r.t. $\alpha$ for fixed $\beta$ from the plots in Figure \ref{fig:Thm4.1}, which illustrates the result of Theorem \ref{thm:monot-property}. Moreover, by setting  $\theta = 3$, Figure \ref{fig:Thm4.1(2)} displays that ${\rm JMES}_{\alpha,\beta}[Y|X] \geq {\rm JMES}_{\alpha,\beta}[X|Y]$ for all $\alpha,\beta\in[0,1)$, which validates the results of Theorem \ref{PairRisks}.
  \begin{figure}[htbp!]
    \subfigure[]{
      \begin{minipage}[t]{0.45\textwidth}
        \centering
        \includegraphics[width=\textwidth]{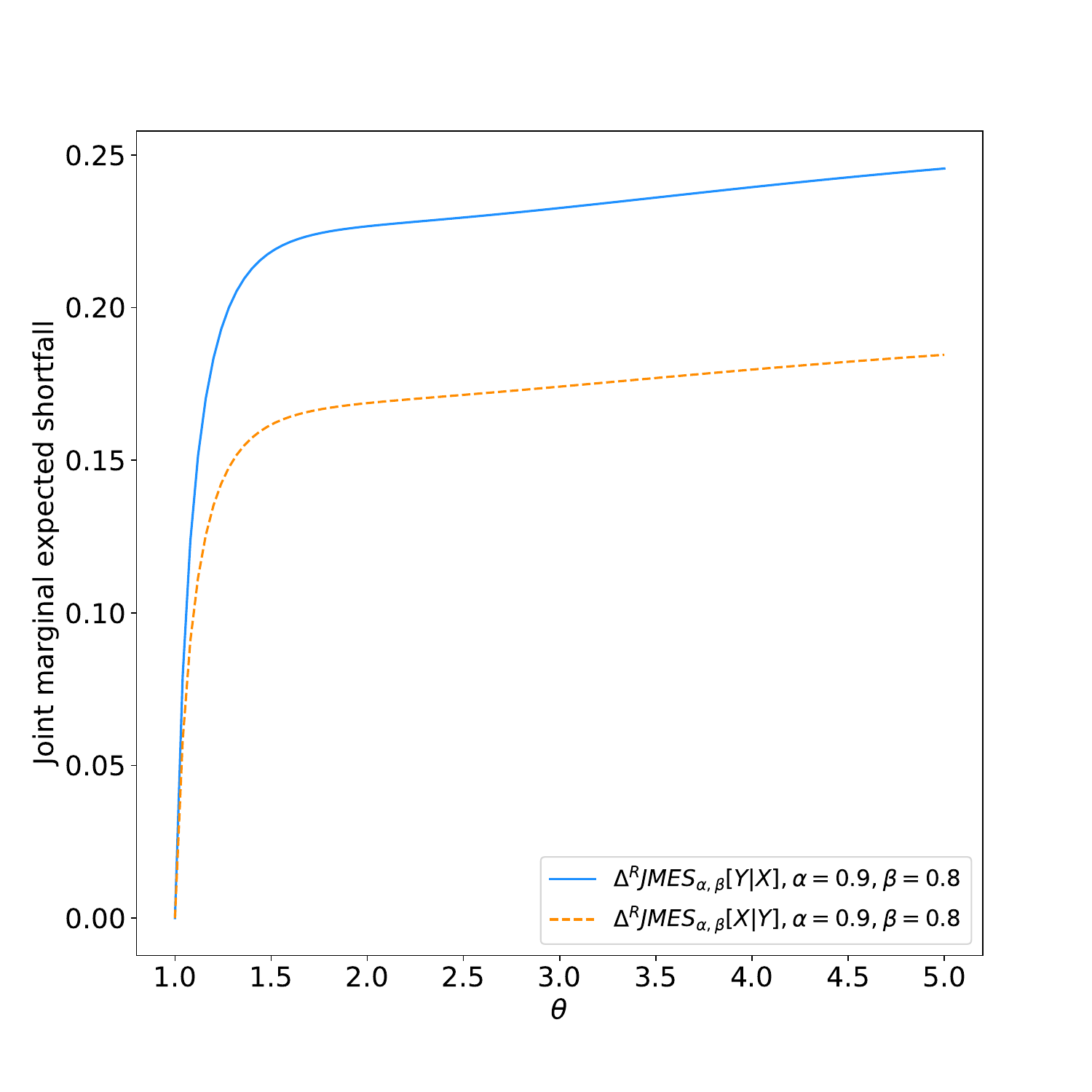}
        \label{fig:Thm4.3}
      \end{minipage}
    }
    \hfill
    \subfigure[]{
      \begin{minipage}[t]{0.45\textwidth}
        \centering
        \includegraphics[width=\textwidth]{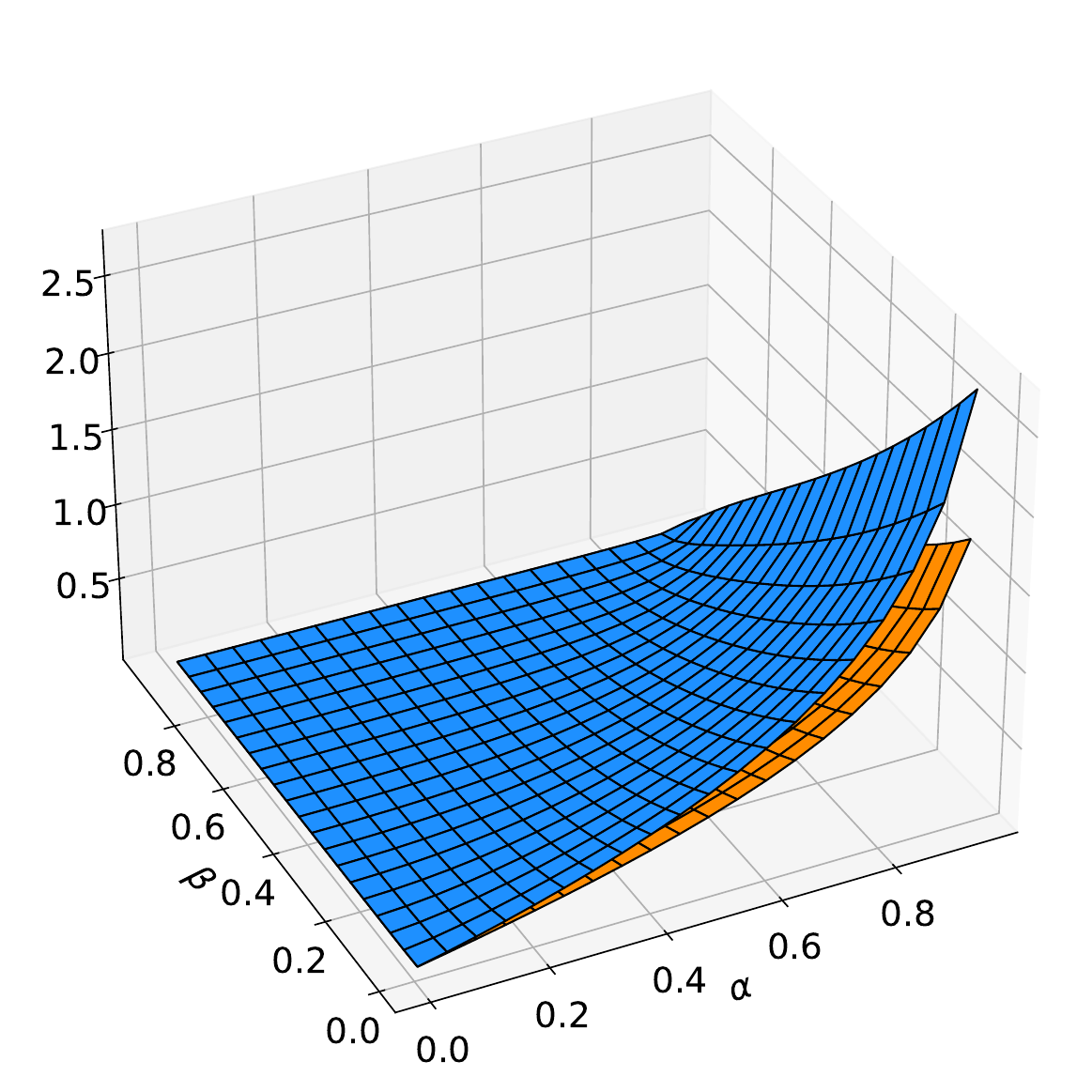}
        \label{fig:Thm4.3(2)}
      \end{minipage}
    }
    \caption{(a) Plot of $\Delta^{\rm R} {\rm JMES}_{\alpha,\beta}[X | Y]$ and $\Delta ^{\rm R}{\rm JMES}_{\alpha,\beta}[Y | X]$ as functions of $\theta$ when $\alpha=0.9, \beta = 0.8$. (b) The graphic plots $\Delta^{\rm R} {\rm JMES}_{\alpha,\beta}[X | Y]$ and $\Delta^{\rm R} {\rm JMES}_{\alpha,\beta}[Y | X]$ as functions of $(\alpha, \beta) \in [0,1) \times [0,1)$. The orange surface corresponds to $\Delta^{\rm R}  {\rm JMES}_{\alpha,\beta}[X | Y]$,  while the blue one corresponds to $\Delta^{\rm R}  {\rm JMES}_{\alpha,\beta}[Y | X]$.}
    \label{fig:pairriskDRJMES}
  \end{figure}
    \item As Figure \ref{fig:Thm4.1} shows, ${\rm JMES}_{\alpha,\beta}[Y|X]$ may not be monotone w.r.t. $\theta$ for some fixed $\alpha$; in other words, stronger positive dependence cannot imply the larger value of $\rm JMES$. We attributed the cause of the dependence inconsistency to two facts. On the one hand, the greater positive dependence between $X$ and $Y$ implies a greater value of expectation $\EE[Y \vert X > {\rm VaR}_{\alpha}[X]]$, instead of a greater value of the tail expectation. On the other hand, it can be verified that ${\rm JMES}_{\alpha,\beta}[Y|X] = {\rm ES}_{\beta}[Y]$ when $\theta = 1$ (independence setting) and ${\rm JMES}_{\alpha,\beta}[Y|X] \to {\rm ES}_{\alpha \vee \beta}[Y]$ when $\theta \to \infty$ (comonotonicity setting), which explains the failure of monotonicity of ${\rm JMES}$ w.r.t. $\theta$. 
        \item Set $X \sim {\rm Gam}(1.5,2.5)$ and $Y \sim {\rm Gam}(2,3)$. Thus, it holds that $X \leq_{\rm disp} Y$. As illustrated in Figure \ref{fig:Thm4.2}, $\Delta{\rm JMES}_{\alpha,\beta}[Y|X] \geq \Delta{\rm JMES}_{\alpha,\beta}[Y|X]$ for $\theta \in (1,5)$, and Figure \ref{fig:Thm4.2(2)} demonstrates that $\Delta{\rm JMES}_{\alpha,\beta}[X|Y]\leq\Delta{\rm JMES}_{\alpha,\beta}[Y|X]$ for all $(\alpha, \beta) \in [0,1) \times [0,1)$ when $\theta = 3$. Hence, the effectiveness of Theorem \ref{contributionpair} is validated.
        \item Set $X \sim {\rm Gam}(2,1.5)$ and $Y \sim {\rm Gam}(1, 1)$. Thus, it holds that $X \leq_{\rm epw} Y$. As shown in Figure \ref{fig:Thm4.3}, $\Delta^{\rm R}{\rm JMES}_{\alpha,\beta}[Y|X] \geq \Delta^{\rm R}{\rm JMES}_{\alpha,\beta}[Y|X]$ for all  $\theta \in (1,5)$. And the conclusion also holds for any $(\alpha, \beta) \in [0,1) \times [0,1)$ as displayed in Figure \ref{fig:Thm4.3(2)} when $\theta =3$. Therefore, the result of Theorem \ref{thm:ratiopair} is supported.
    \end{enumerate}
\end{example}

\begin{figure}[htbp!]
    \subfigure[]{
      \begin{minipage}[t]{0.45\textwidth}
        \centering
        \includegraphics[width=\textwidth]{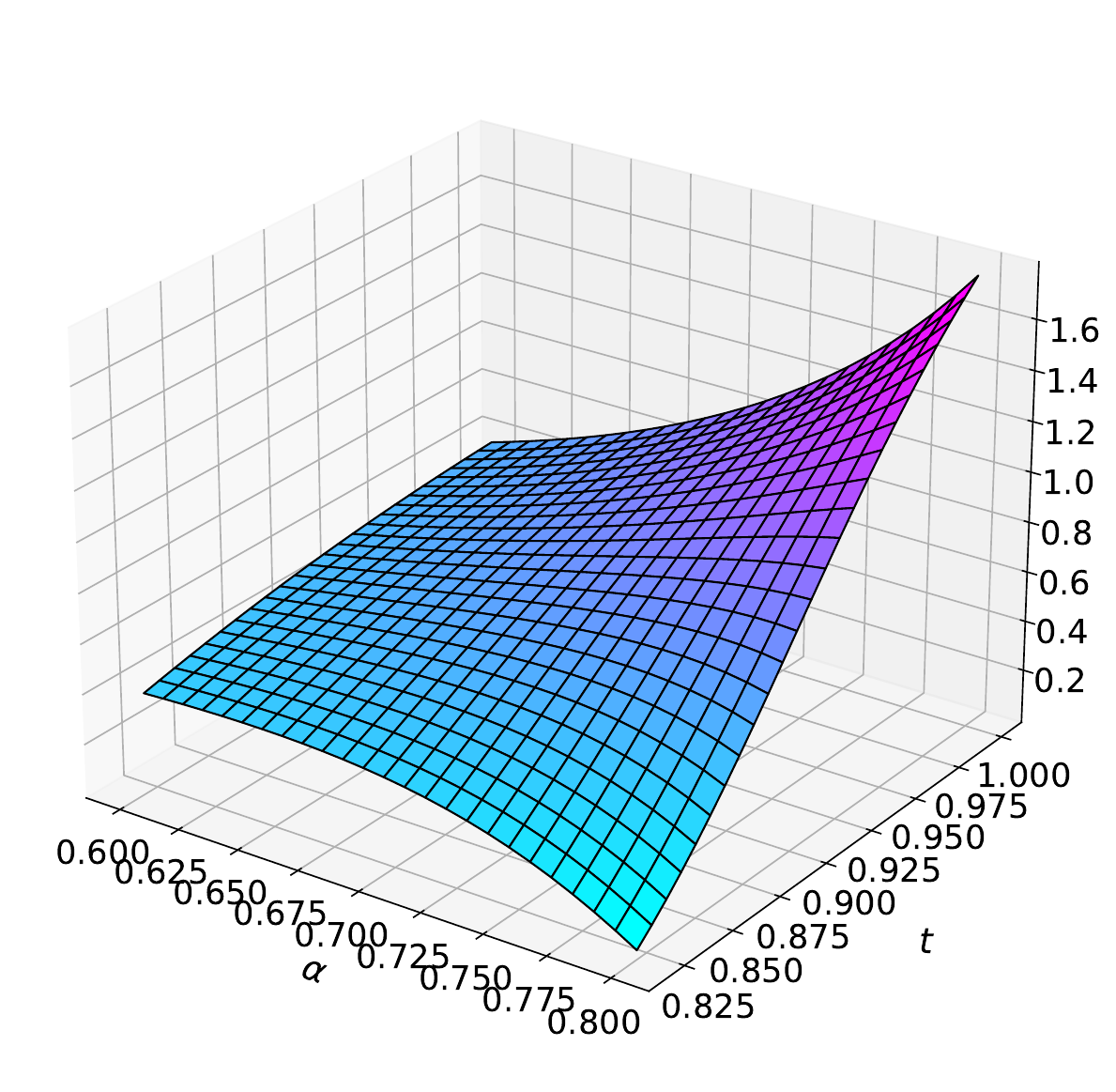}
        \label{fig:Tp2}
      \end{minipage}
    }
    \subfigure[]{
      \begin{minipage}[t]{0.45\textwidth}
        \centering
        \includegraphics[width=\textwidth]{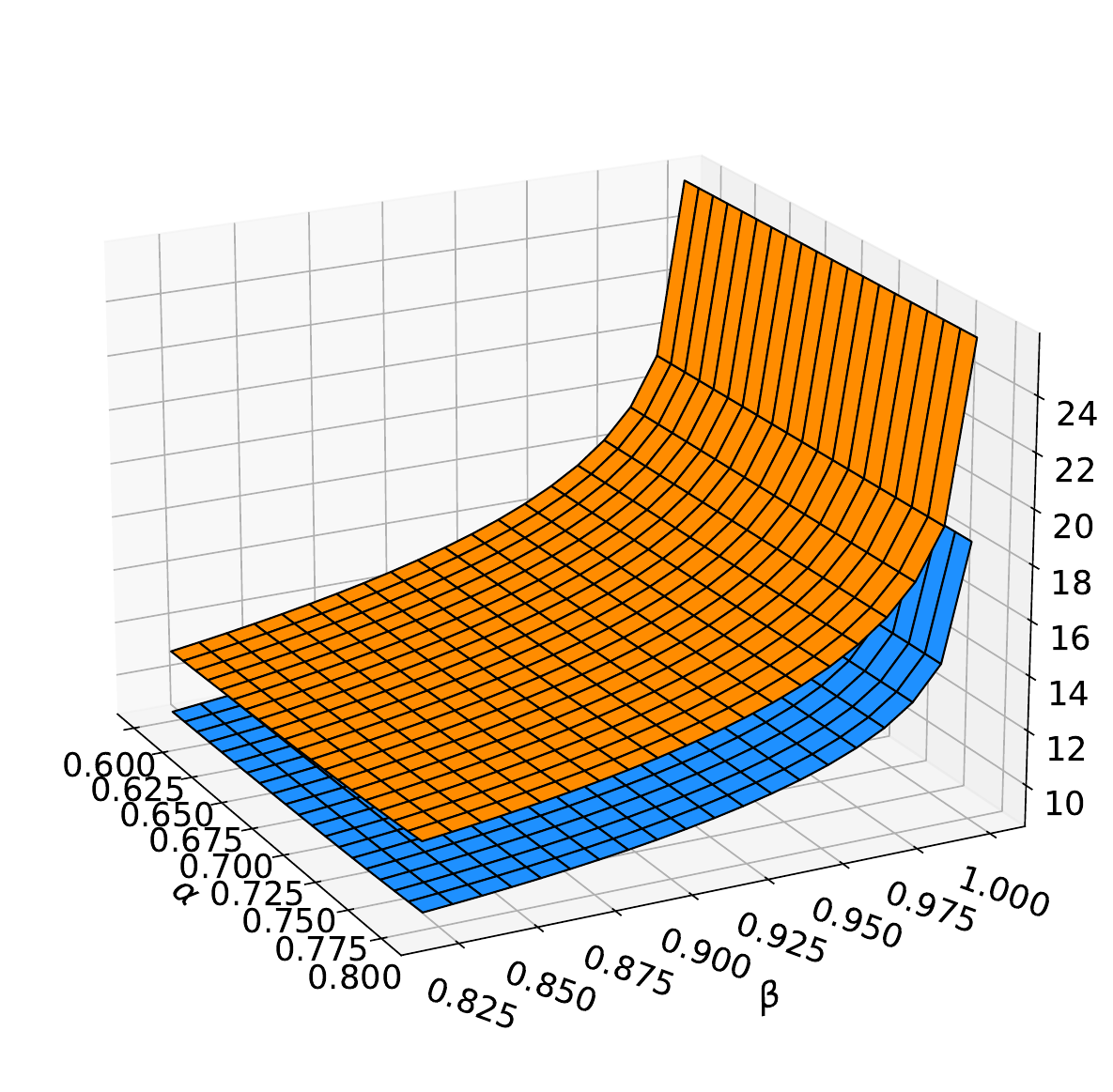}
        \label{fig:Thm5.6.1}
      \end{minipage}
    }
    \hfill
    \subfigure[]{
      \begin{minipage}[t]{0.45\textwidth}
        \centering
        \includegraphics[width=\textwidth]{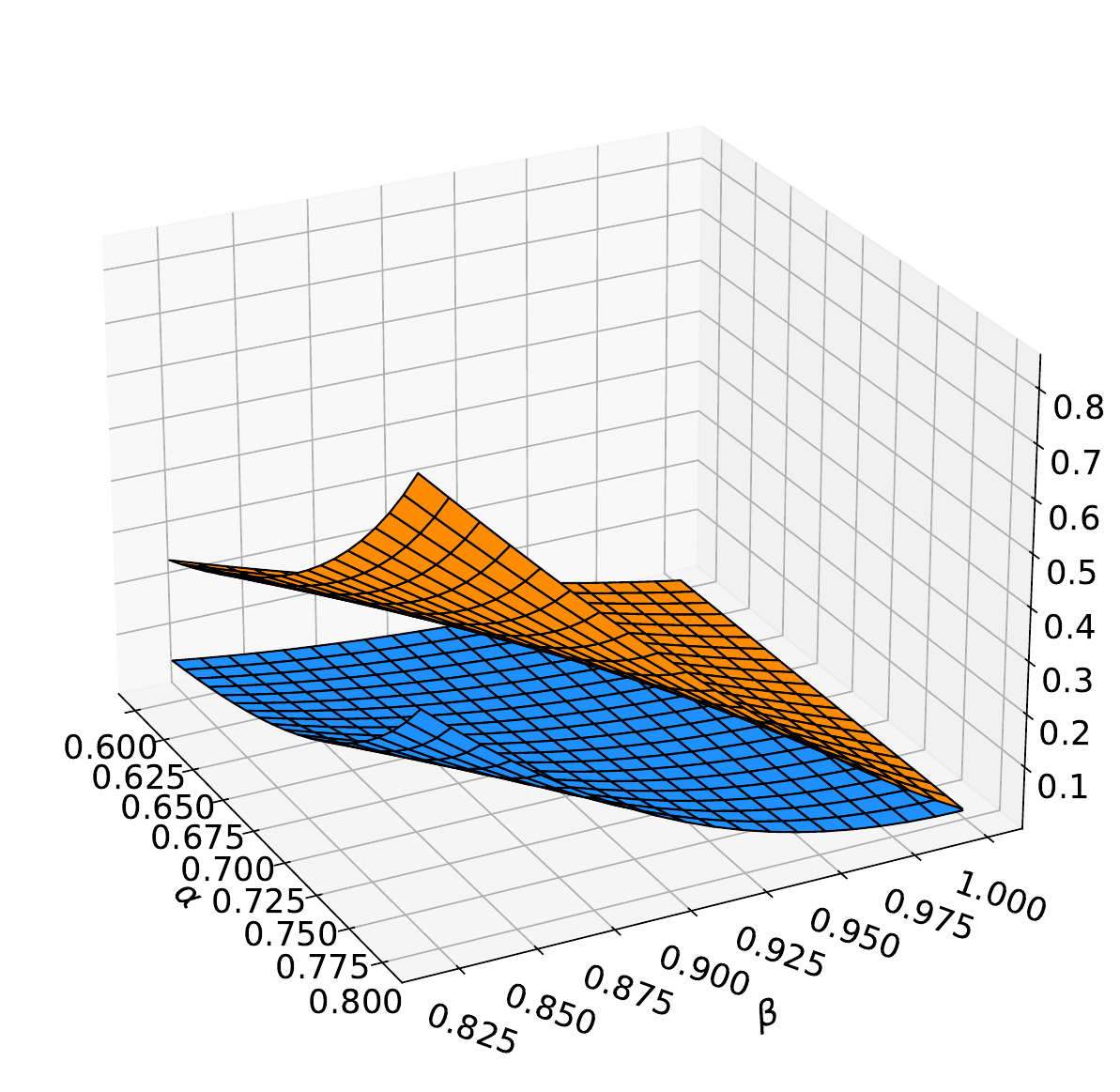}
        \label{fig:Thm5.6.2}
      \end{minipage}
    }
    \hfill
    \subfigure[]{
      \begin{minipage}[t]{0.45\textwidth}
        \centering
        \includegraphics[width=\textwidth]{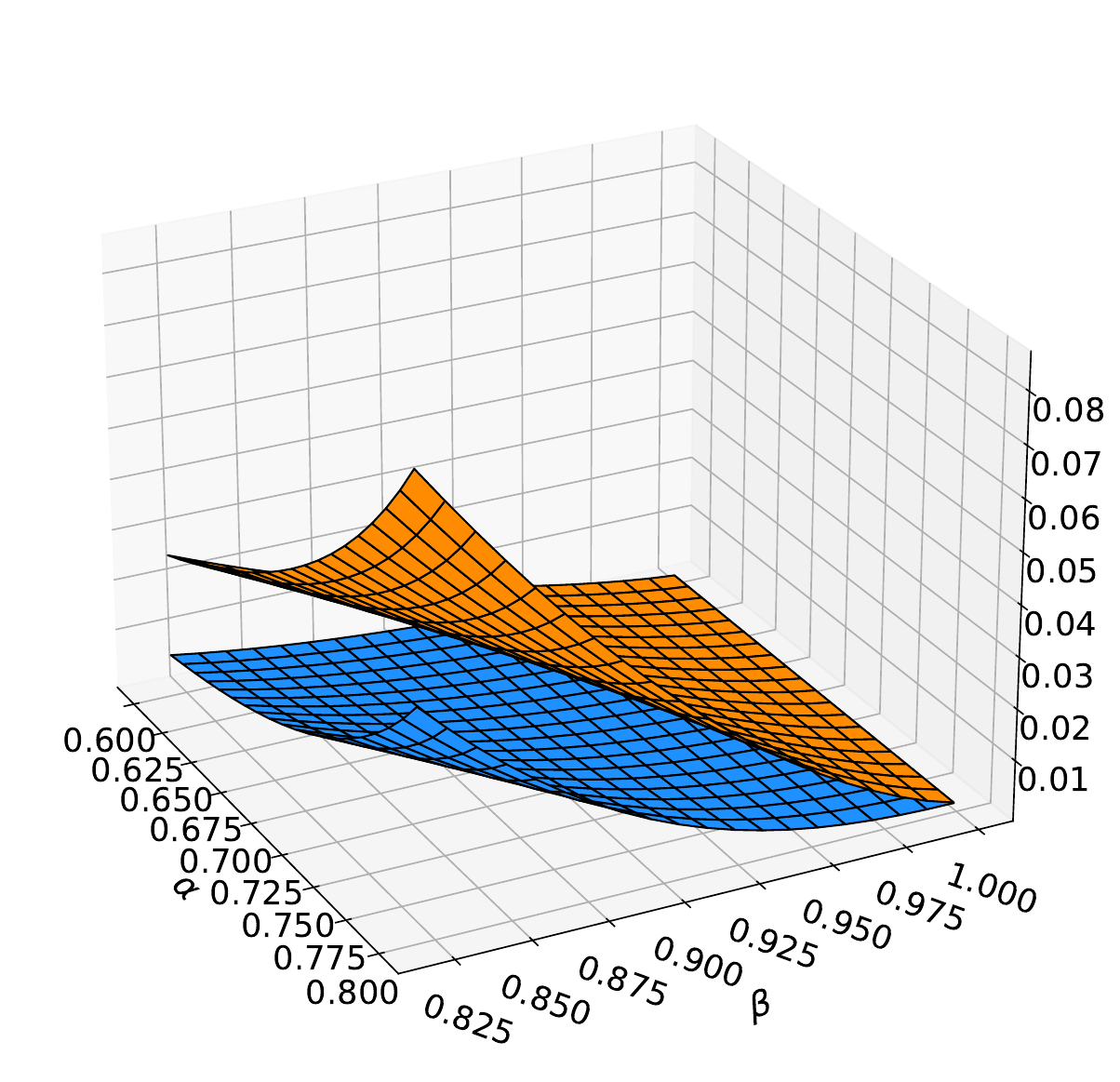}
        \label{fig:Thm5.6.3}
      \end{minipage}
    }
    \caption{(a) Plot of $l'_{\alpha}(t)$. (b) Plots of ${\rm JMES}_{\alpha,\beta}[Y_2 | X_2]$ (the orange surface) and ${\rm JMES}_{\alpha,\beta}[Y_1 | X_1]$ (the blue surface). (c) Plots of $\Delta {\rm JMES}_{\alpha,\beta}[Y_2 | X_2]$ (the orange surface) and $\Delta {\rm JMES}_{\alpha,\beta}[Y_1 | X_1]$ (the blue surface). (d) Plots of $ {\rm\Delta^R JMES}_{\alpha,\beta}[Y_2 | X_2]$ (the orange surface) and $ {\rm\Delta^R JMES}_{\alpha,\beta}[Y_1 | X_1]$ (the blue surface).}
    \label{fig:side_by_side4}
  \end{figure}

\begin{example} \label{exp:birisks} Assume that the random vector $(X_1,Y_1)$ admits a Gumbel copula $C_1$ with parameter $\theta_1 = 3$ and the random vector $(X_2,Y_2)$ admits another Gumbel copula $C_2$ with parameter $\theta_2 = 2$. Figure \ref{fig:Tp2} displays that, for any fixed $\alpha \in [0.6,0.8]$, the function $l'_{\alpha}(t)$ is greater than $0$ for all $t \in [0.82,1)$. Hence, it can be inferred that $l_{\alpha}(t) \geq l_{\alpha}(\beta)$ for all $\beta \in [0.82,1)$ and $t \in [\beta,1]$. Moreover, since the distributions of $X_1$ and $X_2$ play no role in calculating the values of the ${\rm JMES}$-based measures, we only consider the values of $\alpha$ in the following examples. We assume that both $X_1$ and $X_2$ are continuous r.v.'s.
    \begin{enumerate}[(i)]
        \item Set $Y_1 \sim {\rm Gam}(3, 1.5)$ and $Y_2 \sim {\rm Gam}(2, 2.5)$. Thus, it holds that $Y_1 \leq_{\rm icx} Y_2$ but $Y_1 \nleq_{\rm st} Y_2$. Figure \ref{fig:Thm5.6.1} displays that ${\rm JMES}_{\alpha,\beta}[Y_1 \vert X_1] \leq {\rm JMES}_{\alpha,\beta}[Y_2 \vert X_2]$ for all $\alpha\in [0.6, 0.8]$ and $\beta \in [0.82, 1)$. Thus, the result of Theorem \ref{thm:dc-dm3combine}(i) is validated.
        \item Set $Y_1 \sim {\rm Gam}(1.5,2.5)$ and $Y_2 \sim {\rm Gam}(2,3)$. Thus, it holds that $Y_1 \leq_{\rm disp} Y_2$. Figure \ref{fig:Thm5.6.2} plots that $\Delta {\rm JMES}_{\alpha,\beta}[Y_1 \vert X_1] \leq \Delta {\rm JMES}_{\alpha,\beta}[Y_2 \vert X_2]$ for all $\alpha\in [0.6, 0.8]$ and $\beta \in [0.82, 1)$. This agrees with the result of Theorem \ref{thm:dc-dm3combine}(ii).
        \item Set $Y_1 \sim {\rm Gam}(2,1.5)$ and $Y_2 \sim {\rm Gam}(1,1)$. Thus, it holds that $Y_1 \leq_{\rm epw} Y_2$. Figure \ref{fig:Thm5.6.3} displays that $ {\rm\Delta^R JMES}_{\alpha,\beta}[Y_1 \vert X_1] \leq {\rm\Delta^R JMES}_{\alpha,\beta}[Y_2 \vert X_2]$ for all $\alpha,\in[0.6, 0.8]$ and $\beta \in [0.82, 1)$, which illustrates Theorem \ref{thm:dc-dm3combine}(iii).
    \end{enumerate}
\end{example}

\section{An application}\label{realapplication}
\begin{figure}[htbp!]
        \centering
        \subfigure[]{
        \includegraphics[width=7cm]{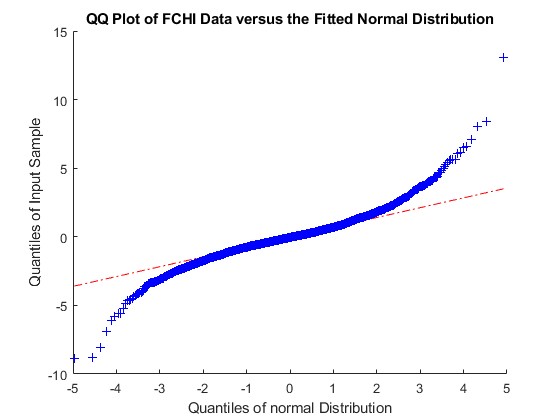}
        }
        \quad
        \subfigure[]{
        \includegraphics[width=7cm]{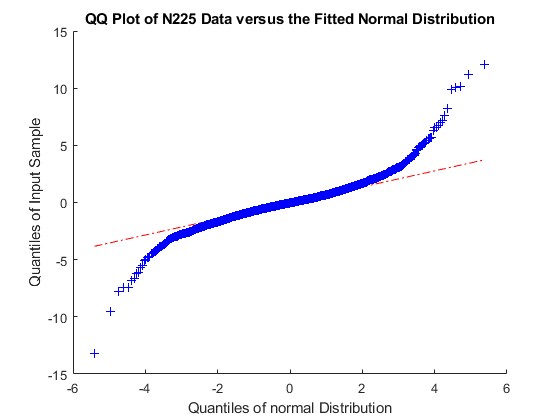}
        }
        \quad
        \subfigure[]{
        \includegraphics[width=7cm]{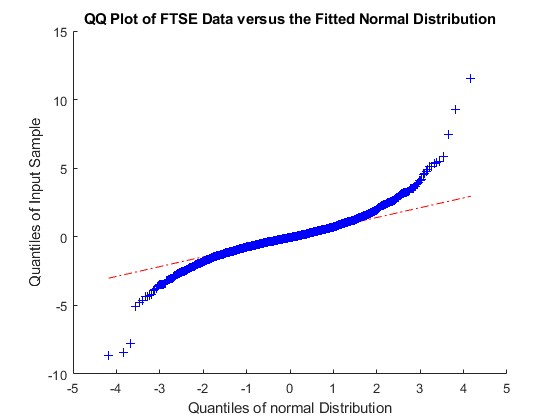}
        }
        \quad
        \subfigure[]{
        \includegraphics[width=7cm]{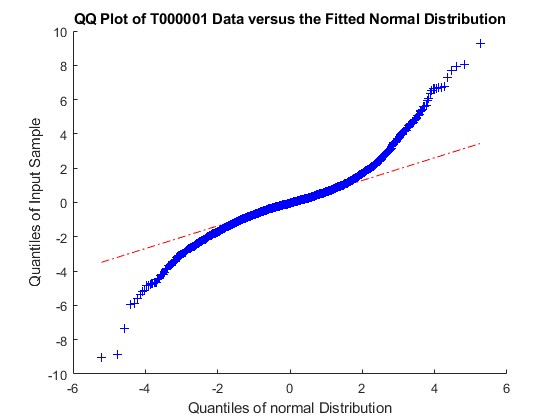}
        }
        \quad
        \subfigure[]{
        \includegraphics[width=7cm]{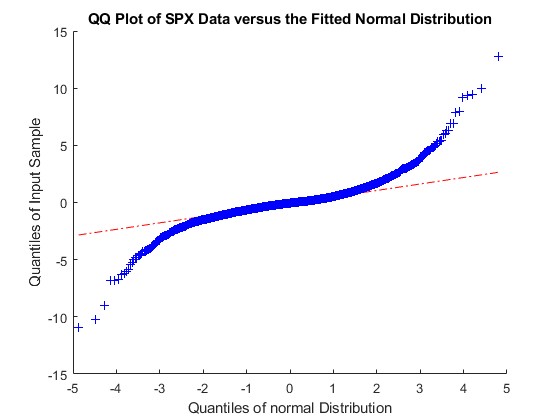}
        }
        \quad
        \subfigure[]{
        \includegraphics[width=7cm]{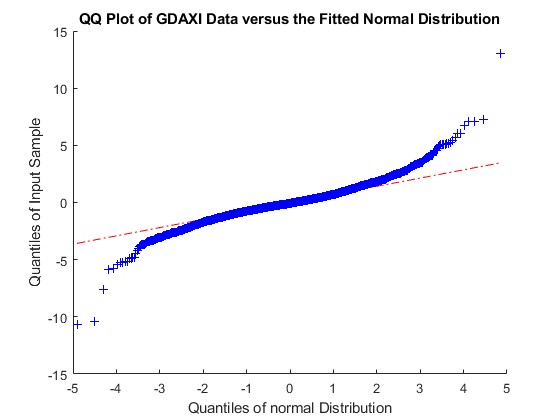}
        }
        \caption{QQ plots of {$L_t$} based on normal distributions for all six stock indexes.}
        \label{fig:NormalQQplot}
        \end{figure}

In this section, we investigate the spillover effects of risk contagion from the U.S. financial market to other major financial markets over the world by applying the JMES and its variety versions of contribution measures as well as some other well-known systemic risk measures. We adopt a database of daily closing prices arising from six main stock indices, including the Standard \& Poor's 500 Index (S\&P 500), the Financial Times Stock Exchange 100 Index (FTSE), the CAC 40 Index (FCHI), the Nikkei Index (N225), the German stock index (GDAXI), and the Shanghai Composite Index (SHZ). The observation period spans from January 1, 2007, to December 31, 2022\footnote{This period covers the 2008 financial crisis and the COVID-19 pandemic which begins at the beginning of 2019.}, and the data has been sourced from \url{choice.eastmoney.com}. To account for the time zone differences between the U.S. stock market and the other five stock markets, we have considered day $t-1$ as the trading day for the U.S. stock market and day $t$ as the corresponding trading day for the UK, France, Japan, Germany, and China stock markets. We compute the daily index losses by taking the logarithm of the closing prices and calculating the first-order differences. In other words, we let
\begin{equation}\nonumber
    L_t=-100 \cdot \log \left(p_t / p_{t-1}\right),
\end{equation}
where $L_t$ represents the negative of 100 times log-losses \footnote{Differing from the majority of economics and finance literature, this paper focuses on risk measurement based on the probability distributions of {losses}. Since we denote {$L_t$} as the stock index's declining changing pattern, the right tail of the distribution of {$L_t$} represents extreme risk.} for the chosen index on day $t$ and $p_t$ denotes the closing price on day $t$. A statistical summary for these loss samples is shown in Table \ref{tab:summary}, from which it can be easily seen that the values of sample skewness and kurtosis deviate from 0. Figure \ref{fig:NormalQQplot} displays the Q-Q plots of $L_t$ of the six selected indices based on normal distribution, which further implies that it is inappropriate to adopt normal distribution to fit the distributions of {$L_t$} for all six indexes. More precisely, the stock market indexes show higher peak and fat tail phenomenon on both sides.

\begin{table}[htbp!]\centering
\setlength{\abovecaptionskip}{0cm}
\setlength{\belowcaptionskip}{0.1cm}
    \caption{Statistical summary of six stock indices.}
    \label{tab:summary}
    \begin{tabular}{ccccccc}
    \toprule
    INDEX & Mean       & Max.    & Min.     & Std.   & Skewness & Kurtosis \\ \midrule
    SPX   & -0.029808  & 12.7653 & -10.9572 & 1.3468 & 0.64053  & 15.537   \\
    FCHI  & -0.020376  & 13.0984 & -8.8678  & 1.3732 & 0.45019  & 9.8693   \\
    FTSE  & -0.0061143 & 11.5125 & -8.6668  & 1.1584 & 0.49841  & 12.0925  \\
    N225  & -0.012221  & 12.1111 & -13.2346 & 1.5007 & 0.46726  & 12.0698  \\
    SHZ   & 0.021988   & 9.2562  & -9.0342  & 1.4577 & 0.51118  & 8.0791   \\
    GDAXI & -0.02459   & 13.0548 & -10.6851 & 1.356  & 0.29072  & 10.5407  \\ \bottomrule
    \end{tabular}
    \centering
\end{table}

Therefore, the importance and feature of extreme tails should be taken into consideration when fitting the marginal distributions of {$L_t$}, especially in the field of quantitative analysis in economics and finance. The Generalized Pareto Distribution (GPD) in Extreme Value Theory (EVT) is a useful tool for estimating the tail of asset log-{loss} distributions, wherein only the data in the tail is used to calibrate the parameters of this distribution. Here we briefly describe the so called Peaks-Over-Threshold (POT) model based on an asymptotic approximation of the excess {loss} distribution by the GPD. Interested readers may refer to \cite{embrechts2013modelling} for more detailed studies.

 \begin{table}[htbp!]\centering
 \setlength{\abovecaptionskip}{0cm}
\setlength{\belowcaptionskip}{0.1cm}
    \caption{Parameters of fitted GPD for both tails of six stock market indices.}
     \label{tab:GDPtail}
        \begin{tabular}{cccccccc}
    \toprule
& & SPX & FCHI & FTSE & N225 & SHZ  & GDAXI\\ \midrule
\multicolumn{1}{l}{\multirow{2}{*}{Right Tail}} & $\beta_R$ & 0.198727 & 0.16111  & 0.135657  & 0.254086  & 0.0537411  & 0.16162  \\
\multicolumn{1}{l}{}& $\xi_R$  &1.00664& 0.887986 & 0.805582 & 0.876857 & 1.2588 & 0.825455 \\
\multirow{2}{*}{Left Tail} & $\beta_L$ &0.327653& 0.12806  & 0.141441  & 0.21226  & 0.10067  & 0.134978  \\
    & $\xi_L$  & 0.665329 & 0.778391 & 0.668388 & 0.724524 & 0.858132 & 0.766569 \\ \bottomrule
    \end{tabular}
    \end{table}

\begin{figure}[htbp!]
        \centering
        \subfigure[]{
        \includegraphics[width=7cm]{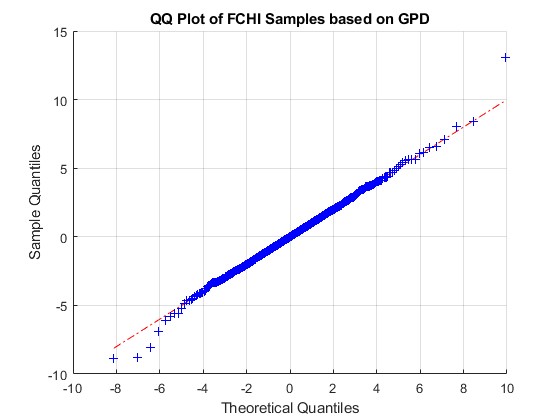}
        }
        \quad
        \subfigure[]{
        \includegraphics[width=7cm]{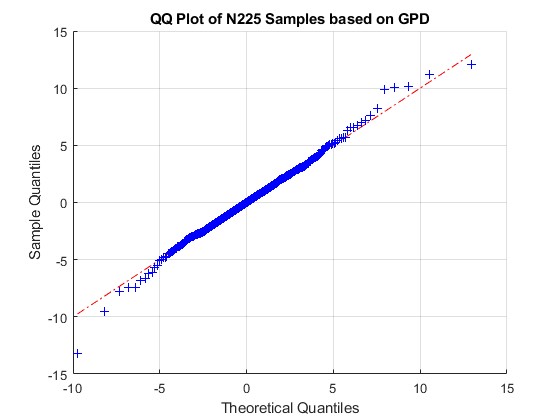}
        }
        \quad
        \subfigure[]{
        \includegraphics[width=7cm]{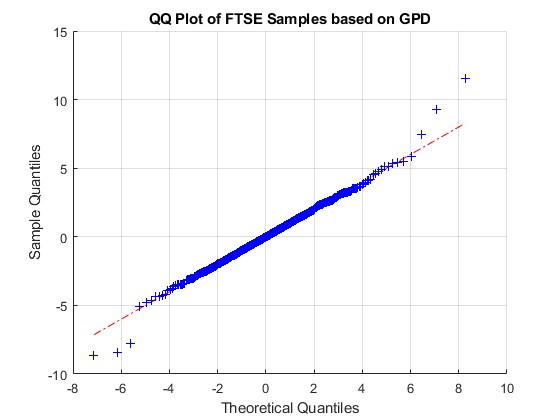}
        }
        \quad
        \subfigure[]{
        \includegraphics[width=7cm]{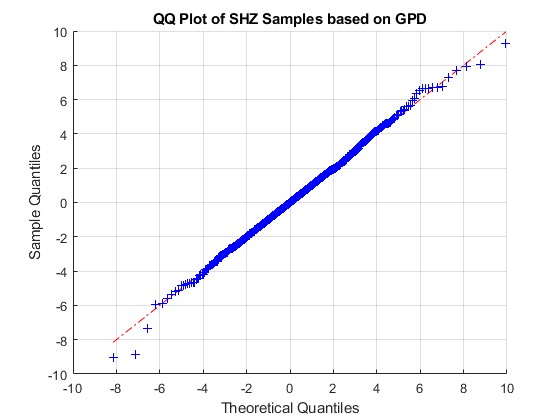}
        }
        \quad
        \subfigure[]{
        \includegraphics[width=7cm]{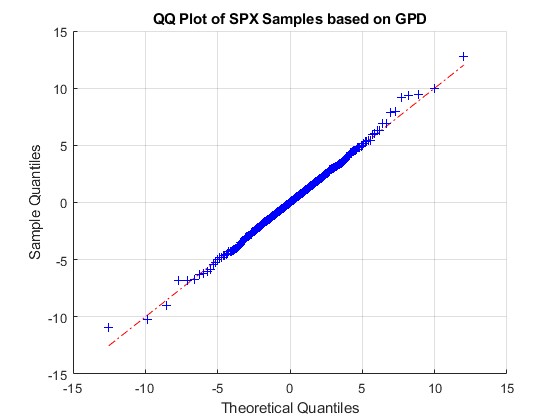}
        }
        \quad
        \subfigure[]{
        \includegraphics[width=7cm]{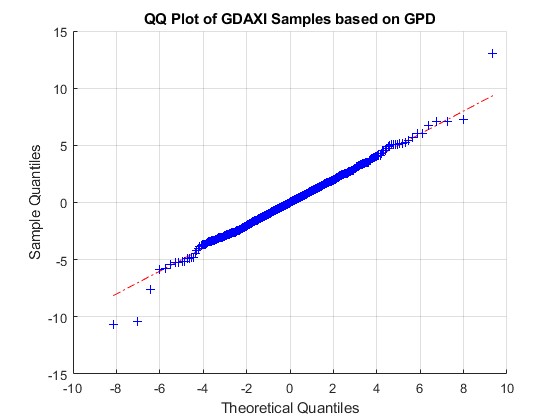}
        }
        \caption{QQ plots of the log-loss samples based on (\ref{GPDdistthree}).}
        \label{fig:QQplot}
        \end{figure}

By definition, the excess distribution of a r.v. ${X}$, representing the risk factor log-{losses}, beyond a fixed threshold $u$, has the following c.d.f.:
\begin{equation} \nonumber
    F_u(x)=\mathbb{P}(X-u \leq x \mid X>u), \quad x \in\mathbb{R}_+,
\end{equation}
where $u<u_R$, and $u_R$ is the right endpoint of the c.d.f. ${F}$ of ${X}$. When ${u}$ is relatively large, an appropriate approximation of ${F}_{u}$ is given by the GPD, say $G_{\xi, \beta}$, where the scale parameter $\beta > 0$ and the shape parameter $\xi \neq 0$ are estimated from the real dataset of excess {losses}. More precisely, the c.d.f. of GPD has the following analytical form:
\begin{equation} \nonumber
    G_{\xi, \beta}(x)=1-\left(1+\xi \frac{x}{\beta}\right)^{-1 / \xi}, \quad x \in\mathbb{R}_+.
\end{equation}
In this work, we employ the GPD to model the upper and lower tails of {$L_t$}, while using empirical distribution for modelling the central portion of the {loss} samples. Hence, the marginal distributions of the indices are assumed to be
\begin{equation}\label{GPDdistthree}
    F(x)= \begin{cases}\frac{N_{u_L}}{N}\left(1-\xi_L \frac{x-u_L}{\beta_L}\right)^{-1 / \xi_L}, & x<u_L, \\ {\rm Ecdf}(x), & u_L \leqslant x \leqslant u_R, \\ 1-\frac{N_{u_R}}{N}\left(1+\xi_R \frac{x-u_R}{\beta_R}\right)^{-1 / \xi_R} ,& x>u_R,\end{cases}
\end{equation}
where $u_L$ and $u_R$ are the upper and the lower thresholds, $N_{u_R}$ and $N_{u_L}$  are the sizes of the samples that are greater than $u_R$ and smaller than $u_L$, respectively, ${\rm Ecdf}(x)$ denotes the empirical distribution function of the {loss} samples in range $[u_L,u_R]$, $\xi_R$ and $\beta_R$, and $\xi_L$ and $\beta_L$ are referred to, respectively, as the shape and scale parameters of the GPD on the right and left tails. As suggested by \cite{dumouchel1983estimating} and \cite{koliai2016extreme}, we use 0.1 for selecting the samples both in the lower tail and upper tail to model the tails of the marginal distributions with GPD. The parameters of the fitted GPD distributions of the six market indices are shown in Table \ref{tab:GDPtail}, and the respective Q-Q plots based on the fitted GPD are provided in Figure \ref{fig:QQplot}. These figures exhibit good fitness of GPD on the distributions of {$L_t$} for all six stock market indices.

\begin{table}[htbp!]\centering
\setlength{\abovecaptionskip}{0cm}
\setlength{\belowcaptionskip}{0.1cm}
    \caption{Parameters of selected copulas for five paired stock indices.}
    \label{tab:tcop-para}
    \begin{tabular}{cccccc}
    \toprule
    INDEX & Copula Family& $\rho$ & $\nu$ & Kendall $\tau$ & $\lambda_U$($\lambda_L$) \\ \midrule
    N225 & Student \emph{t} & 0.5248908&4.05923&0.3517877&0.2641603                          \\
    FTSE & Student \emph{t} & 0.1949571& 3.877734&0.1249135&0.1310733                           \\
    FCHI & Student \emph{t} & 0.1807672&4.784949& 0.1157161&0.0937587                        \\
    GDAXI & Student \emph{t} & 0.177725&5.291066&0.1137475&0.0788018                         \\
    SHZ & Student \emph{t}  & 0.1848658&7.839447&0.1183701&0.0362582                          \\ \bottomrule
    \end{tabular}
    \end{table}

Next, we aim to investigate the dependence structures among various paired stock market indices via copulas. The data are organized in pairs as $(L^X_{t},L_{t-1}^{S\&P500})$, where $X$ denotes one of the aforementioned stock indices except S\&P 500. When at least one market in the paired markets is closed, the data of this day is discarded. We apply the command ``BiCopSelect" in R package ``\texttt{VineCopula}" to select the most appropriate bivariate copula from a wide range class of copulas (including Archimedean copulas like Gumbel copula and elliptical copulas like the Normal copula and the Student \emph{t} copula) based on AIC for each of paired stock indices. The corresponding copula parameters are estimated by applying the method of maximum likelihood estimation. For all paired samples $(L^{\rm FCHI}_{t},L_{t-1}^{\rm S\&P 500})$, $(L^{\rm N225}_{t},L_{t-1}^{\rm S\&P 500})$, $(L^{\rm SHZ}_{t},L_{t-1}^{\rm S\&P 500})$, $(L^{\rm GDAXI}_{t},L_{t-1}^{\rm S\&P 500})$ and $(L^{\rm FTSE}_{t}, L_{t-1}^{\rm S\&P 500})$, the selected copulas are Student \emph{t} copulas defined in (\ref{eq:tcopula}). The respective estimated parameters are reported in Table \ref{tab:tcop-para}\footnote{In Table  \ref{tab:tcop-para}, $\lambda_{U}$ and $\lambda_{L}$, which respectively denote the upper and lower tail dependence coefficients, are defined as
\begin{equation} \nonumber
    \lambda_{U} = \lim_{u \to 1} \frac{1-2u+C(u,u)}{1-u} \quad \text{and} \quad \lambda_{L} = \lim_{u \to 0} \frac{C(u,u)}{u},
\end{equation}
where $C$ is the given copula. For Student \emph{t} copula, it holds that $\lambda_{L}=\lambda_{U}$. Tail dependence coefficients are measures of the probability of extreme values occurring for one r.v. given that another r.v. is beyond an extreme value. Interested readers are referred to Chapter 4.4.4 of \cite{Denuit2005}.}. Based on the respective marginal distributions and the selected copulas, a range of systemic risk measures\footnote{See Appendix \ref{appendix:systemic} for their detailed expressions. Besides, we use ``E'' to denote the Expectation.} are calculated in Table \ref{tab:riskmes1} (with confidence levels $\alpha=0.95$ and $\beta=0.95$), Table \ref{tab:riskmes2} (with confidence levels $\alpha=0.95$ and $\beta=0.97$) and Table \ref{tab:riskmes3} (with confidence levels $\alpha=0.97$ and $\beta=0.95$), and the rankings of each stock index affected by  S\&P 500 based on each of these systemic risk measures are provided in parentheses.

\begin{table}[htbp!] \centering
\setlength{\abovecaptionskip}{0cm}
\setlength{\belowcaptionskip}{0.1cm}
    \caption{Values of some systemic risk measures of the five paired stock indices. $\alpha = 0.95$, $\beta = 0.95$.}
    \label{tab:riskmes1}
    \begin{threeparttable}
    \begin{tabular}{@{}ccccccc@{}}
    
        \toprule
              & ${\rm E}$ & ${\rm VaR}$          & ${\rm ES}$           & ${\rm CoVaR}$            & ${\rm MES}$             & ${\rm JMES}$           \\ \midrule
        N225  & -0.0113         & 2.2407(4)            & 3.6426(4)            & 6.0791(5)                & 1.8926(5)               & 4.2917(5)              \\
        FTSE  & -0.0062         & 1.8134(1)            & 2.8373(1)            & 3.7817(1)                & 0.5554(2)               & 3.2915(1)              \\
        FCHI  & -0.0202         & 2.1457(3)            & 3.3292(3)            & 4.1700(3)                & 0.5881(3)               & 3.8493(3)              \\
        GDAXI & -0.0244         & 2.1193(2)            & 3.2206(2)            & 3.9084(2)                & 0.5504(1)               & 3.6930(2)              \\
        SHZ   & 0.0221          & 2.4226(5)            & 3.8034(5)            & 4.4100(4)                & 0.6616(4)               & 4.2688(4)              \\ \midrule
              & ${\rm\Delta CoVaR}$    & ${\rm\Delta MES}$    & ${\rm\Delta JMES}$   & ${\rm \Delta_m CoVaR}$ & ${\rm \Delta_m MES}$    & ${\rm \Delta_m JMES}$  \\ \midrule
        N225  & 3.8384(5)              & 1.9040(5)            & 0.6491(5)            & 3.2344(5)                & 0.6184(5)               & 1.3573(5)              \\
        FTSE  & 1.9683(2)              & 0.5615(1)            & 0.4542(1)            & 1.7496(3)                & 0.4382(2)               & 0.4043(2)              \\
        FCHI  & 2.0244(4)              & 0.6083(3)            & 0.5200(4)            & 1.7873(4)                & 0.4978(4)               & 0.4310(3)              \\
        GDAXI & 1.7891(1)              & 0.5748(2)            & 0.4724(3)            & 1.5692(1)                & 0.4501(3)               & 0.4015(1)              \\
        SHZ   & 1.9873(3)              & 0.6395(4)            & 0.4654(2)            & 1.6557(2)                & 0.4312(1)               & 0.4429(4)              \\ \midrule
              & ${\rm \Delta^R CoVaR}$ & ${\rm \Delta^R MES}$\tnote{*} & ${\rm\Delta^R JMES}$ & ${\rm \Delta^R_m CoVaR}$ & ${\rm \Delta^R_m MES}$ & ${\rm \Delta^R_m JMES}$ \\ \midrule
        N225  & 1.7131(5)              & 168.3012(5)         & 0.1782(5)            & 1.1370(5)                & 2.5352(2)               & 0.1684(5)              \\
        FTSE  & 1.0854(4)              & 91.2595(4)          & 0.1601(4)            & 0.8610(4)                & 2.6751(3)               & 0.1536(4)              \\
        FCHI  & 0.9435(3)              & 30.0977(3)          & 0.1562(3)            & 0.7501(3)                & 2.7420(5)               & 0.1485(3)              \\
        GDAXI & 0.8442(2)              & 23.5841(1)          & 0.1467(2)            & 0.6708(2)                & 2.6957(4)               & 0.1388(2)              \\
        SHZ   & 0.8203(1)              & 28.9422(2)           & 0.1224(1)            & 0.6011(1)                & 2.0256(1)               & 0.1124(1)              \\ \bottomrule
        \end{tabular}
        \begin{tablenotes}
            \footnotesize 
            \item[*] In this column, the absolute values of ${\rm \Delta^R MES}$ are presented, and the corresponding ranks are based on the absolute values as well.
        \end{tablenotes}
        \end{threeparttable}
    \end{table}
    \begin{table}[htbp!] \centering
\setlength{\abovecaptionskip}{0cm}
\setlength{\belowcaptionskip}{0.1cm}
    \caption{Values of some systemic risk measures of the five paired stock indices. $\alpha = 0.95$, $\beta = 0.97$.}
    \label{tab:riskmes2}
    \begin{threeparttable}
    \begin{tabular}{@{}ccccccc@{}}
        \toprule
              & ${\rm E}$ & ${\rm VaR}$          & ${\rm ES}$           & ${\rm CoVaR}$            & ${\rm MES}$             & ${\rm JMES}$           \\ \midrule
        N225  & -0.0113 & 2.8111(4) & 4.4073(4) & 7.3201(5) & 1.8926(5) & 4.9507(5)              \\
        FTSE  & -0.0062 & 2.2815(1) & 3.3789(1) & 4.5404(1) & 0.5554(2) & 3.7745(1)              \\
        FCHI  & -0.0202 & 2.6743(3) & 3.9594(3) & 5.0707(3) & 0.5881(3) & 4.4331(3)              \\
        GDAXI & -0.0244 & 2.6109(2) & 3.8070(2) & 4.7389(2) & 0.5504(1) & 4.2431(2)              \\
        SHZ   & 0.0221 & 3.0993(5) & 4.5185(5) & 5.3395(4) & 0.6616(4) & 4.9410(4)              \\ \midrule
              & ${\rm\Delta CoVaR}$    & ${\rm\Delta MES}$    & ${\rm\Delta JMES}$   & ${\rm \Delta_m CoVaR}$ & ${\rm \Delta_m MES}$    & ${\rm \Delta_m JMES}$  \\ \midrule
        N225  & 4.5090(5) & 1.9040(5) & 0.5434(5) & 3.8088(5) & 1.3573(5) & 0.5183(5)              \\
        FTSE  & 2.2589(3) & 0.5615(1) & 0.3956(1) & 2.0160(3) & 0.4043(2) & 0.3827(1)              \\
        FCHI  & 2.3964(4) & 0.6083(3) & 0.4737(4) & 2.1278(4) & 0.4310(3) & 0.4548(4)             \\
        GDAXI & 2.1280(1) & 0.5748(2) & 0.4360(3) & 1.8780(1) & 0.4015(1) & 0.4168(3)              \\
        SHZ   & 2.2402(2) & 0.6395(4) & 0.4225(2) & 1.8806(2) & 0.4429(4) & 0.3933(2)              \\ \midrule
              & ${\rm \Delta^R CoVaR}$ & ${\rm \Delta^R MES}$\tnote{*} & ${\rm\Delta^R JMES}$ & ${\rm \Delta^R_m CoVaR}$ & ${\rm \Delta^R_m MES}$ & ${\rm \Delta^R_m JMES}$ \\ \midrule
        N225  & 1.6040(5) & 168.3013(5) & 0.1233(5) & 1.0847(5) & 2.5352(2)  & 0.1169(5)              \\
        FTSE  & 0.9901(4) & 91.2595(4) & 0.1171(3) & 0.7986(4) & 2.6751(3) & 0.1128(3)              \\
        FCHI  & 0.8961(3) & 30.0977(3) & 0.1196(4) & 0.7230(3) & 2.7420(5) & 0.1143(4)              \\
        GDAXI & 0.8150(2) & 23.5841(1) & 0.1145(2) & 0.6564(2) & 2.6957(4) & 0.1089(2)              \\
        SHZ   & 0.7228(1) & 28.9422(2) & 0.0935(1) & 0.5437(1) & 2.0256(1) & 0.0865(1)              \\ \bottomrule
        \end{tabular}
        \begin{tablenotes}
            \footnotesize 
            \item[*] In this column, the absolute values of ${\rm \Delta^R MES}$ are presented, and the corresponding ranks are based on the absolute values as well.
        \end{tablenotes}
        \end{threeparttable}
    \end{table}
    \begin{table}[htbp!] \centering
\setlength{\abovecaptionskip}{0cm}
\setlength{\belowcaptionskip}{0.1cm}
    \caption{Values of some systemic risk measures of the five paired stock indices. $\alpha = 0.97$, $\beta = 0.95$.}
    \label{tab:riskmes3}
    \begin{threeparttable}
    \begin{tabular}{@{}ccccccc@{}}
        \toprule
              & ${\rm E}$ & ${\rm VaR}$          & ${\rm ES}$           & ${\rm CoVaR}$            & ${\rm MES}$             & ${\rm JMES}$           \\ \midrule
        N225  & -0.0113 & 2.2407(4) & 3.6426(4) & 7.1166(5) & 2.2773(5) & 4.5852(5)              \\
        FTSE  & -0.0062 & 1.8134(1) & 2.8373(1) & 4.3049(1) & 0.6654(2) & 3.4523(1)              \\
        FCHI  & -0.0202 & 2.1457(3) & 3.3292(3) & 4.7199(3) & 0.7017(3) & 4.0234(3)              \\
        GDAXI & -0.0244 & 2.1193(2) & 3.2206(2) & 4.3863(2) & 0.6519(1) & 3.8471(2)              \\
        SHZ   & 0.0221 & 2.4226(5) & 3.8034(5) & 4.8543(4) & 0.7680(4) & 4.4027(4)              \\ \midrule
              & ${\rm\Delta CoVaR}$    & ${\rm\Delta MES}$    & ${\rm\Delta JMES}$   & ${\rm \Delta_m CoVaR}$ & ${\rm \Delta_m MES}$    & ${\rm \Delta_m JMES}$  \\ \midrule
        N225  & 4.8759(5) & 2.2886(5) & 0.9425(5) & 4.2719(5) & 1.7419(5) & 0.9119(5)              \\
        FTSE  & 2.4915(3) & 0.6716(1) & 0.6150(2) & 2.2729(3) & 0.5143(2) & 0.5989(2)              \\
        FCHI  & 2.5743(4) & 0.7220(3) & 0.6941(4) & 2.3372(4) & 0.5446(3) & 0.6719(4)              \\
        GDAXI & 2.2670(1) & 0.6763(2) & 0.6265(3) & 2.0471(1) & 0.5030(1) & 0.6041(3)              \\
        SHZ   & 2.4317(2) & 0.7459(4) & 0.5993(1) & 2.1000(2) & 0.5493(4) & 0.5650(1)              \\ \midrule
              & ${\rm \Delta^R CoVaR}$ & ${\rm \Delta^R MES}$\tnote{*} & ${\rm\Delta^R JMES}$ & ${\rm \Delta^R_m CoVaR}$ & ${\rm \Delta^R_m MES}$ & ${\rm \Delta^R_m JMES}$ \\ \midrule
        N225  & 2.1761(5) & 202.3001(5) & 0.2587(5) & 1.5017(5)&  3.2536(2) & 0.2483(5)              \\
        FTSE  & 1.3740(4) & 109.1418(4) & 0.2167(4) & 1.1185(4) & 3.4032(4) & 0.2099(4)              \\
        FCHI  & 1.1998(3) & 35.7183(3) & 0.2085(3) & 0.9809(3)  & 3.4648(5) & 0.2005(3)              \\
        GDAXI & 1.0697(2) & 27.7490(1) & 0.1945(2) & 0.8751(2) & 3.3772(3) & 0.1863(2)              \\
        SHZ   & 1.0037(1) & 33.7578(2) & 0.1576(1) & 0.7625(1) & 2.5122(1) & 0.1472(1)              \\ \bottomrule
        \end{tabular}
        \begin{tablenotes}
            \footnotesize 
            \item[*] In this column, the absolute values of ${\rm \Delta^R MES}$ are presented, and the corresponding ranks are based on the absolute values as well.
        \end{tablenotes}
        \end{threeparttable}
    \end{table}

According to the values displayed in Tables \ref{tab:riskmes1}-\ref{tab:riskmes3}, the following observations can be noted:

\begin{enumerate}[(i)]
    \item The systemic risk measures including ${\rm CoVaR}$, ${\rm MES}$ and ${\rm JMES}$ are all greater than their respective unconditional risk measures such as ${\rm VaR}$, ${\rm E}$ and ${\rm ES}$. This is consistent with our expectation that there is significant risk spillover from the US to the other stock markets, verifying the existence of some potential systemic risk in the global financial market. In addition, the values in Tables \ref{tab:riskmes1}-\ref{tab:riskmes3} reveal that JMES and its corresponding difference and ratio contribution risk measures exhibit relatively smaller variations compared to CoVaR and MES, indicating their more stability in capturing changes in positive dependence. 
    \item For the parameters $\alpha$ and $\beta$, we observe that when one remains constant while the other increases, the value of JMES generally rises. This is consistent with the conclusion drawn in Theorem 3.9 by noting that the $t$-copula satisfies the condition in Theorem 3.9.
    \item The rankings of different stock indices given stress event of S\&P 500 under ${\rm CoVaR}$ are consistent with the ones under ${\rm JMES}$ (for all chosen confidence levels), which indicates that these two conditional risk measures hold a consistent viewpoint regarding the quantification of the importance of systemic risk induced by S\&P 500. This implies that the usage of JMES may perform competitively with CoVaR in stock markets.
    \item Comparing Table \ref{tab:riskmes1} and Table \ref{tab:riskmes2}, we observe that when the value of $\alpha$ remains constant while the value of $\beta$ increases, the rankings of difference contribution risk measures remain largely unchanged, while the rankings of ratio contribution risk measures undergo significant changes. Conversely, comparing Table \ref{tab:riskmes1} and Table \ref{tab:riskmes3}, we find that when the value of $\alpha$ increases while the value of $\beta$ remains constant, the rankings of ratio contribution risk measures remain largely unchanged, while the rankings of difference contribution risk measures undergo significant changes. With an increase in the value of $\beta$, the benchmarks of contribution risk measures also change, leading to greater variations in the ranking of ratio contribution risk measures.
    \item The values of various absolute risk contribution measures ${\rm\Delta CoVaR}$, ${\rm\Delta MES}$, ${\rm\Delta JMES}$, ${\rm \Delta_m CoVaR}$, ${\rm \Delta_m MES}$ and ${\rm \Delta_m JMES}$ of the selected indices are provided in the second row, from which it is not easy to explain the rankings of these measures since the rankings are inconsistent in general. The values of these measures are influenced by two factors: the spillover effect of S\&P 500 on the other stock indices, and the scale of the corresponding distribution of each stock index. This suggests that the contribution measures in the second row are more suitable for cross-time comparison for a selected object, since the scale of a stock index may vary little in a short time, rather than cross-country comparison. 
    \item It can be noted that the stock index Nikkei 225 in Japan is mostly affected by the fluctuations of S\&P 500 under all contribution measures and all chosen significance levels. The strong spillover effect on the Nikkei 225 may be caused by the fact that Japan's trade structure is closely connected with the US. In particular, the data provided by the Ministry of Finance showcases the value of Japan's exports to the US in 2022, which amounted to 18,255 billion JPY. This represents 18.6\% of Japan's total exports, indicating the significance of the US as a trading partner for Japan. The high volume of exports suggests that Japanese companies heavily rely on the US market to sell their goods and services. Additionally, Japan's status as the largest foreign holder of Treasury Securities underscores its financial ties with the US. Therefore, a collapse of the Western markets had a severe negative influence on Japanese exports (\cite{kawai2011japan}). Regarding the complex situation of trade between Japan and the US, the detailed reasons are beyond the scope of this paper. Interested readers are referred to   \cite{karolyi2003financial} and \cite{li2015modelling} for further insights.
    \item Based on the values of ${\rm \Delta^R CoVaR}$, ${\rm \Delta^R_m CoVaR}$, ${\rm \Delta^R JMES}$ and ${\rm \Delta^R_m JMES}$, in a cross-country comparison, the spillover effect of S\&P 500 on the Shanghai Composite Index appears to be the weakest, while its spillover effect on the Nikkei 225 Index is still the strongest. The weak spillover effect on the SPX could be attributed to the relatively strict regulations imposed on the Chinese capital market, which create a ``natural protective barrier'' for the Chinese stock market, providing some short-term resilience against risks.
    \item For the contribution ratio measures ${\rm \Delta^R CoVaR}$, ${\rm \Delta^R_m CoVaR}$, ${\rm \Delta^R JMES}$, and ${\rm \Delta^R_m JMES}$, the rankings of stock indices are substantially consistent with the rankings of their corresponding tail dependence coefficients corresponding to the estimated copula. This matches with economic intuition since higher upper-tail correlation coefficients can indicate a stronger risk spillover effect of S\&P 500 on other stock indices, meaning that when S\&P 500's {losses} are at their risk level, the probability of potential {losses} in other stock indices increases. This aligns perfectly with the good performances and wide applications of the four risk measures mentioned earlier. 
    \item The data in column $\Delta^{\rm R}$MES of the tables seems incongruent in terms of scale with the other columns. In general, using $\Delta^{\rm R}$MES as a practical risk measure becomes precarious when the expectation of risk approaches zero. This is because dividing by the expectation can inflate (or reduce) the value of $\Delta^{\rm R}$MES when the expectation of risk is negative. Additionally, if the expectation of risk is negative, interpreting the value of $\Delta^{\rm R}$MES becomes more challenging. We provide the values of $\Delta^{\rm R}$MES here solely to maintain consistency and continuity in the table data, and make comparisons with with other systemic risk contribution measures such as $\Delta^{\rm R}$CoVaR and $\Delta^{\rm R}$JMES.
\end{enumerate}

\section{Concluding remarks}\label{conclusion}
We introduced three types of JMES-based measures to quantify the (relative) spillover effect of the risk $X$ to the other one $Y$ when it has been known that $Y$ is in distress at some fixed confidence level. The ${\rm JMES}_{\alpha,\beta}[Y|X]$ measure contains ${\rm ES}_{\beta}[Y]$ and ${\rm MES}_{\beta}[Y|X]$ as two special cases. The other two contribution measures, namely ${\rm\Delta JMES}_{\alpha,\beta}[Y|X]$ and ${\rm\Delta^R JMES}_{\alpha,\beta}[Y|X]$, are new compared with the existing ones in the literature. We also analyzed some basic properties of these new measures and established their quantile-based integrals. We have demonstrated that the JMES is neiter identifiable nor elicitable, leading to the conclusion that JMES cannot be backtested alone in practice. We further studied sufficient conditions for comparing these new measures of paired risks or two different bivariate risks in terms of some stochastic orderings and some mild conditions imposed on copula function. Some numerical examples are also provided to validate our theoretical findings. Finally, a real application in analyzing the spillover effect among six selected stock market indices is presented to show the performances of these new measures compared with some traditional ones.

Note that the results of Section \ref{pairedrisks} are built upon the assumption that the copula $C$ is symmetric. Some extensions can be implemented for the situation when $C$ is asymmetric under appropriate conditions, which are omitted here due to the length of the paper. Besides, as a natural extension, one can consider replacing the single risk $X$ with a group of interested risks, say $X_1,\ldots,X_n$, $n\geq2$, to define JMES-based measures under multivariate risk scenarios; see \cite{bernardi2018conditional} and \cite{ortega2021stochastic}. Another interesting problem might be generalizing the JMES-based measures under the setting of CoD-risk measures by considering two joint distress events generated by $X$ and $Y$; see \cite{dhaene2022systemic}. Further, it is also an very interesting topic to investigate whether the JMES together with other risk measures such as VaR is jointly identifiable or jointly elicitable; see \cite{fissler2023backtesting} and \cite{dimitriadis2023osband}. We leave these interesting problems for future studies.

\section*{Acknowledgements}
The authors are very grateful for the insightful and constructive comments and suggestions from two anonymous reviewers, which have greatly improved the presentation of the paper. Yifei Zhang thanks the support from the Special Funds for the Cultivation of Guangdong College Students' Scientific and Technological Innovation (``Climbing Program'' Special Funds, No.~pdjh2023c20101). Yiying Zhang acknowledges the financial support from the National Natural Science Foundation of China (No.~12101336,~72311530687), GuangDong Basic and Applied Basic Research Foundation (No.~2023A1515011806), and Shenzhen Science and Technology Program (No.~JCYJ20230807093312026).

\section*{Disclosure statements}
No potential competing interests were reported by the authors.

\appendix
\section{Supplementary materials for some definitions and notations}
\subsection{Identifiability and elicitability}
In this section, we revisit certain notions and conclusions pertaining to backtesting. Most of the definitions and propositions can be found in \cite{fissler2023backtesting}. Let $\mathcal{F}$ be a class of distribution functions. A function $a: \mathbb{R}^d \rightarrow \mathbb{R}$ is called $\mathcal{F}$-integrable if $\int|a(y)| \mathrm{d} F(y)$ $<\infty$ for all $F \in \mathcal{F}$. If $a$ is $\mathcal{F}$-integrable, we define the $\operatorname{map} \bar{a}: \mathcal{F} \rightarrow \mathbb{R}, \bar{a}(F):=\int a(y) \mathrm{d} F(y)$. Denote the ``action domain'', the space of all possible forecasts, by $\mathrm{A} \subset \mathbb{R}^d$. 
\begin{definition}\label{def:identifiable}
    An $\mathcal{F}$-integrable map $\boldsymbol{V}: \mathrm{A} \times \mathbb{R}^d \rightarrow \mathbb{R}^m$ is an $\mathcal{F}$-identification function for $\boldsymbol{T}$ if $\bar{\boldsymbol{V}}(\boldsymbol{T}(F), F)=\mathbf{0}$ for all $F \in \mathcal{F}$. It is a strict $\mathcal{F}$-identification function for $\boldsymbol{T}$ if, additionally, for all $F \in \mathcal{F}$ and for all $r \in \mathrm{A}, \bar{\boldsymbol{V}}(\boldsymbol{r}, F)=\mathbf{0}$ implies that $\boldsymbol{r}=\boldsymbol{T}(F)$. $\boldsymbol{T}$ is identifiable on $\mathcal{F}$ if there is a strict $\mathcal{F}$-identification function for $\boldsymbol{T}$.
\end{definition}
\begin{definition} \label{def:elicitable}
    An $\mathcal{F}$-integrable map $S: \mathrm{A} \times \mathbb{R}^d \rightarrow \mathbb{R}$ is an $\mathcal{F}$ consistent scoring function for $\boldsymbol{T}$ if $\bar{S}(\boldsymbol{T}(F), F) \leq \bar{S}(\boldsymbol{r}, F)$ for all $\boldsymbol{r} \in A$ and for all $F \in \mathcal{F}$. It is a strictly $\mathcal{F}$-consistent scoring function for $\boldsymbol{T}$ if, additionally, equality only holds for $\boldsymbol{r}=\boldsymbol{T}(F)$. $\boldsymbol{T}$ is elicitable on $\mathcal{F}$ if there is a strictly $\mathcal{F}$-consistent scoring function for $\boldsymbol{T}$.
\end{definition}
\begin{definition}   
    A functional $\boldsymbol{T}: \mathcal{F} \rightarrow \mathrm{A}$ is said to satisfy the CxLS property (i.e. has 
   ``convex level sets at the level of distributions'') on $\mathcal{F}$ if for any $F_0, F_1 \in \mathcal{F}$ such that $\boldsymbol{T}\left(F_0\right)=$ $\boldsymbol{T}\left(F_1\right)=: \boldsymbol{t}$ it holds that $\boldsymbol{T}\left(F_\lambda\right)=\boldsymbol{t}$ for all $\lambda \in(0,1)$ with $F_\lambda:=(1-\lambda) F_0+\lambda F_1 \in \mathcal{F}$.
\end{definition}
\par The following proposition shows that either identifiability or elicitability implies the CxLS property.
\begin{proposition} \label{prop:cxls}
If $\boldsymbol{T}: \mathcal{F} \rightarrow$ A is identifiable on $\mathcal{F}$ or elicitable on $\mathcal{F}$, it satisfies the CxLS property on $\mathcal{F}$.
\end{proposition}

\subsection{Some notations of systemic risk measures}\label{appendix:systemic}
 The contribution and ratio measures employed in Tables \ref{tab:riskmes1}-\ref{tab:riskmes3} are mostly defined in \cite{Girardi2013}, \cite{Adrian2016}, \cite{dhaene2022systemic} and \cite{ZhangIME2023}. Here we only provide a comprehensive review of their expressions for ease of reference. For CoVaR- and MES-based contribution risk measures, we denote
\begin{eqnarray*}
   \Delta {\rm CoVaR}_{\alpha,\beta} [Y \vert X] &=& {\rm CoVaR}_{\alpha,\beta} [Y \vert X] - {\rm VaR}_{\beta}[Y],  \\
  {\rm \Delta_m CoVaR}_{\alpha,\beta} [Y \vert X]  &=& {\rm CoVaR}_{\alpha,\beta} [Y \vert X] - {\rm CoVaR}_{0.5,\beta} [Y \vert X],  \\
    \Delta {\rm MES}_{\alpha} [Y \vert X] &=& {\rm MES}_{\alpha} [Y \vert X] - \mathbb{E}[Y],  \\
   {\rm \Delta_m MES}_{\alpha} [Y \vert X]  &=&  {\rm MES}_{\alpha} [Y \vert X] - {\rm MES}_{0.5} [Y \vert X].
\end{eqnarray*}
Further, their corresponding contribution ratio measures can be defined as follows:
\begin{eqnarray*}
 {\rm \Delta^R CoVaR}_{\alpha,\beta} [Y \vert X]    &=&  \frac{{\rm CoVaR}_{\alpha,\beta} [Y \vert X] - {\rm VaR}_{\beta}[Y]}{{\rm VaR}_{\beta}[Y]}, \\
  {\rm \Delta^R_m CoVaR}_{\alpha,\beta} [Y \vert X]  &=& \frac{{\rm CoVaR}_{\alpha,\beta} [Y \vert X] - {\rm CoVaR}_{0.5,\beta} [Y \vert X]}{{\rm CoVaR}_{0.5,\beta} [Y \vert X]},  \\
    {\rm \Delta^R MES}_{\alpha} [Y \vert X] &=&  \frac{{\rm MES}_{\alpha} [Y \vert X] - \mathbb{E}[Y]}{\mathbb{E}[Y]},\\
    {\rm \Delta^R_m MES}_{\alpha} [Y \vert X] &=&\frac{{\rm MES}_{\alpha} [Y \vert X] - {\rm MES}_{0.5} [Y \vert X]}{{\rm MES}_{0.5} [Y \vert X]}.
\end{eqnarray*}
For the risk contribution and ratio measures corresponding to {\rm JMES}, we denote
\begin{eqnarray*}
   {\rm \Delta_m JMES}_{\alpha,\beta}[Y \vert X] &=& {\rm JMES}_{\alpha,\beta}[Y \vert X] - {\rm JMES}_{0.5,\beta}[Y \vert X],  \\
{\rm \Delta^{\rm R}_m JMES}_{\alpha,\beta}[Y \vert X]    &=& \frac{{\rm \Delta_m JMES}_{\alpha,\beta}[Y \vert X]}{{\rm JMES}_{0.5,\beta}[Y \vert X]}.
\end{eqnarray*}

\section{Proofs of main results}
\subsection{Proof of Lemma \ref{lem:int}}
\proof Noting that $F$ and $G$ are continuous and strictly increasing, $(U,V)\sim C$, and the marginals of $C$ are uniform, it follows that
    \[\mathbb{P}(U>\alpha,V>\beta)=\mathbb{P}(X>{\rm VaR}_{\alpha}[X],Y>{\rm VaR}_{\beta}[Y]).\]
    Note that, for all $y\in\mathbb{R}_+$,
    \begin{equation}\label{equa:JCODdis}
        F_{Y|X>{\rm VaR}_{\alpha}[X],Y>{\rm VaR}_{\beta}[Y]}(y)=\mathbb{P}\left(Y\leq y|X>{\rm VaR}_{\alpha}[X],Y>{\rm VaR}_{\beta}[Y]\right).
    \end{equation}
    Consider the following two cases:
    \begin{itemize}
        \item [(i)] If $y\leq{\rm VaR}_{\beta}[Y]$, then $F_{Y|X>{\rm VaR}_{\alpha}[X],Y>{\rm VaR}_{\beta}[Y]}(y)=0$.
        \item [(ii)] If $y>{\rm VaR}_{\beta}[Y]$, then
        \begin{eqnarray*}
            \lefteqn{F_{Y|X>{\rm VaR}_{\alpha}[X],Y>{\rm VaR}_{\beta}[Y]}(y)}&&\\
            \\
            &=&\frac{\mathbb{P}\left({\rm VaR}_{\beta}[Y]<Y\leq y,X>{\rm VaR}_{\alpha}[X]\right)}{\mathbb{P}\left(X>{\rm VaR}_{\alpha}[X],Y>{\rm VaR}_{\beta}[Y]\right)}\\
            \\
            &=&\frac{\mathbb{P}(Y\leq y,X>{\rm VaR}_{\alpha}[X])-\mathbb{P}(Y<{\rm VaR}_{\beta}(Y),X>{\rm VaR}_{\alpha}[X])}{\mathbb{P}(X>{\rm VaR}_{\alpha}[X],Y>{\rm VaR}_{\beta}[Y])}\\
            \\
            &=&\frac{1}{\mathbb{P}(X>{\rm VaR}_{\alpha}[X],Y>{\rm VaR}_{\beta}[Y])}\left[\mathbb{P}(Y\leq y)-\mathbb{P}(Y\leq y,X\leq {\rm VaR}_{\alpha}[X])\right.\\
            &&\qquad\qquad\qquad\qquad\left.-\mathbb{P}(Y<{\rm VaR}_{\beta}[Y])+\mathbb{P}(Y<{\rm VaR}_{\beta}[Y],X\leq {\rm VaR}_{\alpha}[X])\right]\\
            \\
            &=&\frac{G(y)-\beta-C\left(\alpha,G(y)\right)+C\left(\alpha,\beta\right)}{\overline{C}\left(\alpha,\beta\right)}\\
            &=& 1-\frac{\overline{C}\left(\alpha,G(y) \right)}{\overline{C}\left(\alpha,\beta\right)}\\
            &=&F_{V|U>\alpha,V>\beta}(G(y)).
        \end{eqnarray*}
    \end{itemize}
The fact $F_{Y|X>{\rm VaR}_{\alpha}[X],Y>{\rm VaR}_{\beta}[Y]}^{-1}(p)=G^{-1}(F_{V|U>\alpha,V>\beta}(p))$ for all $p\in(0,1)$ can be derived by considering the equivalence between events $\{ F_{Y|X>{\rm VaR}_{\alpha}[X],Y>{\rm VaR}_{\beta}[Y]}(y)\geq p\}$ and $\{ F_{V|U>\alpha,V>\beta}(G(y))\geq p\}$. When ${\rm VaR}_{\beta}[Y] \geq 0$, combining the above two cases and applying integral by parts and change-of-variable give rise to
        \begin{eqnarray} \label{eq:JMES_representation}
{\rm JMES}_{\alpha,\beta}[Y|X]
            &=&\int_{0}^{{\rm VaR}_{\beta}[Y]}(1-F_{Y|X>{\rm VaR}_{\alpha},Y>{\rm VaR}_{\beta}[Y]}(y))\dif y\nonumber\\
            &&\qquad+\int_{{\rm VaR}_{\beta}[Y]}^{+\infty}(1-F_{Y|X>{\rm VaR}_{\alpha},Y>{\rm VaR}_{\beta}[Y]}(y))\dif y\nonumber\\
            &=&\int_{0}^{{\rm VaR}_{\beta}[Y]} \dif y+\int_{{\rm VaR}_{\beta}[Y]}^{+\infty}(1-F_{Y|X>{\rm VaR}_{\alpha}[X],Y>{\rm VaR}_{\beta}[Y]}(y))\dif y\nonumber\\
            &=&{\rm VaR}_{\beta}[Y]+\int_{0}^{1}\left[F_{Y|X>{\rm VaR}_{\alpha}[X],Y>{\rm VaR}_{\beta}[Y]}^{-1}(1-p)-{\rm VaR}_{\beta}[Y] \right]\dif p\nonumber\\
            &=&\int_{0}^1F_{Y|X>{\rm VaR}_{\alpha}[X],Y>{\rm VaR}_{\beta}[Y]}^{-1}(p)\dif p\nonumber\\
            &=&\int_{0}^1G^{-1}(F_{V|U>\alpha,V>\beta}^{-1}(p))\dif p\nonumber\\
            &=&\int_{\beta}^{1} G^{-1}(t)\dif F_{V|U>\alpha,V>\beta}(t).
        \end{eqnarray}
Hence, the expression of ${\rm JMES}_{\alpha,\beta}[Y|X]$ is derived. When ${\rm VaR}_{\beta}[Y] < 0$, it can be derived that
\begin{equation} \nonumber
    \begin{split}
        {\rm JMES}_{\alpha,\beta}[Y|X]
        &=-\int_{{\rm VaR}_{\beta}[Y]}^0 F_{Y|X>{\rm VaR}_{\alpha},Y>{\rm VaR}_{\beta}[Y]}(y) \dif y \\
        &\quad+\int_{0}^{+\infty}(1-F_{Y|X>{\rm VaR}_{\alpha},Y>{\rm VaR}_{\beta}[Y]}(y))\dif y  \\
        &=\int_{{\rm VaR}_{\beta}[Y]}^{+\infty}(1-F_{Y|X>{\rm VaR}_{\alpha},Y>{\rm VaR}_{\beta}[Y]}(y))\dif y + {\rm VaR}_{\beta}[Y].
    \end{split}
\end{equation}
Repeating the steps in (\ref{eq:JMES_representation}) gives the same desired result. Hence, the proof for the result of ${\rm JMES}$ is finished. Similar arguments can be implemented on the expression and $\Delta{\rm JMES}_{\alpha,\beta}[Y|X]$ and  ${\rm\Delta^R JMES}_{\alpha,\beta}[Y|X]$. \qed
\subsection{Proof of Lemma \ref{deltapositive}}
\proof We only present the proof for $X\uparrow_{\rm RTI}Y$ since the proof for $X\uparrow_{\rm RTD}Y$ is similar. According to the proof of Lemma \ref{lem:int}, we can rewrite
\begin{equation*}
    {\rm \Delta JMES}_{\alpha,\beta}[Y|X]=\int_{0}^1\left[F_{Y_{h_{\alpha,\beta}}}^{-1}(p)-F_{Y_{h_{\beta}}}^{-1}(p)\right]\dif p
\end{equation*}
where $Y_{h_{\alpha,\beta}}:=[Y|X>{\rm VaR}_{\alpha}[X],Y>{\rm VaR}_{\beta}[Y]]$ and $Y_{h_{\beta}}:=[Y|Y>{\rm VaR}_{\beta}[Y]]$ are the distorted r.v.'s induced from $Y$ by the distortion functions $h_{\alpha,\beta}$ and $h_{\beta}$ defined in (\ref{eq:h-def}) and (\ref{eq:h0-def}), respectively.
It is sufficient to show that $F_{Y_{h_{\alpha,\beta}}}^{-1}(p)-F_{Y_{h_{\beta}}}^{-1}(p)\geq0$ for all $p\in [0,1]$. On one hand, $\overline{F}_{Y_{h_{\alpha,\beta}}}(t)=\overline{F}_{Y_{h_{\beta}}}(t)=1$ when $t\leq{\rm VaR}_{\beta}[Y]$. If $t>{\rm VaR}_{\beta}[Y]$, we have
    \begin{equation*}
        \overline{F}_{Y_{h_{\alpha,\beta}}}(t)=\frac{\mathbb{P}\{X>{\rm VaR}_{\alpha}[X],Y>t\}}{\mathbb{P}\{X>{\rm VaR}_{\alpha}[X],Y>{\rm VaR}_{\beta}[Y]\}}=\frac{\overline{C}(\alpha,G(t))}{\overline{C}(\alpha,\beta)}
        \end{equation*}
    and
        \begin{equation*}
            \overline{F}_{Y_{h_{\beta}}}(t)=\frac{\mathbb{P}\{Y>t\}}{\mathbb{P}\{Y>{\rm VaR}_{\beta}[Y]\}}=\frac{1-G(t)}{1-\beta}.
        \end{equation*}
        If $t\leq{\rm VaR}_{\beta}[Y]$, then $\overline{F}_{Y_{h_{\alpha,\beta}}}(t)=\overline{F}_{Y_{h_{\beta}}}(t)=1$.
On the other hand, since $X\uparrow_{\rm RTI}Y$, we then have $$\mathbb{P}(X>{\rm VaR}_{\alpha}[X]|Y>t)\geq\mathbb{P}(X>{\rm VaR}_{\alpha}[X]|Y>{\rm VaR}_{\beta}[Y])$$ whenever $t>{\rm VaR}_{\beta}[Y],$ that is $$\frac{\overline{C}(\alpha,G(t))}{1-G(t)}\geq \frac{\overline{C}(\alpha,\beta)}{1-\beta}.$$
Therefore, $\overline{F}_{Y_{h_{\alpha,\beta}}}(t)\geq\overline{F}_{Y_{h_{\beta}}}(t)$, for all $t\in\mathbb{R}_+$. In other words, $Y_{{h_{\beta}}}\leq_{\rm st}Y_{h_{\alpha,\beta}}$, and thus $F_{Y_{h_{\alpha,\beta}}}^{-1}(p)-F_{Y_{h_{\beta}}}^{-1}(p)\geq0$, which finishes the proof.\qed

\subsection{Proof of Theorem \ref{thm:monot-property}}
\proof  \underline{Proof of (i):} In light of (\ref{eq:JMES_representation}), taking the derivative of ${\rm JMES}_{\alpha,\beta}[Y|X]$ w.r.t. $\beta$ yields that
            \begin{equation} \nonumber
                \begin{split}
                    \frac{\partial}{\partial \beta}{\rm JMES}_{\alpha,\beta}[Y|X] &= G^{-1}(\beta) \frac{\partial_2 \overline{C}(\alpha,\beta)}{\overline{C}(\alpha,\beta)} + \int_{\beta}^{1} G^{-1}(t)\frac{\partial_2 \overline{C}(\alpha,\beta)\partial_2 \overline{C}(\alpha,t)}{\overline{C}^2(\alpha,\beta)}\dif t \\
                    &=\frac{\partial_2 \overline{C}(\alpha,\beta)}{\overline{C}(\alpha,\beta)} \left( G^{-1}(\beta) - \int_{\beta}^{1} G^{-1}(t)\dif F_{V|U>\alpha,V>\beta}(t)\right).
                \end{split}
            \end{equation}
The facts that $\partial_2 \overline{C}(\alpha,\beta) \leq 0$ and $\int_{\beta}^{1} G^{-1}(t)\dif F_{V|U>\alpha,V>\beta}(t) \geq G^{-1}(\beta)$ give rise to $$\frac{\partial}{\partial \beta}{\rm JMES}_{\alpha,\beta}[Y|X]\geq0.$$ Hence, ${\rm JMES}_{\alpha,\beta}[Y|X]$ increases w.r.t. $\beta\in[0,1)$ for any $\alpha\in[0,1)$.

\underline{Proof of (ii):} Arbitrarily choose $0 < \alpha_1 \leq \alpha_2 \leq 1$ and $t$ such that $0 < \beta \leq G(t) \leq 1$. The ${\rm TP_2}$ property of $\overline{C}$ indicates that
\begin{equation} \nonumber
                \frac{\overline{C}(\alpha_1,G(t))}{\overline{C}(\alpha_1,\beta)} \leq \frac{\overline{C}(\alpha_2,G(t))}{\overline{C}(\alpha_2,\beta)},
\end{equation}
which gives rise to
            \begin{equation} \nonumber
                \frac{\PP\Lbrb{X>{\rm VaR}_{\alpha_1}(X), Y>t}}{\PP\Lbrb{X>{\rm VaR}_{\alpha_1}(X), Y>{\rm VaR}_{\beta}(Y)}} \leq \frac{\PP\Lbrb{X>{\rm VaR}_{\alpha_2}(X), Y>t}}{\PP\Lbrb{X>{\rm VaR}_{\alpha_2}(X), Y>{\rm VaR}_{\beta}(Y)}}.
            \end{equation}
Therefore, we have
\begin{equation}\label{equa:stJMES}
  [Y \vert X>{\rm VaR}_{\alpha_1}(X),Y>{\rm VaR}_{\beta}(Y)] \leq_{\rm st} [Y \vert X>{\rm VaR}_{\alpha_2}(X),Y>{\rm VaR}_{\beta}(Y)].
\end{equation}
Taking expectations on both sides of (\ref{equa:stJMES}) obtains the desired result.\qed
\subsection{Proof of Proposition \ref{prop:comono}}
\proof By the nature of comonotonicity, there exists some r.v. $Z$ and continuous and strictly increasing functions $f_1$ and $f_2$ such that $(Y_1,Y_2)\stackrel{\rm d}{=}(f_1(Z),f_2(Z))$\footnote{For example, we can take $Z$ as a standard Uniform r.v., and  $f_1=F_{Y_1}^{-1}$ and $f_2=F_{Y_2}^{-1}$.}; see \cite{Dhaene2002}. Let $f=f_1+f_2$. Clearly, $f$ is also continuous and strictly increasing. Hence, by applying the increasing preservation of VaR and the monotonicity of $f_1$, $f_2$ and $f$ we have
\begin{eqnarray*}
 \lefteqn{{\rm JMES}_{\alpha,\beta}[Y_1|X]+{\rm JMES}_{\alpha,\beta}[Y_2|X]}&&\\
 &=&  {\rm JMES}_{\alpha,\beta}[f_1(Z)|X]+{\rm JMES}_{\alpha,\beta}[f_2(Z)|X]  \\
  &=& \mathbb{E}[f_1(Z)|X>{\rm VaR}_{\alpha}[X],f_1(Z)>{\rm VaR}_{\beta}[f_1(Z)]]\\
  &&\quad+
  \mathbb{E}[f_2(Z)|X>{\rm VaR}_{\alpha}[X],f_2(Z)>{\rm VaR}_{\beta}[f_2(Z)]]\\
  &=& \mathbb{E}[f_1(Z)|X>{\rm VaR}_{\alpha}[X],f_1(Z)>f_1({\rm VaR}_{\beta}[Z])]\\
  &&\quad+
  \mathbb{E}[f_2(Z)|X>{\rm VaR}_{\alpha}[X],f_2(Z)>f_2({\rm VaR}_{\beta}[Z])]  \\
  &=& \mathbb{E}[f_1(Z)|X>{\rm VaR}_{\alpha}[X],Z>{\rm VaR}_{\beta}[Z]]+
  \mathbb{E}[f_2(Z)|X>{\rm VaR}_{\alpha}[X],Z>{\rm VaR}_{\beta}[Z]] \\
  &=& \mathbb{E}[f_1(Z)+f_2(Z)|X>{\rm VaR}_{\alpha}[X],Z>{\rm VaR}_{\beta}[Z]]\\
  &=& \mathbb{E}[f_1(Z)+f_2(Z)|X>{\rm VaR}_{\alpha}[X],f(Z)>f({\rm VaR}_{\beta}[Z])]\\
  &=&\mathbb{E}[Y_1+Y_2|X>{\rm VaR}_{\alpha}[X],Y_1+Y_2>{\rm VaR}_{\beta}[Y_1+Y_2]]\\
  &=&{\rm JMES}_{\alpha,\beta}[Y_1+Y_2|X],
\end{eqnarray*}
which yields the desired result.\qed

\subsection{Proof of Proposition \ref{prop:nine}}
\proof 
By Proposition \ref{prop:cxls}, it suffices to show that $\operatorname{JMES}_{\alpha , \beta}$ violates the CxLS property on any class $\mathcal{F} \subseteq \mathcal{F}^0\left(\mathbb{R}^2\right)$ containing all bivariate normal distributions along with their mixtures.
\par To task this, consider $X, Y \sim N(0,1)$, which are jointly Gaussian and have a correlation $\rho \neq 0$. Let $C_{0}$ denote the bivariate normal copula with parameter $\rho$. Let $\left(X^0, Y^0\right)=(X-1, Y)$ and $\left(X^1, Y^1\right)=(X+1, Y)$ with joint distribution functions $F^0$ and $F^1$, respectively. Notice that $\left(X^0, Y^0\right)$ and $\left(X^1, Y^1\right)$ admit the same normal copula with parameter $\rho$. Moreover, let $F^{0.5}=0.5 F^0+0.5 F^1$ and let $\left(X^{0.5}, Y^{0.5}\right)$ have distribution $F^{0.5}$. On the basis of Lemma \ref{lem:int}, it can be derived that 
\begin{equation} \nonumber
    {\rm JMES}_{\alpha,\beta}[Y^0|X^0] = {\rm JMES}_{\alpha,\beta}[Y^1|X^1] = \int_{\beta}^{1} \Phi^{-1}(t) \dif \overline{h}_{\alpha,\beta}(t),
\end{equation}
where 
\begin{equation} \nonumber
    \overline{h}_{\alpha,\beta}(t) = 1-\frac{\overline{C_{0}}(\alpha,t)}{\overline{C_{0}}(\alpha,\beta)}, \quad \beta < t \leq 1.
\end{equation}
\par Now, let us compute the marginal distributions of $X^{0.5}$ and $Y^{0.5}$. For $X^{0.5}$, it can be derived that $F^{0.5}_{X} = 0.5 F^0_X+0.5 F^1_X$, where $F^0_X$, $F^{0.5}_{X}$ and $F^1_X$ are the respective distribution functions of $X^0$, $X^{0.5}$ and $X^1$. The copula of $\left(X^{0.5}, Y^{0.5}\right)$ is expressed as $C_{0.5}$. 
Based on Lemma \ref{lem:int}, we have 
\begin{equation*}
    {\rm JMES}_{\alpha,\beta}[Y^{0.5}|X^{0.5}] = \int_{\beta}^{1} \Phi^{-1}(t) \dif \overline{h}^*_{\alpha,\beta}(t),
\end{equation*}
where 
\begin{equation*}
    \overline{h}^*_{\alpha,\beta}(t) = 1-\frac{\overline{C_{0.5}}(\alpha,t)}{\overline{C_{0.5}}(\alpha,\beta)}, \quad \beta < t \leq 1.
\end{equation*}
It can be easily established that $C_0$ and $C_{0.5}$ cannot be identical by calculating that $C_0(0.5,0.5) = 0.385$ and $C_{0.5}(0.5,0.5)= 0.320$ when $\rho = 0.75$. Hence, the equality 
\begin{equation} \nonumber
    {\rm JMES}_{\alpha,\beta}[Y^{0.5}|X^{0.5}] = 0.5 \times  {\rm JMES}_{\alpha,\beta}[Y^0|X^0] + 0.5 \times {\rm JMES}_{\alpha,\beta}[Y^1|X^1] 
\end{equation}
cannot be satisfied for all $\alpha, \beta \in (0,1)$. For example, under the setting of $\alpha=0.95$, $\beta=0.95$ and $\rho = 0.75$, we have ${\rm JMES}_{\alpha,\beta}[Y^0|X^0] = 2.190$ while ${\rm JMES}_{\alpha,\beta}[Y^{0.5}|X^{0.5}] = 2.149$. Hence, the CxLS property cannot be established on JMES. Therefore, ${\rm JMES}_{\alpha,\beta}$ is neither identifiable nor elicitable. \qed

\subsection{Proof of Theorem \ref{PairRisks}}
\proof Note that the function $\overline{h}_{\alpha,\beta}(t)$ given in (\ref{eq:JMES-calclt}) is continuous in $t\in[0,1]$. Further, it can be verified that $\overline{h}_{\alpha,\beta}(t)$ is increasing and convex in $t \in (\beta,1]$ since
$$\frac{\dif \overline{h}_{\alpha,\beta}(t)}{\dif t}\overset{{\rm sign}}{=}-\partial_2 \overline{C}(\alpha,t)=1-\partial_2 C(\alpha,t)=1-\mathbb{P}(U<\alpha|V=t)=\mathbb{P}(U>\alpha|V=t)$$
is nonnegative and increasing in $t \in (\beta,1]$ because of $X \uparrow_{\rm SI} Y$. Therefore, it follows that $\overline{h}_{\alpha,\beta}(t)$ in continuous and increasing and convex in the whole support $t\in[0,1]$. By Lemma \ref{lem:icxicv}, and making use of change-of-variable and the condition that $C$ is symmetric, we immediately have
        \begin{eqnarray*}
             {\rm JMES}_{\alpha,\beta}[X|Y]
             =\int_{0}^{1}F^{-1}(t)\dif \overline{h}_{\alpha,\beta}(t)
             \leq\int_{0}^1G^{-1}(t)\dif \overline{h}_{\alpha,\beta}(t)
             ={\rm JMES}_{\alpha,\beta}[Y|X].
        \end{eqnarray*}
Hence, the proof is finished.\qed

\subsection{Proof of Theorem \ref{contributionpair}}
\proof According to the proof of Lemma \ref{lem:int}, we have
\begin{equation*}
    {\rm \Delta JMES}_{\alpha,\beta}[Y|X]=\int_{0}^1\left[F_{Y_{h_{\alpha,\beta}}}^{-1}(p)-F_{Y_{h_{\beta}}}^{-1}(p)\right]\dif p
\end{equation*}
and
\begin{equation*}
    {\rm \Delta JMES}_{\alpha,\beta}[X|Y]=\int_{0}^1\left[F_{X_{h_{\alpha,\beta}}}^{-1}(p)
    -F_{X_{h_{\beta}}}^{-1}(p)\right]\dif p,
\end{equation*}
where $X_{h_{\alpha,\beta}}:=[X|Y>{\rm VaR}_{\alpha}[Y],X>{\rm VaR}_{\beta}[X]]$ and $X_{h_{\beta}}:=[X|X>{\rm VaR}_{\beta}[X]]$ are also the distorted r.v.'s induced from $X$ by (\ref{eq:h-def}) and (\ref{eq:h0-def}) since $C$ is symmetric.

On one hand, it has been verified in the proof of Theorem \ref{PairRisks} that $h_{\alpha,\beta}(p)$ and $h_{\beta}(p)$ are both increasing and concave on $p\in [0,1]$.
Further, for all $p\in [0,1]$, it is easy to check that $h_{\alpha,\beta}(p)\geq h_{\beta}(p)$, for all $p\in [0,1]$. According to Lemma \ref{lem:ineq_delta_invF},  $X\leq_{\rm disp}Y$ implies that
\begin{equation*}
    F_{X_{h_{\alpha,\beta}}}^{-1}(p)-F_{X_{h_{\beta}}}^{-1}(p)\leq F_{Y_{h_{\alpha,\beta}}}^{-1}(p)-F_{Y_{h_{\beta}}}^{-1}(p) ,\quad \text{for all $p\in [0,1]$,}
\end{equation*}
and hence
\begin{eqnarray*}
   {\rm \Delta JMES}_{\alpha,\beta}[X|Y]&=&\int_{0}^1F_{X_{h_{\alpha,\beta}}}^{-1}(t)
   -F_{X_{h_{\beta}}}^{-1}(t)\dif t\\
   &\leq& \int_{0}^1F_{Y_{h_{\alpha,\beta}}}^{-1}(t)-F_{Y_{h_{\beta}}}^{-1}(t)\dif t={\rm \Delta JMES}_{\alpha,\beta}[Y|X],
\end{eqnarray*}
which finishes the proof.\qed

\subsection{Proof of Theorem \ref{JMESconta1a2beta}}
\proof Note that
\begin{equation*}
    {\rm \Delta JMES}_{\alpha_1,\alpha_2,\beta}[Y|X]=\int_{0}^1\left[F_{Y_{h_{\alpha_2,\beta}}}^{-1}(p)-F_{Y_{h_{\alpha_1,\beta}}}^{-1}(p)\right]\dif p
\end{equation*}
and
\begin{equation*}
    {\rm \Delta JMES}_{\alpha_1,\alpha_2,\beta}[X|Y]=\int_{0}^1\left[F_{X_{h_{\alpha_2,\beta}}}^{-1}(p)
    -F_{X_{h_{\alpha_1,\beta}}}^{-1}(p)\right]\dif p,
\end{equation*}
where $X_{h_{\alpha_i,\beta}}:=[X|Y>{\rm VaR}_{\alpha_i}[Y],X>{\rm VaR}_{\beta}[X]]$, for $i=1,2,$ are the distorted r.v.'s induced from $X$ by (\ref{eq:h-def}) since $C$ is symmetric. On one hand, it has been verified in the proof of Theorem \ref{PairRisks} that $h_{\alpha_1,\beta}(p)$ and $h_{\alpha_2,\beta}(p)$ are both increasing and concave on $p\in [0,1]$. Further, it is easy to check that $h_{\alpha_2,\beta}(p)\geq h_{\alpha_1,\beta}(p)$ for all $p\in [0,1]$, due to the assumption that $\alpha_1\leq\alpha_2$ and the ${\rm TP}_2$ property of $\overline{C}$. Thus, according to Lemma \ref{lem:ineq_delta_invF},  $X\leq_{\rm disp}Y$ implies that
\begin{equation*}
    F_{X_{h_{\alpha_2,\beta}}}^{-1}(p)-F_{X_{h_{\alpha_1,\beta}}}^{-1}(p)\leq F_{Y_{h_{\alpha_2,\beta}}}^{-1}(p)-F_{Y_{h_{\alpha_1,\beta}}}^{-1}(p) ,\quad \text{for all $p\in [0,1]$,}
\end{equation*}
and hence
\begin{eqnarray*}
   {\rm \Delta JMES}_{\alpha_1,\alpha_2,\beta}[X|Y]&=&\int_{0}^1F_{X_{h_{\alpha_2,\beta}}}^{-1}(t)
   -F_{X_{h_{\alpha_1,\beta}}}^{-1}(t)\dif t\\
   &\leq& \int_{0}^1F_{Y_{h_{\alpha_2,\beta}}}^{-1}(t)-F_{Y_{h_{\alpha_1,\beta}}}^{-1}(t)\dif t={\rm \Delta JMES}_{\alpha_1,\alpha_2,\beta}[Y|X],
\end{eqnarray*}
which yields the desired result.\qed

\subsection{Proof of Theorem \ref{thm:ratiopair}}
\proof Based on Lemma \ref{lem:int}, we have
    \begin{equation*}
        {\rm \Delta^R JMES}_{\alpha,\beta}[Y|X]=\frac{\int_{0}^1G^{-1}(t)\dif \overline{h}_{\alpha,\beta}(t)}{\int_{0}^1G^{-1}(t)\dif \overline{h}_{\beta}(t)}-1
    \end{equation*}
and
     \begin{equation*}
        {\rm \Delta^R JMES}_{\alpha,\beta}[X|Y]=\frac{\int_{0}^1F^{-1}(t)\dif \overline{h}_{\alpha,\beta}(t)}{\int_{0}^1F^{-1}(t)\dif \overline{h}_{\beta}(t)}-1.
    \end{equation*}
According to the proof of Theorem \ref{PairRisks}, it has been known that both of $\overline{h}_{\alpha,\beta}(t)$ and $\overline{h}_{\beta}(t)$ are increasing and convex on $t\in[0,1]$, and the (generalized) inverse function of $\overline{h}_{\beta}(t)$ is given by
$$\overline{h}^{-1}_{\beta}(t)=\inf \{x \in [0,1] \mid \overline{h}_{\beta}(x) \geq t\}=\beta+(1-\beta)t,\quad t\in [0,1].$$
Therefore, one has
$$\overline{h}_{\alpha,\beta}(\overline{h}^{-1}_{\beta}(t))
=1-\frac{\overline{C}(\alpha,\max\{\beta,\beta+(1-\beta)t\})}{\overline{C}(\alpha,\beta)}=1-\frac{\overline{C}(\alpha,\beta+(1-\beta)t)}{\overline{C}(\alpha,\beta)}$$
is convex in $t\in[0,1]$ due to the PDS property of $C$. Thus, the proof is finished by applying Lemma \ref{lemma:epw}.\qed

\subsection{Proof of Theorem \ref{the:twoJMES}}
\proof  \underline{Proof of (i)}: For ${\rm JMES}_{\alpha,\beta}[Y_1|X_1]$ and ${\rm JMES}_{\alpha,\beta}[Y_2|X_2]$, it is easy to note that the c.d.f.'s of $X_1$ and $X_2$ do not play a role in calculating their values but only dependent on the copula $C$ and the confidence level $\alpha$, which can be noted from Lemma \ref{lem:int} since both of measures have the same distortion function $h_{\alpha,\beta}$. In light of the proof of Theorem \ref{PairRisks}, the condition  $X_1\uparrow_{\rm SI}Y_1$ implies that $\overline{h}_{\alpha,\beta}(t)$ is increasing and convex in $t \in [0,1]$. Since $Y_1\leq_{\rm icx}Y_2$, by Lemma \ref{lem:icxicv}, we have
    \begin{equation*}
        {\rm JMES}_{\alpha,\beta}[Y_1|X_1]=\int_{0}^1G_1^{-1}(t)\dif \overline{h}_{\alpha,\beta}(t)\leq\int_{0}^1G_2^{-1}(t)\dif \overline{h}_{\alpha,\beta}(t)={\rm JMES}_{\alpha,\beta}[Y_2|X_2],
    \end{equation*}
which finishes the proof of this part.

\underline{Proof of (ii)}: According to the proof of Theorem \ref{contributionpair}, we have
\begin{equation*}
    {\rm \Delta JMES}_{\alpha,\beta}[Y_1|X_1]=\int_{0}^1\left[F_{Y_{1,h_{\alpha,\beta}}}^{-1}(p)
    -F_{Y_{1,h_{\beta}}}^{-1}(p)\right]\dif p
\end{equation*}
and
\begin{equation*}
    {\rm \Delta JMES}_{\alpha,\beta}[Y_2|X_2]=\int_{0}^1\left[F_{Y_{2,h_{\alpha,\beta}}}^{-1}(p)
    -F_{Y_{2,h_{\beta}}}^{-1}(p)\right]\dif p,
\end{equation*}
where $Y_{1,h_{\alpha,\beta}}:=[Y_1|X_1>{\rm VaR}_{\alpha}[X_1],Y_1>{\rm VaR}_{\beta}[Y_1]]$, $Y_{1,h_{\beta}}:=[Y_1|Y_1>{\rm VaR}_{\beta}[Y_1]]$, $Y_{2,h_{\alpha,\beta}}:=[Y_2|X_2>{\rm VaR}_{\alpha}[X_2],Y_2>{\rm VaR}_{\beta}[Y_2]]$ and $Y_{2,h_{\beta}}:=[Y_2|Y_2>{\rm VaR}_{\beta}[Y_2]]$ are the distorted r.v.'s induced from $Y_1$ and $Y_2$ by the distortion functions in (\ref{eq:h-def}) and (\ref{eq:h0-def}), respectively.

Since $C$ is of $X_1\uparrow_{\rm SI}Y_1$, it follows from the proof of Theorem \ref{contributionpair} that both $h_{\alpha,\beta}(p)$ and ${h}_{\beta}(p)$ are increasing and concave in $p \in [0,1]$, and $h_{\alpha,\beta}(p)\geq h_{\beta}(p)$ for all $p \in [0,1]$. By using Lemma \ref{lem:ineq_delta_invF}, the condition $Y_1 \leq_{\rm disp} Y_2$ implies that
\begin{equation*}
    F_{Y_{1,h_{\alpha,\beta}}}^{-1}(p)-F_{Y_{1,h_{\beta}}}^{-1}(p)\leq F_{Y_{2,h_{\alpha,\beta}}}^{-1}(p)-F_{Y_{2,h_{\beta}}}^{-1}(p) ,\quad \text{for all $p\in  [0,1]$,}
\end{equation*}
and hence
\begin{eqnarray*}
   {\rm \Delta JMES}_{\alpha,\beta}[Y_1|X_1]&=&\int_{0}^1\left[F_{Y_{1,h_{\alpha,\beta}}}^{-1}(t)
   -F_{Y_{1,h_{\beta}}}^{-1}(t)\right]\dif t\\
   &\geq& \int_{0}^1\left[F_{Y_{2,h_{\alpha,\beta}}}^{-1}(t)-F_{Y_{2,h_{\beta}}}^{-1}(t)\right]\dif t={\rm \Delta JMES}_{\alpha,\beta}[Y_2|X_2],
\end{eqnarray*}
which finishes the proof.

\underline{Proof of (iii)}: Based on Lemma \ref{lem:int}, we have
\begin{equation*}
    {\rm \Delta^R JMES}_{\alpha,\beta}[Y_1|X_1]=\frac{\int_{0}^1G_1^{-1}(t)\dif \overline{h}_{\alpha,\beta}(t)}{\int_{0}^1G_1^{-1}(t)\dif \overline{h}_{\beta}(t)}-1
\end{equation*}
and
 \begin{equation*}
    {\rm \Delta^R JMES}_{\alpha,\beta}[Y_2|X_2]=\frac{\int_{0}^1G_2^{-1}(t)\dif \overline{h}_{\alpha,\beta}(t)}{\int_{0}^1G_2^{-1}(t)\dif \overline{h}_{\beta}(t)}-1.
\end{equation*}
According to the proof of Theorem \ref{thm:ratiopair},  the (generalized) inverse function of $\overline{h}_{\beta}(t)$ is given by $\overline{h}^{-1}_{\beta}(t)=\beta+(1-\beta)t$, for $t\in [0,1]$, and the composite function $\overline{h}_{\alpha,\beta}(\overline{h}^{-1}_{\beta}(t))$ is convex in $t\in[0,1]$ due to the PDS property of $C$. Thus, the proof is finished in accordance with Lemma \ref{lemma:epw} and $Y_1\leq_{\rm epw}Y_2$.\qed


\subsection{Proof of Theorem \ref{thm:diff_copula_same_margin}}
\proof Let $$\overline{h}_{i,\alpha,\beta}(t)=\begin{cases}
        1-\frac{\overline{C}_i(\alpha,t)}{\overline{C}_i(\alpha,\beta)}, & ~ \mbox{for}~~ 1\geq t>\beta\\
        0, & ~ \mbox{for}~~0\leq t\leq \beta
        \end{cases},\quad\mbox{for $i=1,2$.}$$
Note that
\begin{eqnarray*}
 {\rm JMES}_{\alpha,\beta}[Y_2|X_2]-{\rm JMES}_{\alpha,\beta}[Y_1|X_1]   &=& \int_{0}^{1} G^{-1}(t)\dif \overline{h}_{2,\alpha,\beta}(t) -\int_{0}^{1} G^{-1}(t)\dif \overline{h}_{1,\alpha,\beta}(t)  \\
    &=:&  \int_{0}^{1} G^{-1}(t)\dif W(t),
\end{eqnarray*}
where $W(t)=\overline{h}_{2,\alpha,\beta}(t) -\overline{h}_{1,\alpha,\beta}(t)$. Note that the quantile function $G^{-1}(t)$ is nonnegative and increasing, and $W(t)$ is a measure function defined on $[0,1]$ such that
$$W(t)=\begin{cases}
        \frac{\overline{C}_1(\alpha,t)}{\overline{C}_1(\alpha,\beta)}
        -\frac{\overline{C}_2(\alpha,t)}{\overline{C}_2(\alpha,\beta)}, & ~ \mbox{for}~~ 1\geq t>\beta;\\
        0, & ~ \mbox{for}~~0\leq t\leq \beta.
        \end{cases}$$
Since $\overline{C}_i(\alpha,t)$ is ${\rm TP}_2$ in $(i,t)\in\{1,2\}\times[0,1]$, it follows that
\begin{equation}\label{rr2:equa1}
  \frac{\overline{C}_1(\alpha,t)}{\overline{C}_1(\alpha,\beta)}
      \leq\frac{\overline{C}_2(\alpha,t)}{\overline{C}_2(\alpha,\beta)},\quad \mbox{for any $1\geq t\geq\beta$}.
\end{equation}
Thus, we have
\begin{equation*}
  \int_t^1 \dif W(u)=W(1)-W(t)=\begin{cases}
       W(1)+\frac{\overline{C}_2(\alpha,t)}{\overline{C}_2(\alpha,\beta)}-
       \frac{\overline{C}_1(\alpha,t)}{\overline{C}_1(\alpha,\beta)}, & ~ \mbox{if}~~ 1\geq t>\beta;\\
        W(1)-W(\beta), & ~ \mbox{if}~~0\leq t\leq \beta.
        \end{cases}
\end{equation*}
Observe that $W(1)=\overline{h}_{2,\alpha,\beta}(1)-\overline{h}_{1,\alpha,\beta}(1)=0$ and $W(\beta)=\frac{\overline{C}_2(\alpha,\beta)}{\overline{C}_2(\alpha,\beta)}-
       \frac{\overline{C}_1(\alpha,\beta)}{\overline{C}_1(\alpha,\beta)}=0$. Hence, according to (\ref{rr2:equa1}), it must hold that $\int_t^1 \dif W(u)\geq0$ for all $t\in[0,1]$. Therefore, by using Lemma \ref{lemma:BP}, it follows that
\begin{equation*}
   {\rm JMES}_{\alpha,\beta}[Y_2|X_2]-{\rm JMES}_{\alpha,\beta}[Y_1|X_1]   =  \int_{0}^{1} G^{-1}(t)\dif W(t)\geq0,
\end{equation*}
which yields the desired result.\qed

\subsection{Proof of Theorem \ref{thm:dc-dm3combine}}
\proof Without loss of generality, we assume $X_1 \uparrow_{\rm SI} Y_1$. Let $(X',Y')$ be a random vector with marginal c.d.f.'s $(F_1,G_1)$ and copula $C_2$. For (i), we have $Y' \leq_{\rm icx} Y_2$. In light of Theorem \ref{the:twoJMES}, it follows that ${\rm JMES}_{\alpha,\beta}[Y'|X']\leq{\rm JMES}_{\alpha,\beta}[Y_2|X_2]$. On the flip side, the result of Theorem \ref{thm:diff_copula_same_margin} implies that ${\rm JMES}_{\alpha,\beta}[Y_1|X_1]\leq{\rm JMES}_{\alpha,\beta}[Y'|X']$. Combining these two inequalities finishes the proof. For the results in (ii) and (iii), the proofs are routine and thus omitted. \qed

\bibliographystyle{mystyle}
{\small\bibliography{SystemicRisk}}

\end{document}